\documentclass[11pt]{article}
\usepackage[pdftex]{graphicx}
\DeclareGraphicsExtensions{.pdf}
\pdfoutput=1

\usepackage{subfigure}
\usepackage{makeidx}
\usepackage{bm}
\usepackage{amsfonts,amsmath,dsfont}
\usepackage{rotating}

\newfont{\tenmsb}{msbm10 scaled\magstep1}

\let\ssection=\section\renewcommand{\section}{\setcounter{equation}{0}\ssection}

\font\BBBig=cmr10 scaled\magstep4
\font\small=cmr9

\def\parag{\hfil\break} 
\def\kikezd{\parag\underbar}
\def\Thm{\kikezd{Theorem\,}}
\def\IR{{\mathds{R}}} 
\def\IC{{\mathds{C}}} 
\def\IZ{{\mathds{Z}}} 
\def\IS{{\mathds{S}}}
\def\II{{\mathds{1}}}
\def\IT{{\mathds{T}}}
\def\IQ{{\mathds{Q}}}
\def\cQ{{\cal{Q}}}

\def\cA{{\cal{A}}}

\def\adP{{{\rm ad\,}{\cal P}}}
\def\cC{{\cal{C}}}

\def\S{{S}}

\def\smallover#1/#2{\hbox{$\textstyle\frac{#1}{#2}$}} %
\def\smallcirc{{\,\raise 0.5pt \hbox{$\scriptstyle\circ$}\,}}
\def\2{{\smallover1/2}}
\def\={{\!=\!}}
\def\Ort{{\rm O}}
\def\ort{{\rm o}}
\def\UN{{\rm U}}
\def\un{{\mathfrak{u}}}
\def\SO{{\rm SO}}
\def\SU{{\rm SU}}
\def\su{{\mathfrak{su}}}
\def\so{{\mathfrak{so}}}
\def\SP{{\rm Sp}}

\def\p{{\partial }}
\def\dAlembert{\vcenter {
    \hbox {\vrule height8pt width0.4pt depth0.0pt
           \vrule height8pt width7.2pt depth-7.6pt
           \vrule height8pt width0.4pt depth0.0pt
           \kern-8pt
           \vrule height0.4pt width8pt depth0.0pt
          \,}}}

\def\bp{\bar{\partial}}
\def\bz{{\overline{z}}}
\def\bw{{\overline{w}}}
\def\and{{\qquad\hbox{\small and}\qquad}}

\def\rot{{\rm curl\ }}

\def\bJ{{\bm{J}}}
\def\bA{{\bm{A}}}
\def\bB{{\bm{B}}}
\def\bD{{\bm{D}}}

\def\bb{{\bm{b}}}
\def\ba{{\bm{a}}}
\def\bx{{\bm{r}}}
\def\bL{{\bm{L}}}
\def\bM{{\bm{M}}}
\def\bS{{\bm{S}}}

\def\bnabla{\mbox{\boldmath$\nabla$}}

\def\cH{{\cal{E}}}
\def\cP{{\cal{P}}}
\def\cY{{\cal{Y}}}

\def\Dir{
{D\mkern-2mu\llap{{\raise+0.5pt\hbox{\big/}}}\mkern+2mu}
}  

\def\oQ{{\stackrel{\smallcirc}{Q}}}
\def\oIQ{{\stackrel{\smallcirc}{\IQ}}}
\def\oW{{\stackrel{\smallcirc}{W}}}
\def\oalpha{{\stackrel{\smallcirc}{\alpha}}}
\def\beq{\begin{equation}}
\def\eeq{\end{equation}}
\def\beqa{\begin{eqnarray}}
\def\eeqa{\end{eqnarray}}
\def\nn{\nonumber}
\def\barr{\left(\begin{array}}
\def\earr{\end{array}\right)}
\newcommand{\gk}{\mathfrak{k}}

\newcommand{\gh}{\mathfrak{h}}
\newcommand{\gt}{\mathfrak{t}}
\def\tr{{\,\rm Tr\,}}
\def\wK{{\widetilde{K}}}

\newcommand{\LG}{\mathfrak{g}}
\def\barray{\begin{array}}
\def\earray{\end{array}}
\def\nn{\nonumber}

\newcommand{\Tr}{\mathop{\mathrm{Trace}}}

\begin{document}
\setlength{\baselineskip}{15pt}

\title{{\BBBig Topology,\\[6pt]
 and (in)stability 
\\[4pt]
 of non-Abelian monopoles}
\\[16pt]
}
 
\author{
Peng-Ming ZHANG\footnote{e-mail: zhpm-at-impcas.ac.cn}\quad and\quad
Peter A. HORVATHY 
\footnote{e-mail: horvathy-at-univ-tours.fr}
\\[8pt]
Laboratoire de Math\'ematiques et de Physique Th\'eorique,
\\[4pt]Tours (France)
\\[4pt]
and
\\[5pt]
Institute of Modern Physics
\\[4pt]
Chinese Academy of Sciences, Lanzhou (China)
\\[12pt]
John RAWNSLEY\footnote{e-mail: J.Rawnsley-at-warwick.ac.uk}
\\[8pt]
Mathematics Institute
\\[4pt]
University of Warwick, Coventry (England)
\\[16pt]
}

\maketitle

\begin{abstract}
The stability problem of non-Abelian monopoles 
with respect to ``Brandt--Neri--Coleman type'' variations reduces to that of a pure gauge 
theory on the two-sphere.
Each topological sector admits exactly one stable monopole charge, and
each unstable monopole admits $\displaystyle{2\sum
 \left(2|q|-1\right)}$
negative modes, where the sum goes over the negative eigenvalues $q$ of an operator related 
to the non-Abelian charge $\IQ$ of Goddard, Nuyts and Olive. 
An explicit construction for 
 the [up-to-conjugation] unique stable charge, as well as the negative modes  of the Hessian at any other charge
 is given. The relation to loops in the residual group is explained. From the global point of view,
the instability is associated with  energy-reducing two-spheres,
which, consistently with the Morse theory, generate the homology
of the configuration space. Our spheres  are tangent to the negative modes
at the considered critical point, and 
may indicate possible decay routes of an unstable monopole as a cascade into lower lying critical points. 
\end{abstract}
\vskip15mm

\newpage
\null\newpage
\tableofcontents
\newpage

\section{Introduction: stability}

Magnetic monopoles arise as exact solutions of spontaneously broken Yang--Mills--Higgs theory \cite{tHooftPolyakov,GORPP,COL,HPAMonop,Konishi}, see Section \ref{monopoles} for an outline. It has been pointed out by Brandt and Neri \cite{BN} and emphasized by Coleman \cite{COL}, however, that 
most such solutions are unstable when
the  residual gauge group $H$ is non-Abelian. 

This review, which heavily draws on previous work of two of us with
late L. O'Raifeartaigh, \cite{HoRa, HORR88}, is devoted to the study of various aspects of ``Brandt--Neri--Coleman'' monopole instability.
Further related contributions can be found in \cite{GOCS,FrHab,NAUH}.

\subsection{Local aspects: the Hessian}

The intuitive picture behind the stability problem is that of the Morse theory \cite{Morse}. The Yang--Mills--Higgs energy functional, $\cH$, can be viewed as a ``surface" above the (infinite dimensional) ``manifold of static field  configurations'' $\cC$. 
Static solutions (like  monopoles) of the
 Yang--Mills--Higgs field equations are \textit{critical points} of $\cH$ i.e. points where
 the gradient of $\cH$ vanishes,
\beq
\delta\cH=0.
\label{1stvar}
\eeq 

\begin{figure}
\begin{center}\includegraphics[scale=0.6]{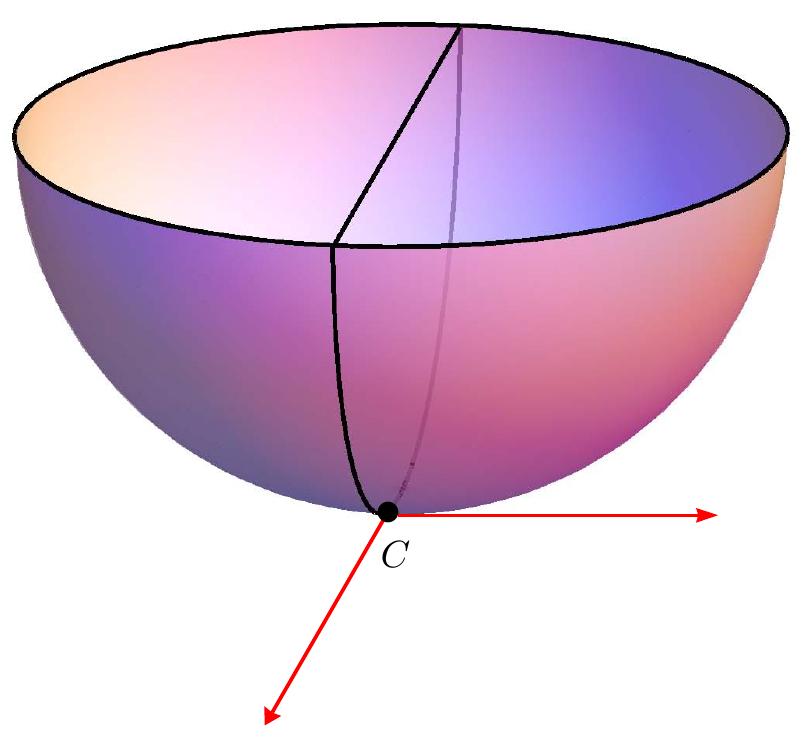}
\hskip14mm
\includegraphics[scale=0.75]{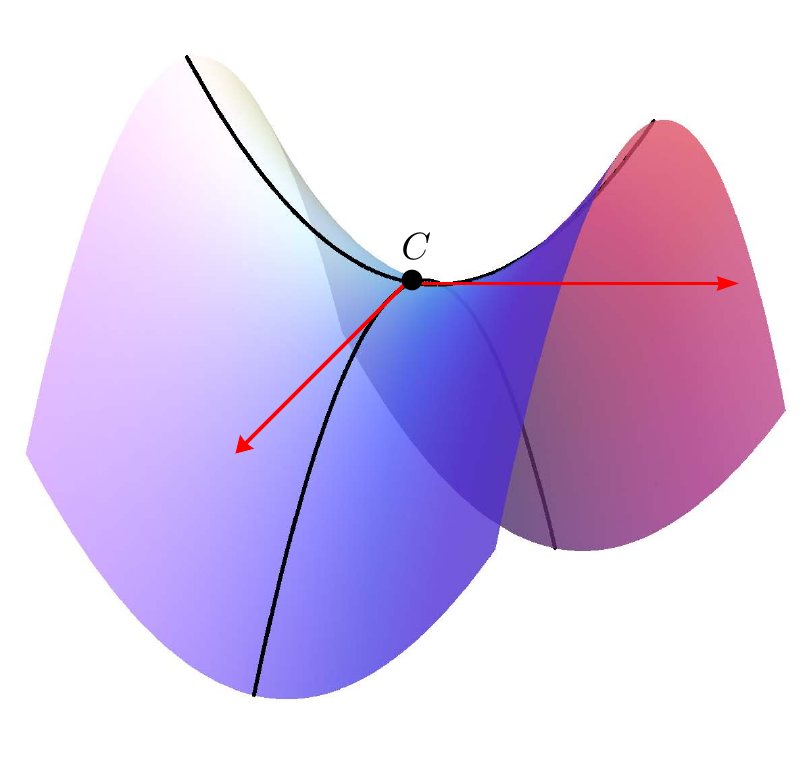}\vspace{-5mm}
\end{center}
\caption{\it The energy functional is a ``surface'' over the ``infinite dimensional manifold'' of static, finite-energy field configurations. Monopoles are critical points whose (local) stability depends on the shape of the surface in the neighborhood of the critical point. For example, the critical point on Fig. \ref{cupsaddle}a is stable, while that on Fig. \ref{cupsaddle}b is unstable.}
\label{cupsaddle}
\end{figure}

\begin{figure}
\begin{center}\vskip2mm
\includegraphics[scale=0.8]{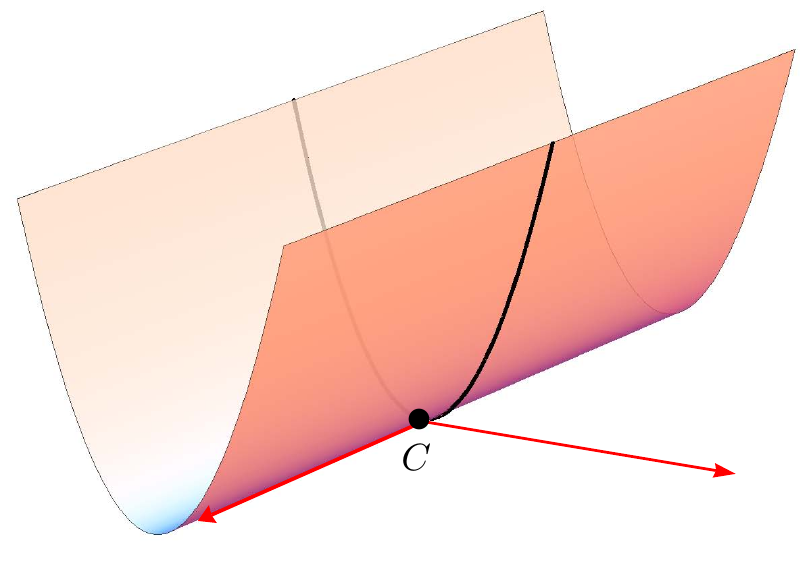}\vspace{-8mm}
\end{center}
\caption{\it The energy functional can also have a flat ``bottom'',  whose tangent vectors are zero-modes.
Other modes are positive. 
}
\label{flat}
\end{figure}

These critical points can be  local minima [or maxima], or
saddle points and can also be degenerate, meaning that it belongs
 to a submanifold with constant  value of $\cH$.
The theoretically possible ``landscapes'' are, hence, as depicted on  Figs. \ref{cupsaddle} and \ref{flat}.

The nature of the critical point can be tested by considering small oscillations around it: for a minimum, 
represented by the bottom of a ``cup'' (Fig.\ref{cupsaddle}a), all oscillations would 
\textit{increase} the energy. Such a configuration is classically \textit{stable}.

For a saddle point (Fig.\ref{cupsaddle}b) some oscillations would {increase} the energy; these are the {stable modes}. Some other
ones would instead \textit{decrease} the energy:
there exist \textit{negative modes}.

A critical point can also  be degenerate, meaning that one may have \textit{zero modes}, i.e. oscillations which leave the energy unchanged, cf. Fig.\ref{flat}.

The intuitive picture is that if one puts a ball into a critical point, it will roll down along energy-reducing  directions, --- except when it is a (local) minimum and no such directions exist.

How can we determine the type of a critical point?
In finite dimensions, we would use differential calculus: a critical point is where all first partial derivatives vanish. Then the behavior of oscillations depends on the matrix of second derivatives
called the \textit{Hessian},
\beq
\delta^2\cH=\left[\p_i\p_j\cH\right].
\eeq
$\delta^2\cH$ defines  a symmetric quadratic form
 which is positive or negative definite if it is a [local] minimum
 or maximum, indefinite for a saddle point and degenerate  if
it has energy-preserving deformations.
All this can be detected by looking at the eigenvalues of 
$\delta^2\cH$: are they all positive, or both positive or negative, or do we have zero eigenvalues.

The number of negative modes, called the \textit{Morse index} of the critical point
under investigation, is denoted by $\nu$.

In  Section \ref{monopoles}
below we will apply this analysis to non-Abelian monopoles, which are critical points of the static Yang--Mills Higgs energy 
(\ref{YMHEner}). Then eqn. (\ref{1stvar}) requires the vanishing of the first variation and yields the static Yang--Mills--Higgs field equations.

For monopoles of the 't~Hooft--Polyakov type 
\cite{tHooftPolyakov,GORPP,COL,HPAMonop,Konishi} 
finite energy requires that on the ``sphere at infinity'' $\IS^2$ [meaning for large distances]
 the original gauge group, $G$,
breaks down to the so-called \textit{``residual gauge group''}, $H$. Then \textit{finite-energy YMH configurations}
monopoles fall into \textit{topological sectors} labelled
by elements of the \textit{first homotopy group} of $H$, 
\beq
\hbox{topological sectors}\sim\pi_1(H),
\eeq
see \cite{GORPP,COL,HPAMonop,Arafune,Schwarz,Taubes81,HRCMP,
Lubkin}.
Each \textit{monopole solution} admits, furthermore, a \textit{constant non-Abelian charge vector} $\IQ$
introduced by Goddard, Nuyts and Olive (GNO) \cite{GNO}.
The GNO charge is quantized in that 
\beq
\exp[4\pi i\IQ]=1,
\label{GNOquant}
\eeq
and then the topological sector of the monopole
is the homotopy class in $H$ of the loop
\beq
h(t)=\exp[4\pi i\,\IQ\,t],
\qquad
0\leq t\leq 1.
\label{hloop}
\eeq

Then the clue is that for certain type of variations referred to as of the ``Brandt--Neri type'', the stability problem reduces to
that of a \textit{pure Yang--Mills theory on the sphere at infinity with group $H$} \footnote{This is just a special case of YM on a Riemann surface, studied by Atiyah and Bott \cite{ABOTT}. See also
\cite{Speight} for a recent contribution.}.
Then the problem boils down
to studying  the GNO charge, $\IQ$. 
Below we show indeed

\Thm (Goddard-Olive -- Coleman) \cite{GOCS,COL}:~{\it For a  't~Hooft--Polyakov monopole each topological sector contains exactly one stable charge $\oIQ$.}

\vskip3mm
The proof will be deduced from the formula which counts the number of negative modes,
\beq
\nu=\displaystyle{2\sum_{q} \left(2|q|-1\right)},
\label{negmodcount}
\eeq
where the (half-integer) $q$ are the eigenvalues of definite sign of the
$GNO$ charge $\IQ$ \cite{FrHab,HoRa,HORR88,NAUH}.
It follows that a monopole is
stable if its only eigenvalues are $0$ or $\pm1/2$ \cite{BN,GOCS,COL}. All other monopoles correspond to saddle points, cf. Fig.\ref{cupsaddle}b.

The number of instabilities, (\ref{negmodcount}), is conveniently counted by the
so-called \textit{Bott diagram} \cite{Bott}, see Section \ref{negmodes}.

\begin{figure}
\begin{center}
\includegraphics[scale=0.8]{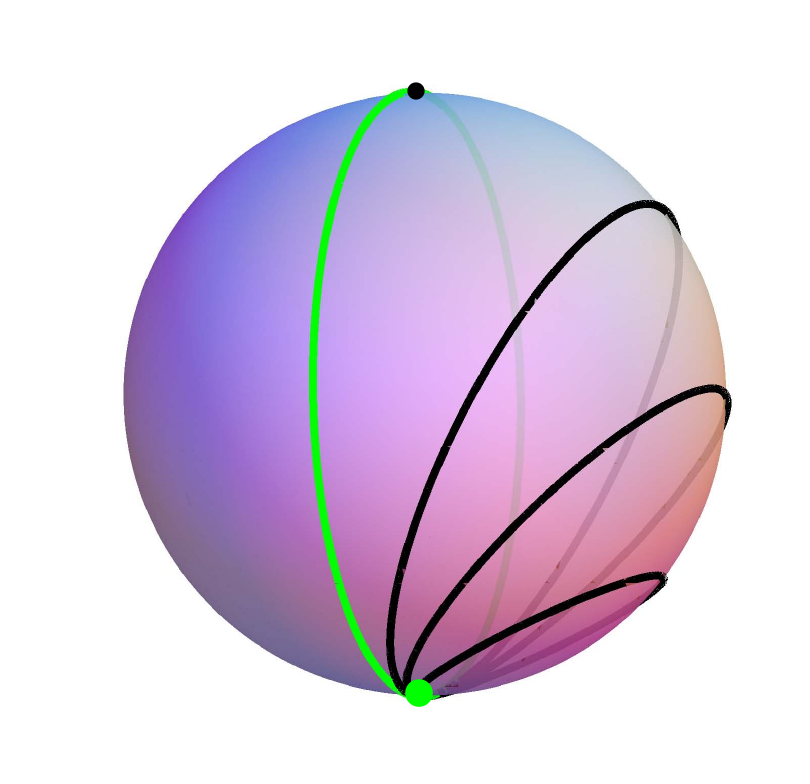}
\end{center}\vspace{-10mm}
\caption{\it An elastic string wound around a sphere
is in unstable equilibrium and shrinks to a point if it is perturbed.}
\label{shrink}
\end{figure}

\vskip2mm
Another intuitive way of understanding monopole instability, put forward by  Coleman \cite{COL}, is by thinking of them as of \emph{elastic strings} \cite{COL}: 
\emph{monopoles decay just like strings shrink}, namely to the shortest one allowed by the topology, see Fig. \ref{shrink}. 

Remarkably, this analogy can be made rigorous. Indeed,
choosing a $1$-parameter family of loops $\theta\to\gamma_{\varphi}(\theta),\, 0\leq\varphi\leq2\pi$ 
 sweeping through the two-sphere
 such that $\gamma_{0}
=\gamma_{2\pi}$ is a point, \textit{parallel transport}
along $\gamma_{\varphi}$,
\beq
h^A(\varphi)={\cal P}\left(\exp\oint_{\gamma_{\varphi}}\bA\right)\,,
\label{1.8}
\eeq
associates
a \emph{loop in the residual group $H$} to any YM potential $\bA$ on $\IS^2$.

The energy of a loop in $H$ can be defined (Sec. \ref{loops}) and a variational calculus,
analogous to YM on $\IS^2$, can be developed. Remarkably, the map (\ref{1.8}) carries monopoles i.e. critical points of the YM functional into geodesics, which are critical points of the loop-energy functional. Furthermore,
the number of instabilities is also the same, namely (\ref{negmodcount}).

The map (\ref{1.8}), which has been used before \cite{Lubkin,GORPP,COL,GNO} for describing the topological sector
of the monopole, contains much more information, however: as a matter of fact, it
 puts  {all} homotopy groups of finite-energy YM configurations on $\IS^2$ and of loops in
$H$ in (1-1) correspondence \cite{Singer}: it is a homotopy equivalence.

\subsection{Global aspects: Morse theory}

\begin{figure}
\begin{center}
\includegraphics[scale=1.05]{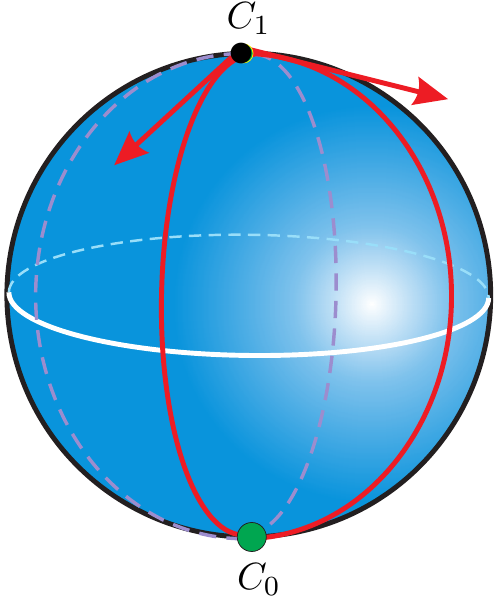}\qquad\qquad
\includegraphics[scale=0.7]{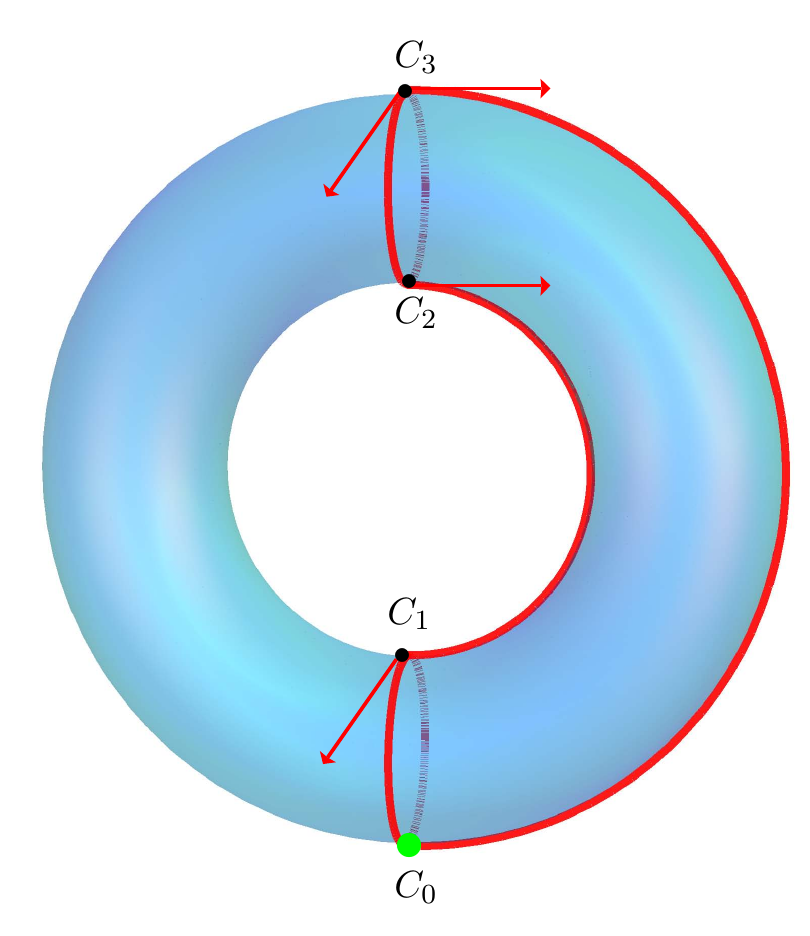}
\end{center}\vspace{-4mm}
\caption{\it Global aspects of instability. A ball put to the top of a sphere (Fig.\ref{SphereTorus}a) or to that of a torus
(Fig.\ref{SphereTorus}b) rolls down to some lower-lying critical point and ultimately arrives at the stable configuration.
The index-$\nu=2$ critical points correspond to non-vanishing classes in the second homology group,
 $H_2$. 
 }
\label{SphereTorus}
\end{figure}

A ball placed at the top of a torus will roll down to another critical point (Fig. \ref{SphereTorus}b). If this is again an unstable configuration, it will continue to roll until it arrives at a stable position.  
But what happens to an unstable monopole?
 By analogy, one can expect that, for sufficiently slow motion, an unstable monopole will preserve its identity and move (semi-)classically so as to decrease  its energy. 
 Although it cannot leave its topological sector since this would require infinite energy, it can
 go into another state in the same sector, because such configurations are separated by finite energy barriers \cite{HORR88}.

Describing the ``landscape'' of static YM configurations can  provide us 
therefore with useful information on the (possible) fate of unstable monopoles.
It is tempting to think, in fact, that our energy-reducing spheres might indicate the possible \textit{decay routes} of the monopole.

The stability problem can also be investigated from the
\textit{global} point of view.
Following the Morse theory \cite{Morse}, for a ``perfect Morse function'' 
(of which the energy functionals of both YM on $\IS^2$ and of loops in $H$ are examples), the appearance of a
critical point corresponds  to a sudden \textit{change of the topology} of the underlying space. 

Looking at the torus (Fig.\ref{SphereTorus}b), one starts with the bottom with
``energy" [identified with the height] $\cH=0$. Then, for low ``energies",  the section $\cH\leq\cH_1$ of the
surface is contractible.

 Arriving at the first critical
point, $C_1$,  however, with energy $\cH=\cH_1$,
the topology changes  as 
\textit{non-contractible loops} arise\footnote{According to the Morse theory, the correct notion is \textit{homology}, rather than homotopy. The first homotopy and homology groups are the same. 
However, $H_2=\pi_2=\IZ$ for $\IS^2$, but $H_2=\IZ$ and $\pi_2=0$ for the torus.}. Climbing higher,
we reach another critical point $C_2$ at $\cH=\cH_2$,
 and the topology changes once again with the arising of a different class of non-contractible loops.

Reaching the top, $C_3$ of the torus, $\cH=\cH_3$, 
 yet another sudden topology change takes place: we get a closed  non-contractible two-surface --- namely the torus itself.

\begin{figure}
\begin{center}\vspace{2mm}
\includegraphics[scale=1]{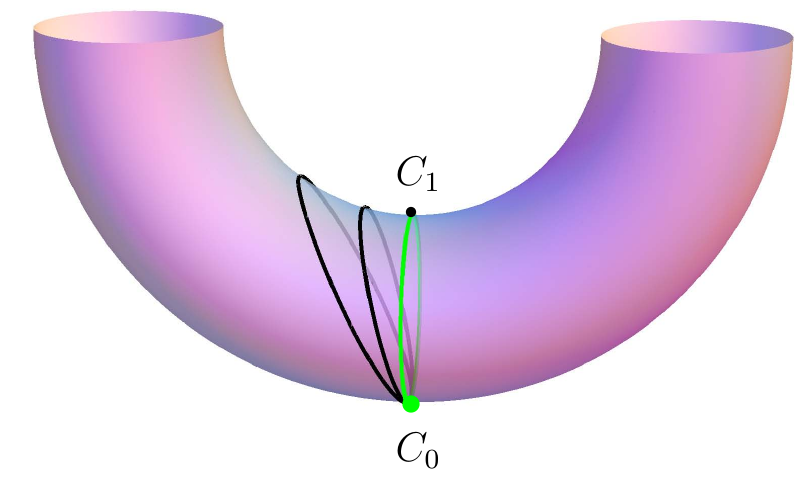}
\vspace{2mm}
\end{center}\vspace{-8mm}
\caption{\it If one has a family of non-contractible loops all passing through the bottom, $C_0$, then the ``mountain pass lemma'' says that the infimum $C_1$ of the highest points on
each of the loops is, under suitable conditions, an
(unstable) critical point.}
\label{MountainPass}
\end{figure}

In field theories,
 saddle-points are often associated with \emph{non-contractible loops} of field configurations \cite{MaWSTop,Taubes83,FHloops,Sibner}. The strategy is provided by the so-called ``mountain-pass lemma'': consider first the highest point on each non-contractible loop; then the infimum of these tops
 will be a critical point, see Fig.\ref{MountainPass} \footnote{Plainly, the
 mountain-pass lemma is only valid under appropriate conditions as the compactness of the underlying manifold, etc. \cite{MaWSTop}.
 For YM over a compact Riemann surface, the required conditions are satisfied \cite{ABOTT} --- as they
 are also for Prasad-Sommerfield monopoles \cite{BPSmon}.}.

 It is easy to see that there are no non-contractible loops in our case. There exist, however, non-contractible \emph{spheres}, and in Sec. \ref{spheres}
 we ``hang'' indeed $k$ \emph{energy reducing two-spheres} at a given unstable configuration of Morse index $\nu=2k$ with their bottoms at
certain other, lower-energy configurations. The  tangent vectors at their top yield the required number of negative modes (cf. Fig. \ref{SphereTorus}). 
A handy choice of the $\gamma_{\varphi}(\theta)$'s allows also to recover the loop-negative modes (explicitly constructed in Sec. 7) as images of the YM-modes.

\begin{figure}
\begin{center}\vskip-15mm
\includegraphics[scale=.16]{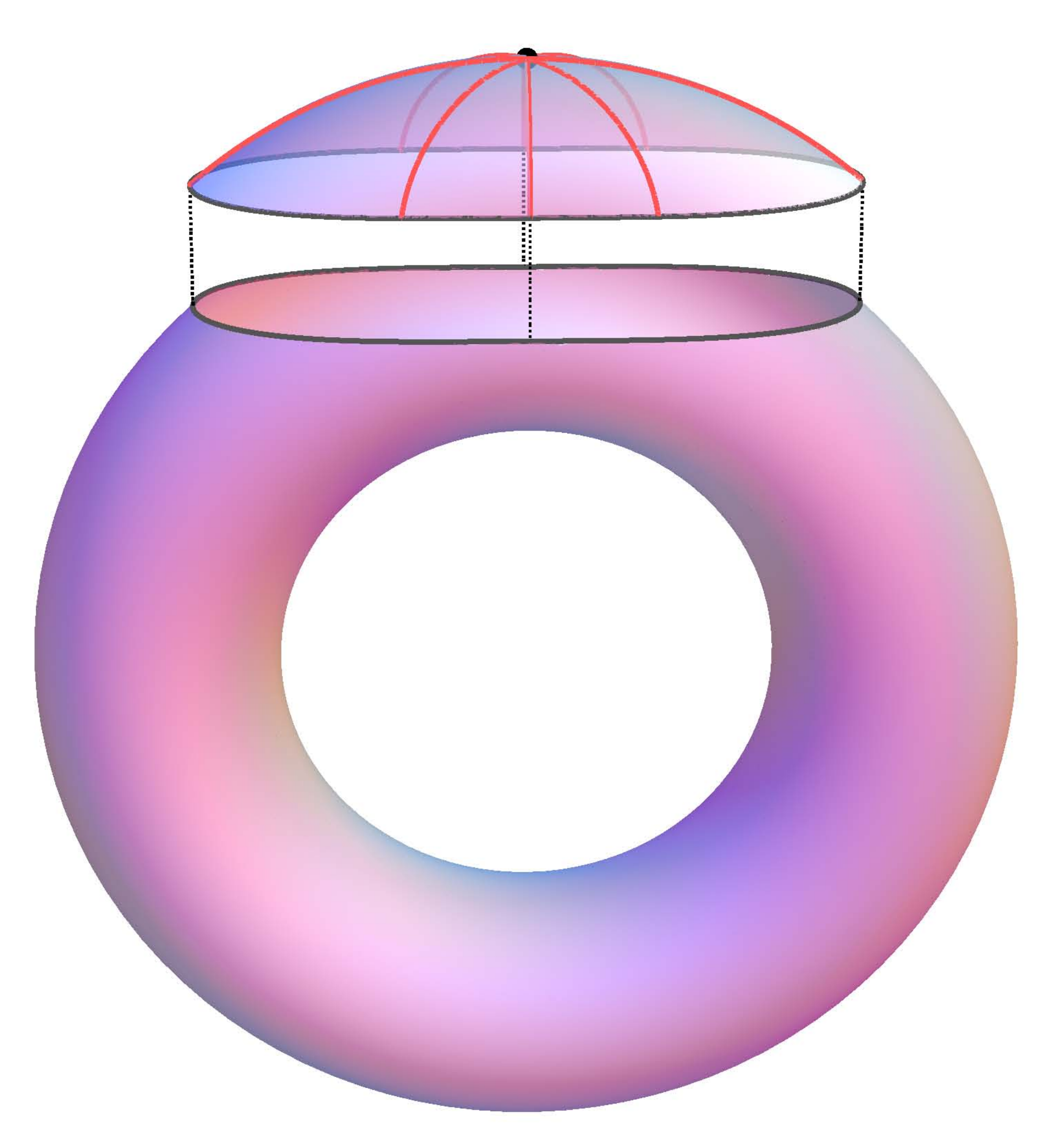}
\end{center}\vskip-6mm
\caption{\it Flowing down from a critical point of Morse index $\nu$ along the negative-mode directions yields a $\nu$-dimensional ``cap'' which, when glued to the lower-energy part of the configuration space, forms a closed $\nu$-dimensional surface
which generates a non-trivial homology class in $H_{\nu}$.}
\label{MorseConstr}
\end{figure}

The  relation of a critical point to topology change
is understood 
intuitively as follows \cite{Morse}. Flowing down along the negative-mode directions, provides us with a small $\nu$-dimensional ``cap'' which, when glued to the lower-energy part of configuration space, forms a closed, $\nu$-dimensional surface cf. Fig.\ref{MorseConstr}. 

\newpage
\section{Monopoles in Unified Gauge Theories}\label{monopoles}

\subsection{Electric charge quantization  and the Dirac Monopole}\label{Dmonop}

The first explanation for the quantization of the electric charge was put forward by Dirac
\cite{Dirac1931}, who, in 1931, posited the
existence of a radial magnetic field,
\beq
\bB=\frac{g}{r^2}\,\hat{\bx},
\label{DmonopB}
\eeq
where the real constant $g$ is the magnetic
charge,
\beq
g=\frac{1}{4\pi}\int_{\IS^2}\bB\cdot d\bS.
\eeq
 Note for further reference that  (\ref{DmonopB}) is in fact
\beq
B_i={g}\,\2\epsilon_{ijk}\omega_{jk},
\eeq
where
\beq
\omega=\2\omega_{ij}dx^i\wedge dx^j=
\2\,\frac{\bx\cdot d\bx\times d\bx}{r^3}\,.
\eeq
is the surface form of the two-sphere.
The electromagnetic two-form
$
F=\2 F_{\mu\nu}dx^\mu\wedge dx^\nu
$
where
$
F_{0i}=0,\;
F_{ij}=\epsilon_{ijk}B_k=g\,\omega_{ij}
$
satisfies therefore the vacuum Maxwell equations \footnote{written in coordinates as
$ 
\epsilon_{ijk}\p_iF_{jk}=0,
\;
\p_iF_{ij}=0
$.}
\beq
dF=0,
\qquad 
d\star F=0
\eeq
 everywhere except at the origin. 
 
The unusual feature of this field is that $F$ does
not derive from a globally well-defined vector potential. Assuming $\bB=\rot\bA$ i.e.
$F=dA$ for some $1$-form $A=A_idx^i$ would indeed
yield a contradiction. On the one hand, Stokes' theorem would require
$$
\int_{\IS^2}F=\int_{\IS^2}dA=\int_{\emptyset}A=0,
$$
since the closed two-surface $\IS^2$ has no boundary. Direct evaluation yields, however,
$$
\int_{\IS^2}F=4\pi g\neq0,
$$
a contradiction.

The clue of Dirac has been that this fact has no
physical consequence provided the electric and magnetic
charges, $e$ and $g$, satisfy a suitable quantization condition. 

At the purely classical level, the vector potential
plays no role. It does play a role at the quantum level, 
though, as it  can be understood 
as follows. $dF=0$ does imply the existence of a vector potential in each contractible subset of space.
Restricting ourselves to the surface of the unit
two-sphere, vector potentials can be found on the
``upper'' and ``lower" hemispheres,
\beqa
U_N&=&\Big\{(\theta,\varphi)\,\Big|\, 0\leq\theta<\pi/2+\epsilon,\, 0\leq\varphi\leq2\pi\Big\}
\\
U_S&=&\Big\{(\theta,\varphi)\,\Big|\, \pi/2-\epsilon<\theta\leq\pi\, 0\leq\varphi\leq2\pi\Big\},
\eeqa
e.g., in polar coordinates,
\beq
A_\theta^{\pm}=0,\quad
A_\varphi^{\pm}=g(\pm1-\cos\theta)
\label{Dvecpot}
\eeq
in a suitable gauge. Here the $\pm$ sign
refers to the N and S hemispheres, respectively.

In each coordinate patch we can describe our particle
by a wave function, $\psi^+$ and $\psi^-$, respectively,
each satisfying the minimally coupled Schr\"odinger
equation 
\beq
i\hbar\p_t\psi^{\pm}=-\frac{\bD^2}{2m}\psi^{\pm},
\quad
\bD=\bnabla-ie\bA.
\label{Schr}
\eeq
In order to have a well-defined physical system,
the two descriptions must be gauge-related, 
i.e., 
\beq
\psi^+(x)=h(x)\psi^-(x)
\label{gaugerelated}
\eeq
for some $\UN(1)$-valued function $h(x)$  called the \emph{transition function}, defined in the
overlap $U^+\cap U^-$. Then 
\beq
(e/\hbar)(\bA^+-\bA^-)=\frac{dh}{ih}\,
\quad\Rightarrow\quad
D^+\psi^+=h(D^-\psi^-).
\label{vpwd}
\eeq
if $
\psi^+=h\psi^-,
$
so that the descriptions (\ref{Schr}) are indeed equivalent.
For (\ref{Dvecpot}) the condition (\ref{vpwd})  yields
\beq
h(x)=e^{i(2eg/\hbar)\varphi},
\label{Dtransfunc}
\eeq
$x=(\varphi,\theta),\,0\leq\varphi\leq2\pi,\,
\pi-\epsilon<\theta<\pi+\epsilon$,
whose periodicity provides us with the celebrated
\emph{Dirac quantization condition} \footnote{From now on we work in units where $\hbar=1$.}
\beq
2eg=n\,\hbar,\quad n\in\IZ.
\label{DiracQuant}
\eeq
 Equivalently,
\beq
\exp\big[i\oint e(A^+-A^-)\big]=1
\label{WYcirccond}
\eeq
for any closed loop.

This same condition can also
be expressed by saying that the \emph{non-integrable phase factor}
\beq
\exp\big[i\oint_\gamma eA\big]
\label{nipf}
\eeq
must have a \emph{gauge-independent meaning} for any
closed loop. 
Here mathematicians recognize the expression
 for holonomy \cite{Steenrod,Kobayashi,BottTu} obtained by \emph{parallel transport}; this is indeed the
 starting point of Wu and Yang's ``integral formulation'' of gauge theory \cite{WuYa1}
 \footnote{The same conditions can also be
derived in a path integral framework \cite{WuYa2,WYHPA}.}.

\kikezd{The bundle picture} \cite{Sniatycki,Petry,TrautmanHopf,HPADirac,TrautmanDG}

The above result can be reformulated in a geometric language: (\ref{WYcirccond}) is the \emph{necessary and sufficient condition for the existence of a principal
$\UN(1)$ bundle $\cY$ with connection $\cA$.}
\cite{Steenrod,Kobayashi,TrautmanDG}.
Locally i.e. over a coordinate patch $U$,
\beq
\cY\Big|_U=\Big\{(x,z)\in
U\times \UN(1)\Big\},
\qquad
\cA=A+\frac{dz}{iz},
\label{Ypm}
\eeq
The right action of $h\ni\UN(1)$ on $\cY$ is locally
$h~:\zeta=(x,z)\to \zeta\cdot h=(x,zh)$.

 Covering the whole base manifold
(here $\IS^2$) with such patches, with $U^+$ and $U_-$ for example,
condition (\ref{WYcirccond}) guarantees that
\beq
h(x)=\exp\big[ie\int_\gamma(\bA^+-\bA^-)\cdot dx\big],
\eeq
where $\gamma$ is \emph{any} path from a
reference point $x_0$ to $x$ in $U^+\cap U^-$
is well-defined in that it only depends on
the initial and final points but not on the path $\gamma$ itself.
In fact,
$
h(\varphi)=\exp\big[i2eg\varphi\big]=e^{in\varphi}
$ 
 cf. (\ref{Dtransfunc}) and  the local gauge potentials are related as in (\ref{vpwd}).
 
Conversely, if 
$s: U\to \cY$ is a section of the bundle, then
$
A=s^*\cA
$
is a local vector potential; choosing another section yields a gauge related potential.

The wave function can also be lifted to the bundle. Setting, in a local patch,
\beq
\Psi(\zeta)=\psi(x)z,
\eeq
provides us with an \emph{equivariant} function on the bundle $\cY_n$,
$
\Psi(\zeta\cdot h)=h\,\Psi(\zeta)\,,
\; h\in \UN(1).
$

The most convenient way to realize the general theory
is to use the Hopf fibration \cite{WYHPA,TrautmanHopf,HPADirac,TrautmanDG}: 
dropping the irrelevant radial variable, we
consider
$\IS^3$ as sitting in $\IC^2$,
\beq
\IS^3=\Big\{\zeta=\left(\barray{c}z_1\\ z_2
\earray\right)\in\IC^2\;\Big|\,
|z_1|^2+|z_2|^2=1\Big\}.
\eeq
$\UN(1)$ acts on $\IS^3$ as
$
\zeta\to\zeta\cdot h=\left(\barray{c}z_1h\\ z_2h\earray\right).
$
 The subgroup
$\IZ_n=\big\{\exp 2\pi i k/n|0\leq k\leq n-1\big\}$
acts on $\IS^3$ and the quotient
\beq
\cY_n=\IS^3/\IZ_n
\eeq
is therefore a principal $\UN(1)$ bundle with base 
$\IS^2\subset\IR^3$. The projection $\cY_n\to\IS^2\subset\IR^3$ is given by
\beq
\pi([\zeta])_i=\overline{\zeta}\sigma_i\zeta,
\qquad i=1,2,3,
\eeq
where the $\sigma_i$ are the $2\times2$ Pauli
matrices and $\overline{\zeta}=\left(
\bz_1, \bz_2\right)$.
Then
\beq
\cA_n=\frac{n}{i}\,\overline{\zeta}\,d\zeta=\frac{n}{i}\big(
\bz_1dz_1+\bz_2dz_2\big)
\eeq
is a connection form on $\cY_n$ with curvature
\beq
d\cA=n\omega,
\eeq i.e.,
$n$-times the surface form of the two-sphere.
Thus $\cY_n$ has Chern class $c(\cY_n)=n$
\cite{Kobayashi,BottTu}.

Parametrizing the two-sphere with Euler angles $(\theta,\varphi,\chi)$, 
\beq
\zeta\sim\left(\barray{c}\exp[\smallover{i}/{2}(\varphi+\chi)]\cos\theta
\\[6pt] 
\exp[-\smallover{i}/{2}(\varphi-\chi)]\sin\theta/2\big)\earray\right),
\eeq
\beq
s_+(\theta,\varphi)=\left(\barray{c}\cos(\theta/2)
\\[6pt]
i\exp[i\varphi]\sin(\theta/2)
\earray\right)
\eeq
is a local section of our bundle such that $A^+=s_+^*\cA$.

\subsection{Unified gauge theories}

The success of unifying weak and electric interactions into a single theory by extending the
$\UN(1)$ gauge group of electromagnetism into
$\UN(2)$ (locally $\un(1)_e\oplus\su(2)_w$) \cite{WS}
generated an outburst of interest in attempting further unification by including strong interactions,
\beq
\gh=\un(1)_e\oplus\su(2)_w\oplus\su(3)_s
\eeq
obtained from some Grand Unified gauge group $G$ as residual symmetry after spontaneous symmetry breaking by the Higgs mechanism. 
The most attractive ones of 
these Grand Unification Theories (GUT)s start with
$G=\SU(5)$, or $\SO(10)$ \cite{GUT}.

GUTs made seemingly unnecessary Dirac's experimentally never-confirmed hypothesis, as they provided an alternative explanation of electric charge quantization.
Around 1974, however,
`t~Hooft, and Polyakov \cite{tHooftPolyakov} found that Unified Gauge Theories admit \textit{finite-energy
particle-like exact solutions} which for large distances
behave as Dirac monopoles. Their original results
show this in the context of $G=\SU(2)$, 
later extended to more general gauge groups
\cite{GUTmonopoles}.

\vskip2mm
Below we present a brief outline of
non-Abelian gauge theories and of
their monopole solutions.

Let $G$ be a compact simple Lie group. A (static) Yang--Mills--Higgs theory is given by the following ingredients:

\begin{enumerate}
\item
A (static and purely magnetic)  \textit{Yang--Mills gauge field} is a connection
form on a  $G$-bundle $P$ over space, $\IR^3$. Locally, such a connection form is given by a $1$-form $A_i$ with values in $\LG$, the Lie algebra of $G$, $A_i=(A_i^{\ a}),\ a=1,\dots, \dim (\LG)$,
represented by hermitian matrices, $\xi^\dagger=\xi$. 
The Yang--Mills field strength $F_{ij}$
is the curvature of the connection, and the \textit{magnetic field}, $\bB=(B_i)$, is its dual,
\beq
F_{ij}=\p_iA_j-\p_jA_i+ie[A_i,A_j],
\qquad
 B_i^a=\2\epsilon_{ijk}F_{jk}^a,
\eeq
respectively,
where   $[\,\cdot\,,\,\cdot\,]$ is the  commutator in the Lie algebra
$\LG$ of $G$, and $e$ is a coupling constant. 

\item
A \textit{Higgs field} 
$\Phi$ is a scalar function on $P$ which takes its values in some linear
representation space $V$ of $G$, $G: V\ni\xi\to g\cdot \xi\in V$. 
$\Phi$ is equivariant, $\Phi(p\cdot g)=g\cdot\Phi(p)$ where $p\cdot g$ denotes the right action of $G$ on $P$. The 
infinitesimal [Lie algebra] action on $V$ is denoted by $\eta\cdot\xi$.

A frequent choice  is  $V=\LG$ the Lie algebra, and then the actions are
the adjoint ones, 
\beq
{\rm Ad}_g(\xi)=g^{-1}\xi g
\quad\hbox{and}\quad
{\rm ad}_\eta(\xi)=-i[\eta,\xi],
\eeq
 respectively.
In what follows, we shall mostly consider the adjoint case, $V=\LG$, when $\Phi=(\Phi^a),\,
a=1,\dots,{\rm dim}\,\LG$.

\item
A \textit{Higgs potential} $U$ is a non-negative
 invariant function on $V$, $U\geq0$,
$U(g\cdot\Phi)=U(\Phi)$. Then the absolute minima
of $U(\Phi)$ lie in some orbit ${\cal O}\simeq G\cdot\Phi_0$ of $G$. This assumption that
$G$ acts transitively means the orbit
is identified with ${\cal O}\simeq G/H$ where
$H$ is the stability subgroup of $\Phi_0,\, H\cdot\Phi_0=\Phi_0$. In the physical context  $H$ will be referred to as the {\it residual group}.

In the simplest case $G=\SU(2)$, the most frequent choice is
\beq
U(\Phi)=\frac{\lambda}{4}\big(1-||\Phi||^2\big)^2
\label{Higgspot}
\eeq
where, for an adjoint Higgs, $||\Phi||^2=\tr(\Phi^2)=\Phi^a\Phi^a$.
\end{enumerate}
A (static and purely magnetic) finite-energy 
Yang--Mills--Higgs configuration is such that the energy,
\beq\barray{lll}
\cH&=&
\displaystyle\int_{\IR^3}\Big\{
\frac{1}{2}{\tr}(\bB^2)+\frac{1}{2}{\tr}(\bD\Phi)^2+U(\Phi)\Big\}d^3\bx
\earray
\label{YMHEner}
\eeq
is finite. Here  
\beq
D_i\Phi=\p_i\Phi+ie[A_i,\Phi]
\label{covder}
\eeq 
is the covariant derivative where, for simplicity,
we restricted ourselves to an adjoint Higgs field.
 
The energy (\ref{YMHEner}) is invariant w.r.t.  \textit{gauge transformations},
\beq
A_j\to gA_jg^{-1}-\frac{i}{e}g\p_jg^{-1},
\qquad
\Phi\to g\cdot\Phi = g\Phi g^{-1},
\eeq
which imply that $\bB\to g\bB g^{-1}$, $\bD\Phi \to g\cdot\bD\Phi$.

Monopoles arise as finite-energy solutions of the associated variational Yang--Mills--Higgs
equations \cite{tHooftPolyakov,GORPP,COL,HPAMonop}.
Restricting ourselves, for notational simplicity, to the adjoint case, 
the field equations read
\beqa
\bD\times\bB&=& ie[\bD\Phi,\Phi], 
\label{YMHeq1}
\\[6pt]
\bD^2\Phi&=&\frac{\delta U}{\delta\Phi}.
\label{YMHeq2}
\eeqa

\subsection{Finite-energy configurations}\label{finiteenergy}

In this subsection we shall not require that the fields satisfy the 
field equations, but only that they be of finite energy, i.e., such that the integral in
(\ref{YMHEner}) converges. One reason for this is to emphasize that the most important spontaneous symmetry breakdown, namely that of the Higgs potential, comes from the finite energy and not from the field equations.

We shall consider the three terms in  (\ref{YMHEner}) in turn. 
It will be convenient to use the radial gauge $\bx\cdot\bA=0$.

\goodbreak
\kikezd{Pure gauge term $\tr \bB^2$}

For sufficiently smooth gauge fields the finite energy condition imposed by this term is, with some abuse of notation,
\beq
\bA(\bx)\to\frac{\bA(\theta,\varphi)}{r},
\qquad
\bB(\bx)\to\frac{\bb(\theta,\varphi)}{r^2}
=\frac{b(\theta,\varphi)}{r^2}\,\hat{\bx},
\label{asgaugefields}
\eeq
where $\theta,\,\varphi$ denote the polar angles \footnote{$\bA=(A^a_i)$ is a 
Lie algebra valued vector potential with $a$ and $i$  Lie algebra and resp. space indices. 
Similarly, $\bB=(B^a_i)$ and $\bb=(b^a_i)$ are Lie algebra valued
vectors. The last equality in
(\ref{asgaugefields}) decomposes the Lie algebra valued asymptotic
magnetic field $\bb$ 
into a Lie algebra-valued scalar $b(\theta,\varphi)/r^2$
 times the radial direction $\hat{\bx}=\bx/r$ 
\cite{HPAMonop}.}.
 Note that (\ref{asgaugefields}) only involves the gauge field.

\kikezd{Higgs potential $U(\Phi)$}

The finite energy condition for this term is 
$r^{2}\,U(\Phi)\sim 0$ as $r\to\infty$. A necessary condition for this is that $U\to0$. But $U\geq0$ is assumed to be a Higgs potential i.e. one whose minima lie on some non-trivial group orbit $G/H$, where $H$ is an appropriate subgroup of $G$, called the ``residual gauge group''.

At large distances the Higgs field 
 takes therefore its values in the orbit $G/H$ and may only depend 
non-trivially on the polar angles:
$\Phi(r,\theta,\varphi)\to \Phi(\theta,\varphi)$ as $r\to\infty$ (again with some abuse of notation). Then 
the asymptotic values of
$\Phi$ define a map of ``the two-sphere at infinity'' 
 $\IS^2_\infty$ parametrized  with the polar angles $(\theta,\varphi)$
 into the orbit $G/H$,
\beq
\Phi~:~\IS^2_\infty\to G/H.
\eeq 
The asymptotic values of the Higgs field define thus a homotopy class in $\pi_2(G/H)$. Since this class
can not be changed by smooth deformations, the ``manifold'' of finite-energy
configurations splits into \emph{topological sectors},
labelled by $\pi_2(G/H)$ \cite{GORPP,COL,HPAMonop,Arafune,Schwarz,Taubes81,HRCMP}.

As 
\beq
\pi_2(G/H)\simeq \pi_1(H)
\eeq
 for any [simply connected] Lie group $G$, the topological sectors can be labelled also by
classes in $\pi_1(H)$; the first homotopy group of the residual group. Indeed, on the upper and respectively on the lower hemispheres $N$ and $S$ of $\IS^2$,
$$
\Phi(\theta,\varphi)=
\left\{\barray{lll}g_N(\theta,\varphi)\Phi(E)
\quad&\hbox{in} &N 
\\[12pt]
g_S(\theta,\varphi)\Phi(E)\quad&\hbox{in} &S
\earray\right.,
\label{lifts}
$$ 
where (the ``east pole'') $E$ is an arbitrary point in the overlap.
\beq
h(\varphi)=g_N^{-1}(\varphi)g_S(\varphi)
\label{3.3}
\eeq
where $\varphi$ is the polar angle on the equator of $\IS^2$
is a loop in $H$ which represents the topological sector. (\ref{3.3}) is contractible in $G$ \cite{COL,GORPP}.

For any compact and connected Lie group $H$, $\pi_1(H)$ is Abelian so has a free part and a torsion part
$$
\pi_1(H)=\IZ^p\oplus\IT,
$$
where $p$ is the dimension of the centre of $\gh$
and $\IT$ is a finite Abelian group \cite{HRCMP},
see Section \ref{Liealgebra} below. In fact,
$\IT$ is isomorphic to $\pi_1(K)$ where $K$ is the compact and semisimple subgroup of $H$ generated by
$\gk=[\gh,\gh]$.

The free part $\IZ^p$ provides us with $p$ integer
``quantum'' numbers $m_1,\dots, m_p$. They can be calculated 
as surface integrals as follows. To the asymptotic physical Higgs field
$\Phi$ in {any} representation and to each vector $\zeta$ from the centre of the Lie algebra $\gh$, we can associate an auxiliary adjoint ``Higgs'' field $\Psi$ defined by
\beq
\Psi_\zeta(\theta,\varphi)=g(\theta,\varphi)\,\zeta\, g^{-1}(\theta,\varphi),
\label{Psizeta}
\eeq
 where $g(\theta,\varphi)$
is  any of those ``lifts'' in (\ref{3.3}). 
Although the lifts in (\ref{lifts}) are ambiguous,
$gh$ works if $g$ does when $h$ belongs to $H$,
$\Psi(\theta,\varphi)$ is
well-defined, because $\zeta$ belongs to the centre
of $\gh$.
The projection of the charge lattice
$\Gamma_Q$ into the centre is a $p$-dimensional lattice there, generated over the integers by $p$ vectors $\zeta_1,\dots,\zeta_p,$,
\beq
 \zeta=\sum_{i=1}^p n_i\zeta_i,
\quad
n_i\in \IZ
\eeq 
Note  that the $\zeta_a$ are not in general charges themselves, nor are they normalized. 

 The above construction associates then
an adjoint ``Higgs'' field $\Psi_a$ to each 
generator $\zeta_a$, and the quantum numbers $m_a$ are calculated according to \cite{Arafune,HRCMP}
\beq
m_a=\frac{1}{4\pi|\zeta_a|^3}\int_{\IS^2}
B_a\bigl(\Psi_a,\,
\bigl[\p_\theta\Psi_a,\p_\varphi\Psi_a \bigr]\bigr)\, d\varphi\wedge d\theta,
\qquad
a=1,\dots,p,
\label{Higgscharge}
\eeq
where 
 $\Psi_a$ is the auxiliary Higgs field
(\ref{Psizeta}) for $\zeta=\zeta_a$, $B_a$ is a multiple $(\alpha_a,\alpha_a)B/2$ of 
the Killing form $B$ of $G$ and $\alpha_a$ is a simple root determined by $\zeta_a$. 

The mathematical content of this theorem is that
the free part of $\pi_2({\cal O})$ has the same rank as the dimension of the
second de Rham cohomology, 
\beq
\pi_2({\cal O})\otimes\IR\simeq H_2({\cal O})\otimes\IR\simeq H^2_{dR}({\cal O}),
\eeq which is in turn generated by the pull-backs to the orbit ${\cal O}\simeq(G/H)$
 of the canonical symplectic forms of the coadjoint orbits of
the basis vectors $\zeta_k$ 
\cite{Schwarz,Taubes81,HRCMP}. 

For a matrix group,
$B$ can be replaced by the trace, for example, for 
$G=SU(n)$ we have 
\beq
B(\eta,\zeta) = 2n\Tr(\eta\zeta)
\qquad\hbox{and}\qquad
(\alpha_k,\alpha_k)=2
\eeq
for all simple root $\alpha_k$.
The charge (\ref{Higgscharge}) is in fact the same as
\beq
m_a=\frac{1}{2\pi |\zeta_a|}\int_{\IS^2}\!\Psi_a^*\Omega_a,
\label{Higgschargebis}
\eeq
where $\Omega_a$ is the canonical symplectic form of the coadjoint orbit 
${\cal O}_a=Ad^*_G\,\zeta_a$ identified
with an adjoint orbit using $(\alpha_a,\alpha_a)B/2$.
$\IT$, the finite part of $\pi_1(H)$, has no similar expression.

The physically most relevant case is when $\IT=0$
and the sectors are  
described by a single integer quantum number $m$.
This happens when the Lie algebra $\gh$ of $H$ has a $1$-dimensional centre generated by a single vector $\zeta$ and the
semisimple subgroup $K$ is simply connected.

Another case 
of [mostly pedagogical] interest \cite{COL} is
$H=\SO(3)$, for which $p=0$ and $\IT=\IZ_2$.

The homotopy classification is not merely convenient, 
but is {mandatory} in that
 the classes are separated by infinite
energy barriers \cite{HORR88}.

Note that since not only $U\to0$ but $r^3U\to0$ one has \cite{HORR88}, 
\beq
\Phi(\bx)\to\Phi(\theta,\varphi)+\eta(r,\theta,\varphi),
\eeq
 where
$r\eta(r,\theta,\varphi)\to0$ as $r\to\infty$\footnote{
 A notable exception
to this observation is the Bogomolny--Prasad--Sommerfield (BPS)
limit of vanishing potential, $U=0$, for which the Bogomolny condition $\bB=\bD\Phi$
implies \cite{BPSmon} that 
\beq
\Phi(\bx)\to\Phi(\theta,\varphi)+\frac{b(\theta,\varphi)}{r}+
\Ort(1/r^2)
\qquad\hbox{as}\qquad
r\to\infty.
\label{3.7}
\eeq}.

We stress that monopole topology only depends on the Higgs field and not on the gauge field \cite{Arafune,HRCMP}.

\kikezd{The cross-term $(\bD\Phi)^2$}

This final term involves both $\Phi$ and $\bA$ and it hence provides the connection between the [asymptotic] Higgs field $\Phi(\theta,\varphi)$
 and the
gauge field $\bb(\theta,\varphi)$ and thus puts a topological constraint on the gauge field. 
This constraint may be expressed as a quantization condition as follows: the finite energy
condition is easily seen to be $r^2(\bD\Phi)^2\to0$ and thus
$\Phi$ is covariantly
constant on $\IS^2_\infty$,
\beq
D_i\Phi\equiv\p_i\Phi+iA_i\cdot\Phi=0
\label{covconst}
\eeq
[and hence also $D_i\Psi\equiv\p_i\Psi+i[A_i,\Phi]=0$]
where, with some abuse of notations, we switched to fields and  covariant derivative on the sphere at infinity involving their asymptotic values.

Then the topological quantum numbers $m_a$ can be expressed as
\cite{HRCMP}
\beq
m_a=\frac{1}{2\pi|\zeta|}\int_{\IS^2}\!\!dS\,\tr(\Psi_ab),
\qquad
a=1,\dots,p.
\label{3.9}
\eeq

Equation (\ref{3.9}) shows that in general it is not the gauge field
$B$ itself, but only its \textit{projection onto the centre}
that is quantized. Note that the quantization of 
$\displaystyle\int\!\!\tr(\Psi_ab)$
is again \emph{mandatory}, since the value of $\tr(\Psi_aB)$
cannot be changed without violating at least one of the finite-energy conditions $r^2V\to0$ or $r^3(\bD\Phi)^2\to0$ and
thus passing through an infinite energy barrier.

Notice that the value of (\ref{3.9}) is actually independent of the choice of the Yang--Mills potential $\bA$ as long as $\Phi$ is covariantly constant \cite{HRCMP}.

Note also that (\ref{covconst}) i.e. $\bD\Phi=0$  implies that one can choose a gauge such that the gauge potential, $\bA$, takes its values in the residual Lie algebra $\gh$.

\begin{figure}
\begin{center}
\includegraphics[scale=.65]{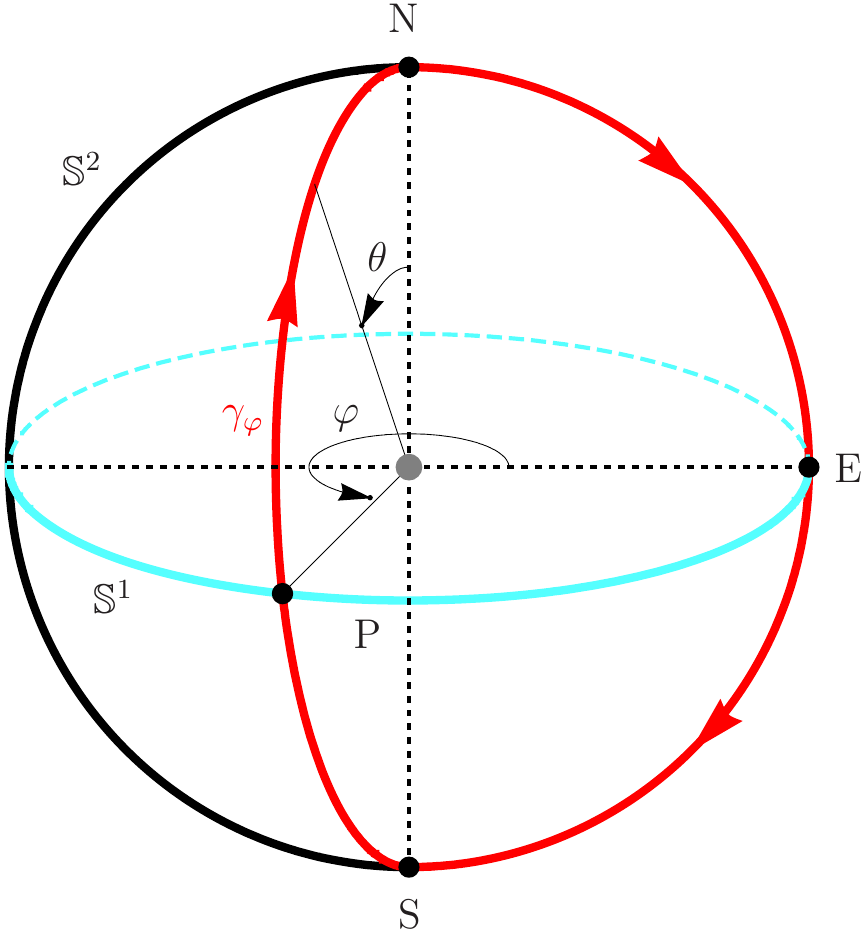}
\end{center}\vspace{-6mm}
\caption{\it To any value of the angle $\varphi$ is associated a
loop $\gamma_\varphi$ which sweeps though the entire 2-sphere as $\varphi$ varies from $0$ to $2\pi$. Since $\gamma_0=\gamma_{2\pi}$ is the zero loop, we get a loop of loops. Parallel transport along $\gamma_\varphi$, (\ref{paralleltransport}), provides us hence with a loop 
$h^A(\varphi)$ in the residual group $H$.}
\label{Hloop}
\end{figure}

Then a loop representing the homotopy sector can be found by parallel transport \cite{Lubkin,GORPP,COL}. Let us indeed cover 
$\IS^2_\infty$ with a $1$-parameter family of loops $\gamma_{\varphi}(\theta)$,
e.g., by choosing $\gamma_{\varphi}$ to start from the north pole $N$, follow the meridian
at angle $\varphi=0$ through the ``east pole'' $E$ down to the south pole $S$ and return then to the north pole along the meridian at angle $\varphi$, see Fig.\ref{Hloop}.
The loop
\beq
h^A(\varphi)={\cal P}\left(\exp\oint_{\gamma_{\varphi}}\bA\right)
\label{paralleltransport}
\eeq
[where ${\cal P}$ means path-ordering]
then represents  the topological sector. 
 Other choices of the $1$-parameter family of paths
$\gamma_{\varphi}$ would lead to homotopic loops $h^A$.

The calculation of the topological ``quantum''
numbers will be greatly simplified in
Sec.\ref{solutions} devoted to solutions of the
field equations.

\section{Lie  algebra structure and Lie
group topology}\label{Liealgebra}

Interrupting our investigations of monopoles,
we digress on the first homotopy of a 
compact Lie group using
some knowledge of its Lie algebra structure \cite{Humphreys}.

Let us indeed consider a compact simple matrix Lie algebra $\gk$ and
choose a Cartan subalgebra $\gt$. A root $\alpha$ is a linear function on the complexified Cartan algebra 
$\gt^{\IC}$, and to each $\alpha$ is associated a vector $E_\alpha$ (the familiar step operator) from
$\gk^{\IC}$ which satisfies, with any vector $H$ from
$\gt^{\IC}$, the relation\footnote{
Alternatively, we can consider the real combinations 
$X_\alpha=E_\alpha+E_{-\alpha}$ and $Y_\alpha=-i(E_\alpha-
E_{-\alpha})$ which satisfy $[H,X_\alpha]=iq_\alpha Y_\alpha$,
$[H,Y_\alpha]=-iq_\alpha X_\alpha$.}
\beq
\big[H,E_\alpha\big]=q_{\alpha}\,E_\alpha\,, 
\quad
q_{\alpha}=\alpha(H).
\label{2.1}
\eeq
There exists a set of primitive roots $\alpha_i,\, i=
1,\dots,r$ ($r =$ rank) such that every positive root is a linear combination of the $\alpha_i$ with non-negative integer coefficients
i.e. $\alpha=\sum m_i\alpha_i$ for all $\alpha$.

If $\alpha$ is a root, let us define the vector $H_\alpha$ in $\gt^{\IC}$ by 
\beq
\alpha(X)=\tr(H_\alpha X).
\eeq
With  suitable normalization  we have 
\beq
(E_\alpha)^{\dagger}
=E_{-\alpha}
\qquad
\tr(E_\alpha,E_{-\alpha})=1,
\qquad
[E_\alpha,E_{-\alpha}]=H_\alpha\,.
\label{cLiealgrel}
\eeq
For each
root $\alpha,\, H_\alpha$ and the $E_{\pm\alpha}$'s form  therefore [complexified]
 $\so(3)^c$ subalgebras of $\gk$.

\kikezd{Lattice of primitive charges, $\Gamma_P$.}

The \textit{primitive charges} $P_i$ are defined by
\beq
P_i=\frac{2H_i}{\tr(H_i^2)}
\quad
\hbox{where}\quad H_i=H_{\alpha_i}.
\label{primitcharge}
\eeq
The primitive charges form a natural (non-orthogonal) basis for the
Cartan algebra and by adding the $E_\alpha$'s we get a basis for the Lie algebra $\gk^{\IC}$. 
 The integer combinations
$\sum_in_iP_i$ of the primitive charges form an $r$-dimensional lattice $\Gamma_P$ sitting in the Cartan algebra.

\vskip10mm\goodbreak
\kikezd{Co-weight lattice, $\Gamma_W$.}

Let us introduce next another basis for the Cartan algebra with elements
$W_i$ dual to the primitive roots,
\beq
\alpha_i(W_j)=\tr(H_iW_j)=\delta_{ij},
\qquad
i,j=1,\dots,r.
\label{2.3}
\eeq
Comparing (\ref{2.3}) with the conventional definition \cite{Singer}
of primitive weights, for which there is an extra factor $(\alpha_i,\alpha_i)/2$ in front of the $\delta_{ij}$, one sees that the $W_i$'s are just re-scaled weights. They are called 
\textit{co-weights} \cite{GOCS} and it is evident that they can be normalized so as to coincide with the conventional weights (by choosing 
$(\alpha_i,\alpha_i)=2$) for all groups whose roots are all of the same length, i.e. all groups except
$\SP(2r), \SO(2r+1)$, $G_2$ and $F_4$.

The integer combinations $\sum m_iW_i$ form another lattice we denote by $\Gamma_W$.

 Since
$\alpha(P_i)$ is always an integer, the $W$-lattice actually contains the primitive-charge lattice, 
\beq
\Gamma_P\subset\Gamma_W.
\eeq 
The \emph{root planes} of $\gk$ are those vectors $X$ in the Cartan algebra for which $\alpha(X)$ is an integer, i.e., those vectors which have integer eigenvalues in the adjoint
representation. The root planes intersect in the points of the $W$-lattice. 

Let us stress that 
the weights and primitive charges only depend on the
Lie algebra, and not on its global group structure.
 Now we define a third lattice, which \emph{does}
depend on the global structure.

\kikezd{Charge lattice, $\Gamma_Q$.}

Denote by $\wK$ the (unique) compact, simple, and simply connected Lie group generated by $\gk$. Any other group $K$ whose Lie algebra is
$\gk$ is then of the form $K=\wK/C$, where $C$ is a subgroup
of $Z=Z(\wK)$, the centre of $\wK$. $Z$ is finite and Abelian, so $C$ is always discrete. Since $\wK$ is simply connected, $C$ is just $\pi_1(K)$, the first homotopy group of $K$.

The primitive charges satisfy the quantization condition
 $\widetilde{\exp}2\pi iP_i=1$  (exponential in $\wK$) and thus also in any
 representation of $\wK$ i.e. in any other group $K$
 with the same Lie algebra. For any set $n_i$, $i=1,\dots,r$ of integers,
\beq
\exp\left[2\pi it\sum n_iP_i\right],
\qquad
0\leq t\leq 1,
\label{2.4}
\eeq
(exponential in $K$) is hence a contractible loop in all representations. Since any loop is homotopic to one of the form $\exp2\pi i tP,\ 0\leq t\leq1$ where $P$ is a constant vector in $\gk$, we conclude that the lattice
$\Gamma_P$ consists of the generators of \emph{contractible loops}.

More generally, let us fix a group $K$ (i.e., a representation of $\wK$) and define a general \textit{charge} $Q$ to be an element of the Cartan algebra such that
\beq
\exp[2\pi iQ]=1
\quad\hbox{in}\quad
K,
\label{chargdef}
\eeq
so that $\exp[2\pi itQ],\ 0\leq t\leq 1$, is a loop,
and any loop is homotopic to one of this form,
 as said above \footnote{Warning: by historical reasons, there is a slight discrepancy between the mathematical formalism adopted here and the physical one in Section \ref{solutions}: a ``GNO charge'',
 $\IQ$, is the \emph{half} of ``charge" noted
$Q$ in this Section, $2\IQ=Q$. This comes from the $2\pi$ resp. $4\pi$ between the definitions
(\ref{chargdef}) and (\ref{GNOquant}).}.

Those $Q$'s satisfying the quantization condition (\ref{chargdef}) 
[with ``$\exp$'' meant in $K$]
form the \emph{charge lattice} denoted by $\Gamma_Q$. It depends on the global structure, but it always contains
$\Gamma_P$, the lattice of contractible loops. $\Gamma_P$ and $\Gamma_Q$ are actually the same for the covering group $\wK$.
More generally, two loops $\exp[2\pi itQ_1]$ and
 $\exp[2\pi itQ_2]$ are homotopic if and only if $Q_1-Q_2$ belongs to $\Gamma_P$, so that $\pi_1(K)$ is the quotient of the lattices $\Gamma_Q$ and $\Gamma_P$.

On the other hand, the charge lattice $\Gamma_Q$ is contained 
in the $W$-lattice $\Gamma_W$, because for any root $\alpha$ and charge $Q$,
\beqa
1&=&\big(\exp[2\pi iE_\alpha]\big)
\big(\exp[2\pi iQ]\big)\big(\exp[-2\pi iE_\alpha]\big)
=
\exp\left[2\pi i\big(e^{2\pi iE_\alpha}Q e^{-2\pi iE_\alpha}\big)
\right]\nn
\\[8pt]
&=&e^{2\pi i\alpha(Q)}\exp[2\pi iE_{\alpha}]
=e^{2\pi i\alpha(Q)},
\nn
\eeqa
and hence $\alpha(Q)$ is an integer.

The three lattices introduced above satisfy therefore the relation
\beq
\Gamma_P\subset\Gamma_Q\subset\Gamma_W.
\label{2.6}
\eeq
In general, $\widetilde{\exp}[2\pi iW_j]$ is not unity in the fundamental representation of $\wK$. It is however unity in the
adjoint representation. 
\beq
\widetilde{\exp}[2\pi iW_j]=z_j
\label{2.7}
\eeq
belongs therefore to the \emph{centre} of $\wK$. Hence the two lattices
$\Gamma_P$ and $\Gamma_W$ coincide for the adjoint group\footnote{Note that 
the correspondence $W_j\sim z_j$ is one-to-one only
for $\SU(N)$ since for the other groups there are $r$ $W$'s
but less than $r$ elements in the centre.}.

\goodbreak
On the other hand, the correspondence $W\sim z$ can be made one-to-one by restricting the
$W$'s to those ones, $\oW$'s (say), for which the geodesics
$\widetilde{\exp}[2\pi i\oW t]\ (0\leq t\leq1)$ are geodesics of \emph{minimal length} from $1$ to $z$ i.e. for which $\tr W^2$
is minimal for each $z\in Z$. (Since the weights $W$ are all of different lengths and are unique up to conjugation, the $\oW$
for each $z\in Z$ will be unique up to conjugation). Such co-weights
$\oW$ are called \emph{minimal vectors} or \emph{minimal co-weights} \cite{GOCS}, and a simple intuitive way to find them
(indeed an alternative way to introduce them) is as follows.

In terms of roots $\alpha$, the $0,\pm1$ property (which 
will be crucial for the stability investigation \cite{BN,COL,GOCS} )
may be expressed by saying that for any positive root $\alpha$,
\beq
\alpha(\oW)=0,\pm1.
\label{2.11}
\eeq
If one considers in particular the expansion of the highest root $\theta$ in terms
of the primitive roots $\alpha_i$, $\theta=\sum h_i\alpha_i,\
h_i\geq1$, and applies (\ref{2.11}), one sees that $\alpha_i(\oW)$ can be non-zero for only one primitive root, $\oalpha_i$ (say), and that the coefficient $\stackrel{\smallcirc}{h}_i$ of $\oalpha_i$ must be unity \cite{GOCS,HoRa}.
This result provides us with a simple, practical method of identifying the $\oW$'s in terms of primitive weights, namely as the duals to those primitive roots for which the coefficient in
the expansion of $\theta$ is unity \cite{GOCS,HoRa}.

The $W$-lattice containing the charge lattice, together with the root planes, form the \emph{Bott diagram} \cite{Bott} of $K$.
Those vectors satisfying the ``minimality'' or ``stability'' condition (\ref{2.11}) either lie in the
centre or belong to the root plane which is the closest to the centre.

Magnetic monopoles belong to topological classes, described by the
first homotopy group of the residual symmetry group, $H$.
Now for any compact and connected Lie group $H$, $\pi_1(H)$
is of the form
\beq
\pi_1(H)=\IZ^p\oplus\IT,
\label{3.4}
\eeq
where $p$ is the dimension of the centre $Z$ of $H$
and $\IT$ is a finite Abelian group \cite{HRCMP}. In fact,
$\IT$ is isomorphic to $\pi_1(K)$, where $K$ is the compact and semisimple subgroup of $H$ generated by
$\gk=[\gh,\gh]$.
The free part $\IZ^p$ provides us with $p$ integer
``quantum'' numbers $m_1,\dots, m_p$. 

On the other hand, (twice) the ``GNO charge'' of a monopole mentioned in the previous Section  is a ``
charge'' in the sense defined here, and belongs therefore to the
charge lattice $\Gamma_Q$.

 The projections of the charge lattice
$\Gamma_Q$ into the centre is a $p$-dimensional lattice there, generated over the integers by $p$ vectors 
$\Psi_1,\dots\Psi_p$. 
The simplest and physically most relevant case is when the homotopy group
$\pi_1(H)$ is described by a single integer quantum number $m$. This happens when the Lie algebra $\gh$ of $H$ has a $1$-dimensional centre generated by a single vector $\Psi$ and the
semisimple subgroup $K$ is simply connected.

Now the fundamental statement  in Refs. \cite{COL,GOCS} says

\Thm (Goddard-Olive --- Coleman):~\textit{For any compact Lie group $H$, each topological sector contains an [up to conjugation] unique stable charge
$\oQ$.}
\vskip2mm

Curiously, Coleman  \cite{COL}, stated this theorem generally, but only proved
it for $H=\SO(3)$, which appears spurious. Incredibly, his
proof already contains the germ of the general proof \cite{HoRa,HORR88},
though.
The strategy is to reduce the problem to the adjoint group by factoring
out the centre; this leaves us with the semisimple part alone, $\gk$.
Now any semisimple Lie algebra can be decomposed into a sum of simple
Lie algebras, $\gk = \gk_1+\dots+\gk_s$, and the minimal charge ---
which, for a simple adjoint group, is the same as a minimal co-weight ---
can be checked by inspection using the list of simple Lie algebras, see e.g.
\cite{Humphreys}.

\vskip2mm
The examples below may help to understand the 
general theory outlined above.

 \goodbreak 
\kikezd{Example 1:~$H=\SO(3)$}

\begin{figure}
\begin{center}
\includegraphics[scale=0.75]{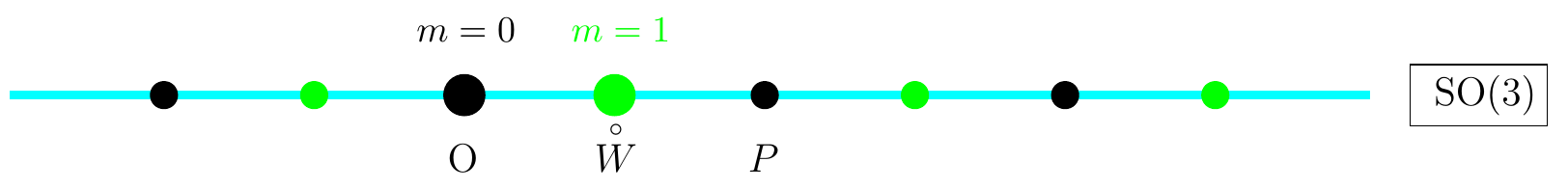} 
\end{center}\vspace{-5mm}
\caption{{\it  The Bott diagram of $\SO(3)\simeq\SU(2)/\IZ_2$,
 the adjoint group of $\SU(2)$.
 The minimal co-weight $\oW=\oW_1$ is now a charge, so that $\Gamma_W=\Gamma_Q$. $\pi_1(\SO(2))\simeq\IZ_2$,
$\oW_0=0$ and $\oW$ are the minimal charges of the two topological sectors. 
} }
\label{SO3Bott}
\end{figure}

The simplest non-trivial example is when the residual Lie algebra is
$\gh=\su(2)\simeq\so(3)$, see Fig.\ref{SO3Bott}. Then the residual gauge group can be 
either $H=\SU(2)$ which is simply connected and has therefore
trivial topology. The unique minimal charge is the vacuum, $\oQ=0$.
 
Non-trivial topology can, however, be obtained by changing the global structure by factoring out the centre, 
$Z\simeq\IZ_2$, yielding $H=\SO(3)$.  Then we have two topological
sectors, labelled by $m=0$ and $m=1$.

In detail, the Cartan algebra of $\su(2)$ consists of 
traceless diagonal
matrices generated by $\sigma_3$. The
only positive root $\alpha$ is the difference of the diagonal entries, 
\beq
H_\alpha=\sigma_3=\barr{cc}1&0\\0 &-1\earr,
\quad
E_+=\sigma_+=\barr{cc}0&1\\ 0&0\earr,
\quad
E_-=\sigma_-=\barr{cc}0&0\\ 1&0\earr,
\label{su2c}
\eeq
generate the complexified Lie algebra $\su(2)^c$.
The unique minimal co-weight is
\beq
\oW=\2\sigma_3=\2\barr{cc}1&0\\ 0&-1\earr.
\label{su2W}
\eeq
The root lattice consists of integer multiples of $\oW$.
The unique primitive charge is 
\beq
P=2\oW=\sigma_3.
\eeq

The centre of $\SU(2)$ is $Z\simeq\IZ_2=\{\II_2,-\II_2\}$ and the
adjoint group, $\SO(3)\simeq\SU(2)/\IZ_2$, has two topological sectors, represented by the curves in $\SU(2)$
\beq
\gamma_0(t)=\exp[2\pi i\sigma_3t]
\quad\hbox{and}\quad
\gamma_1(t)=\exp[\pi i\sigma_3t],
\eeq
$0\leq t\leq1$, respectively. Note that while $\gamma_0$ is a loop in 
$\SU(2)$, $\gamma_1$ is only ``half of a loop'', as it  ends in $-\II_2$.
Factoring out the centre, both curves project to loops in
$\SO(3)$; $\gamma_0$ projects into a contractible one, but
$\gamma_1$ represents the non-trivial class $[1]$.
Conversely, when lifted to $\SU(2)$, all contractible in $\SO(3)$ loops end 
at $\II_2$ and the lifts of loops in the non-trivial class end at $-\II_2$.
The stable charges of the respective sectors are
\beq
\oQ{}^{(0)}=0
\qquad\hbox{and}\qquad
\oQ{}^{(1)}=\oW=\2\sigma_3.
\eeq
Then any charge is
\beq
Q=\oQ{}^{(m)}+n{P}=\left\{\barray{cl}
n\sigma_3 &\hbox{trivial sector}
\\[8pt]
(\2+n)\sigma_3,
 &\hbox{nontrivial sector}
 \earray\right.
\label{genSO3charge}
\eeq
where $n$ is some integer.

\goodbreak
\kikezd{Example 2:~ $H=\UN(2)$}

Another simple case of considerable interest is that of when the little group $H$ of the Higgs field is $H=\UN(2)$. The Lie algebra is decomposed into  centre plus the semisimple part,
$\gh=\un(1)\oplus\su(2)$, which is indeed used to describe
electroweak interactions.

\begin{figure}
\begin{center}
\includegraphics[scale=.55]{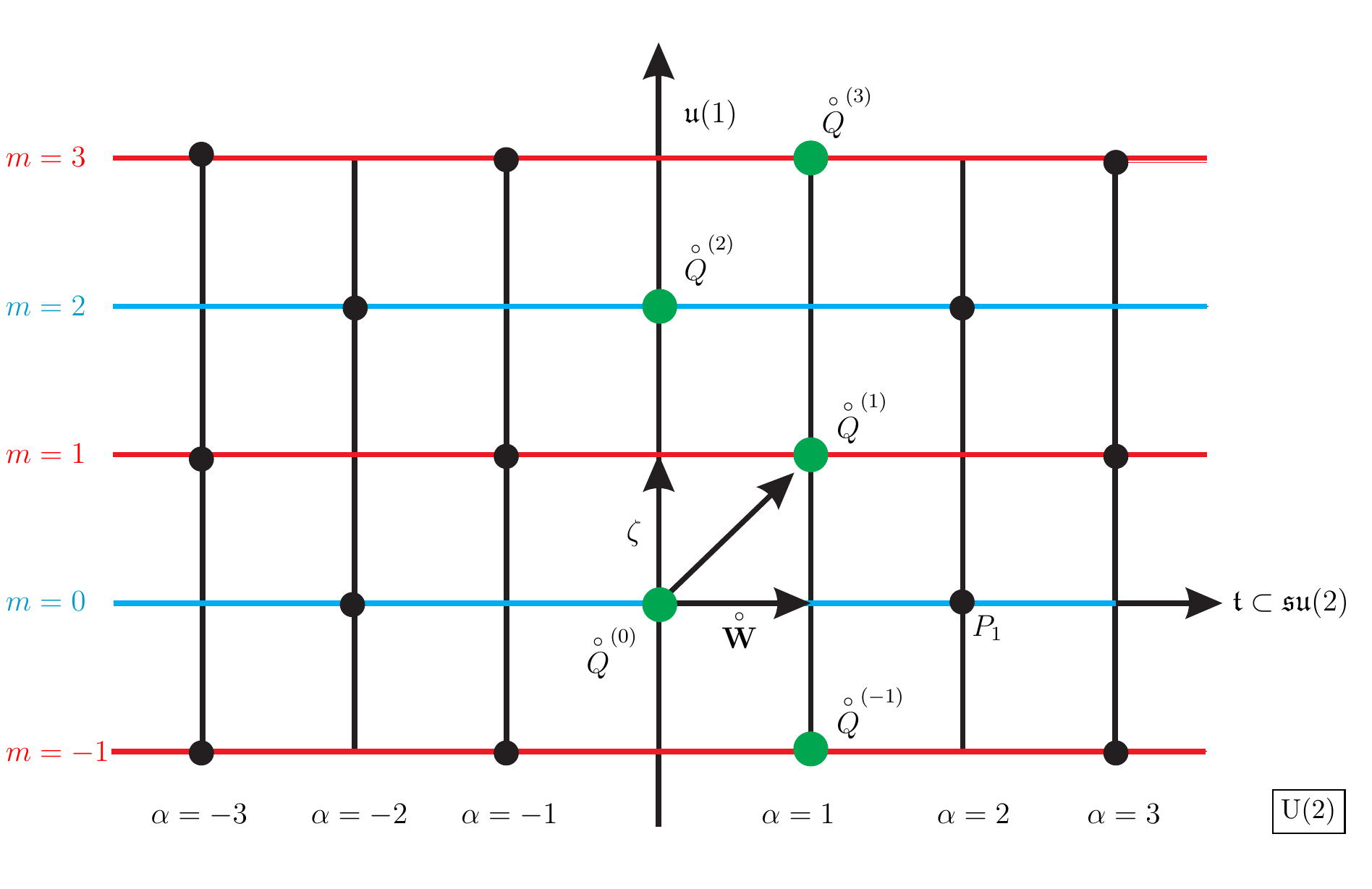}
\caption{{\it Bott diagram of $\UN(2)$. The horizontal axis represents the Cartan algebra of $\su(2)$, and the vertical axis is the centre generated by $\zeta={\rm diag}(\2,\2)$. The co-weight lattice $\Gamma_{W}$ consists of
the intersections all (horizontal and vertical) lines, generated by $\zeta$ and the minimal vector $\oW\equiv W_1=\sigma_3/2$.  
$P=P_1=2\oW=\sigma_3$ is the unique primitive charge. 
The root ``planes'' are vertical lines which intersect the horizontal axis at integer multiples of $\oW$.  
The charges belonging to the charge lattice $\Gamma_{Q}$ and represented by dots are $Q=m\zeta+\oW_{[m]}+nP_{1}$,
 where $m$ is the
topological quantum number, $[m]=m$} (mod $2$), $W_{0}=0$.
{\it Those charges in the same horizontal lines are the topological sectors labelled by $m$.
The pattern is periodic in $[m]$.
In each topological sector, the minimal charge 
$\oQ{}^{(m)}\!=m\zeta+\oW_{[m]}$  is the one which is the closest to the centre. 
}}
\end{center}
\label{U2Bott}
\end{figure}

  The Cartan algebra consists of diagonal matrices (combinations of $\sigma_3$ and of the unit matrix $\II_2$). 
In fact,
 $H\in\gt,\, E_{\pm}$ and the primitive weight $\oW=\oW_1$ are as in
(\ref{su2c}).
The only primitive vector, $\oW_1$, is also a minimal one. In fact,
$\exp 2\pi i\oW_1=-\II_2$. $P_1=2\oW_1=\sigma_3$ generates the charge lattice of $K=\SU(2)$ which is also the topological zero-sector
of $\UN(2)$. The topological sectors are labelled by a single integer $m$, defined by projecting onto the centre,
\beq
Q_{||} 
=m\,{\rm diag}(\2,\2)\equiv m\,\zeta.
\eeq 
Note that $\zeta={\rm diag}(1/2,1/2)$ generates the centre, see Fig. \ref{U2Bott}. Note that only $2\zeta$ is a charge, $\exp[4\pi i\zeta]=1$.
The [up to conjugation] unique minimal charge of the sector
$m$ is
\beq
\oQ{}^{(m)}=m\zeta+\oW_{[m]}=\left\{
\begin{array}{lll}
{\rm diag\ }(k,k)
&\hbox{for} &m=2k
\\[8pt]
{\rm diag\ }(k+1,k)
&\hbox{for} &m=2k+1
\end{array}\right.
\label{9.2}
\eeq
where $[m]$ is $m$ modulo $2$ and $\oW_0=0$ by convention.
Any other charge of Sector $m$ is
\beq
Q{}^{(m)}=\oQ{}^{(m)}\!+\;{n}P_1=
\oQ{}^{(m)}\!+\;{n}\sigma_3
=\oQ{}^{(m)}\!+\;{\rm diag }(n,-n).
\label{9.3}
\eeq
\begin{figure}
\begin{center}
\includegraphics[scale=0.6]{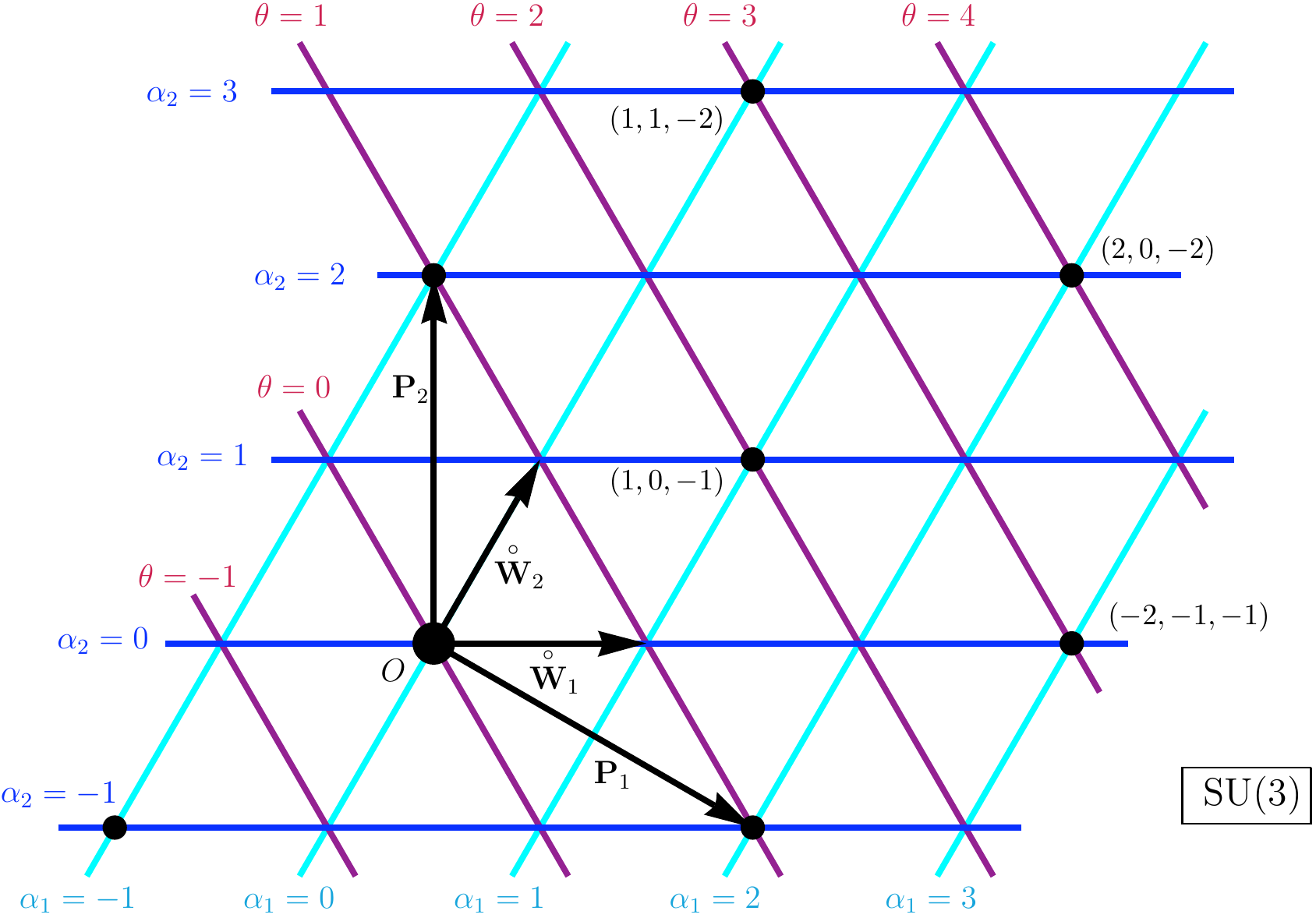}
\end{center}\vspace{-4mm}
\caption{\it The Bott diagram of $\SU(3)$, generated by the 
minimal co-weight vectors 
$\oW_1=\smallover1/3{\rm diag}(2,-1,-1)$ and
$\oW_2=\smallover1/3{\rm diag}(1,1,-2)$. $P_1={\rm diag}(1,-1,0)$ and $P_2={\rm diag}(0,1,-1)$ are the primitive charges and the two primitive roots are $\alpha_1=\tr(P_1\,\cdot\;)$
and $\alpha_2=\tr(P_2\,\cdot\;)$. 
The three families of root planes, labelled by the values of $\alpha_1,\,\alpha_2$ and of the highest root $\theta$, intersect at an angle $\pi/3$. Charges and primitive charges
coincide, $\Gamma_Q=\Gamma_P$.}
\label{SU3Bott}
\end{figure}
\goodbreak
\kikezd{Example 3:~$\SU(3)/\IZ_3$}

$\SU(3)=\widetilde{K}$ is simply connected; its centre is $\IZ_3$, with elements
\beq
z_0=\hbox{\rm diag}\big(1,1,1\big),
\;
z_1=\hbox{\rm diag}\big(e^{4\pi i/3},e^{-2\pi i/3}, e^{-2\pi i/3}\big),
\;
z_2=\hbox{\rm diag}\big(e^{2\pi i/3},e^{2\pi i/3}, e^{-4\pi i/3}\big).
\label{SU3Z}
\eeq
The co-weights are now charges, so that
\beq
\Gamma_W=\Gamma_Q.
\eeq
The adjoint group of $\SU(3)=\widetilde{K}$ is $\SU(3)/{\rm Z}$ has therefore three homotopy classes labelled by the $z_m,\, m=1,2,3$  Each of such loop is homotopic to one of the form
$ 
\widetilde{\exp}\left[2\pi i Qt\right]
\,
0~\leq~t~\leq~1$,
\beq
 \widetilde{\exp}\left[2\pi i Q\right]=z_m,
\eeq
where $\widetilde{\exp}$ means the exponential in the covering group
$\widetilde{K}=\SU(3)$. A loop which has class $[m]\in\IZ_3$ ends at
$z_a,\,a=0,1,2$, and is homotopic to one with charge
\beq
Q=\oQ{}^{(m)}+n_1P_1+n_2P_2= \oW_m+n_1P_1+n_2P_2
\label{mQ}
\eeq
The unique minimal charge of class $[m]$ is
$\oQ{}^{[m]}=\oW_{m}$.

\begin{figure}
\begin{center}\vspace{-2mm}
\includegraphics[scale=0.6]{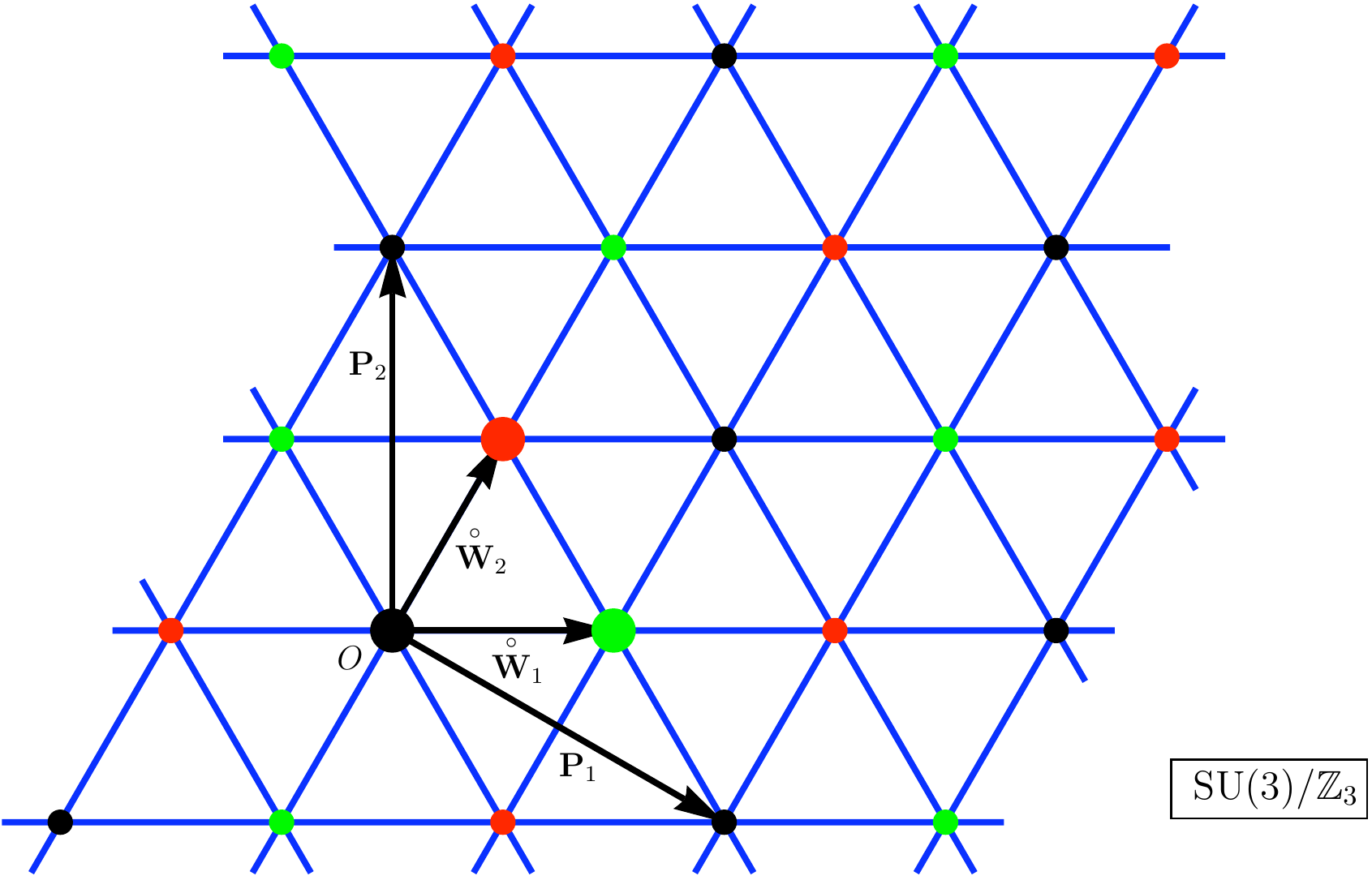}\vspace{-2mm}
\end{center}
\caption{{\it  The Bott diagram of $\SU(3)/\IZ_3$, the adjoint group of $\SU(3)$. The root lattice is that of $\su(3)$, with identical co-weights and primitive charges, the $W$'s and $P$'s. The only difference is that the minimal $W$'s are now charges,
so that $\Gamma_W=\Gamma_Q$. $\pi_1(\SU(3)/\IZ_3)\simeq\IZ_3$ and
$\oW_m,\ m=0,1,2,\, \oW{}_0=0$ are the minimal charges of the three topological sectors. A  charge in Sector $m=0,1,2$ is of the form
$Q=\oW_m+n_1P_1+n_2P_2$.
} 
}
\label{SU3ZBott}
\end{figure}

\kikezd{Example 4:~$H=\UN(3)$}

A physically  relevant example is when the Higgs little
group is $H=\UN(3)$ i.e. locally $\su(3)_c\oplus\un(1)_{em}$, the symmetry of strong and electromagnetic interactions.

\begin{figure}
\begin{center}
\includegraphics[scale=.8]{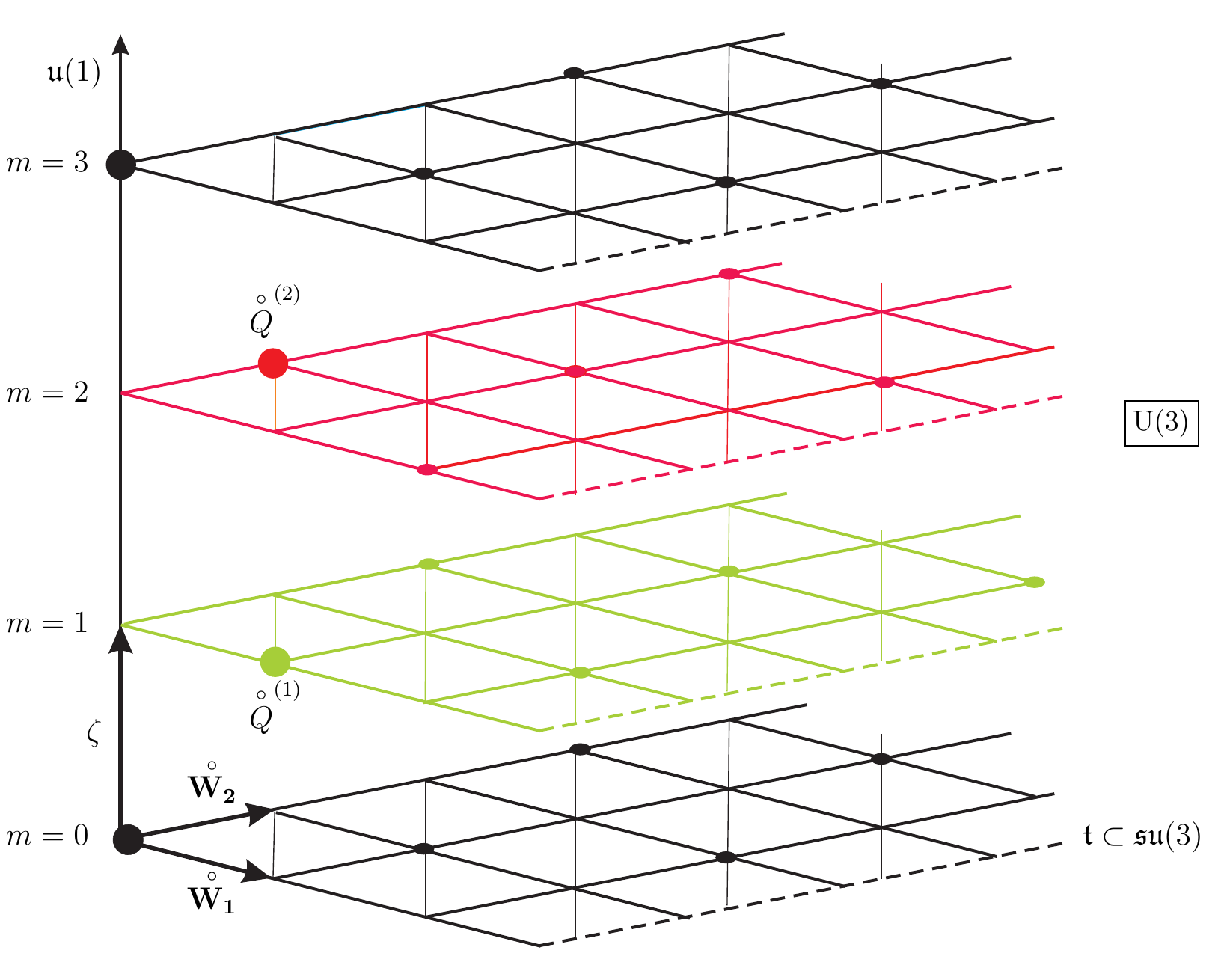}
\caption{\it Bott diagram of $\UN(3)$. The vertical axis represents the centre generated by $\zeta={\rm diag }(1/3,1/3,1/3)$, and the horizontal plane $\gt$ is the Cartan algebra of $\SU(3)$, shown in more detail on Fig. \ref{SU3Bott}. The charges are
$m\,\zeta+\oW_{[m]}+n_1P_1+n_2P_2$, where $P_1$ and $P_2$ are
the primitive charges of $\SU(3)$. 
 The horizontal planes are the topological sectors. Sector $m$ is obtained from the vacuum sector by shifting by $m\,\zeta+\oW_{[m]}$. They also coincide with the $m$ mod(3) sectors of the adjoint group, $\SU(3)/\IZ$.
 In each sector, the unique minimal charge
  $\oQ{}^{(m)}$, is the one which is closest to the centre.
 The diagram is periodic in $m$ mod 3.}
 \end{center}
\label{U3Bott}
\end{figure}

The diagram is now three-dimensional, the central $\un(1)$ being the vertical axis on Fig. 
\ref{U3Bott},
and $\gt$ being the horizontal plane. The primitive roots are 
\beq
\alpha_1(X)=X_1-X_2
\qquad\hbox{and}\qquad
\alpha_2(X)=X_2-X_3
\eeq 
($X={\rm diag}(X_1,X_2,X_3)$). The corresponding weight vectors, 
\beq
\oW_1={\rm diag\ }(2/3,-1/3,-1/3)\qquad\hbox{and}\qquad
\oW_2={\rm diag\ }(1/3,1/3,-2/3)
\eeq 
are also minimal vectors:
they exponentiate to the elements in the
$Z=\IZ_3$-centre of $\SU(3)$.

The highest root is $\theta=\alpha_1+\alpha_2$, and the charge lattice
of $K=\SU(3)$ is generated by 
\beq
Q_1={\rm diag}(1,-1,0)
\qquad\hbox{and}\qquad
Q_2={\rm diag}(0,1,-1).
\eeq

The topological sectors are labelled by an integer $m$.
In fact, the projection of Sector $m$ onto the centre is
\beq
z(Q^{(m)})=
m\,\zeta=
m\,{\rm diag }(\smallover1/3,\smallover1/3,\smallover1/3).
\eeq
The unique minimal charge in sector $m$ is
\beq
{\oQ}{}^{(m)}=m\zeta+\oW_{[m]}=
\left\{\begin{array}{l}
{\rm diag}(k,k,k)
\\[6pt]
{\rm diag}(k+1,k,k)
\\[6pt]
{\rm diag}(k+1,k+1,k)
\end{array}\right.
\quad
\hbox{for}
\quad
m=
\left\{\begin{array}{l}
3k
\\[6pt]
3k+1
\\[6pt]
3k+2
\end{array}\right.
\label{9.9}
\eeq
where $[m]$ means $m$ modulo $3$. Any other charge is
\beq
Q={\oQ}^{(m)}+Q'=
{\oQ}^{(m)}+n_1P_1+n_2P_2=
{\oQ}^{(m)}+{\rm diag}(n_1,n_2-n_1,-n_2).
\label{9.10}
\eeq

\kikezd{Example 5:~$H=\big(\UN(1)\times{\rm Sp}(4)\big)/\IZ_2$}

\begin{figure}
\begin{center}\hspace{-4mm}
\includegraphics[scale=1.1]{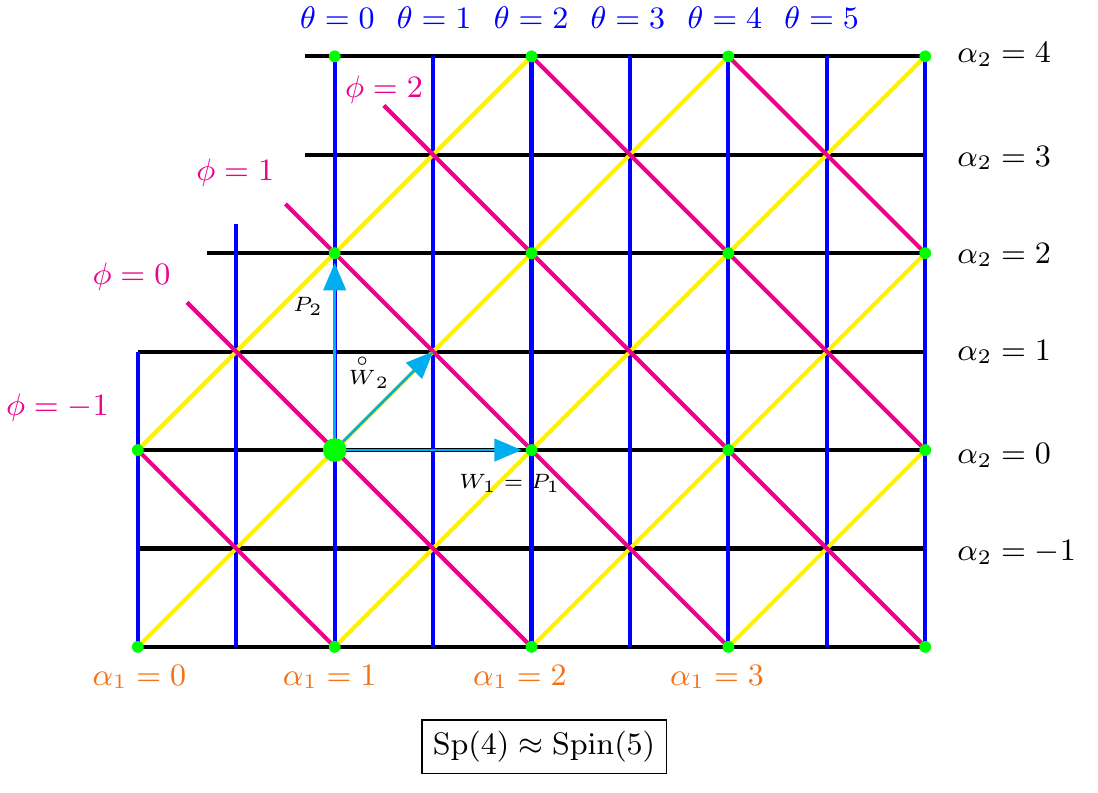}\vspace{-4mm}
\end{center}
\caption{\it Bott diagram of\, ${\rm Sp}(4)\simeq
 {\rm Spin}(5)$, the double covering of $\SO(5)$. The primitive
charges are $P_1={\rm diag}(1,0,-1,0)$ and $P_2={\rm diag}(0,1,0,-1)$. The two co-weights $W$'s are 
$W_1={\rm diag}(1,0,-1,0)$ and  
$\oW_2={\rm diag}(\2,\2,-\2,-\2)$ out of which only $\oW_2$ is minimal. There are $4$ families of root planes, namely $\alpha_1$ and $\alpha_2$ associated with the two co-weights, augmented with
$\varphi=\alpha_1+\alpha_2$ and the highest root 
 $\theta=2\alpha_1+\alpha_2$. 
}
\label{Figure7}
\end{figure}

To have a simple example where not all primitive weights are minimal, let us assume that the residual group is
$$
H=\big(\UN(1)\times{\rm Sp}(4)\big)/\IZ_2.
$$
Then $\gk={\rm sp}(4)\simeq\so(5)$, and $\widetilde{K}$ is 
${\rm Spin}(5)$, the
double covering of $\SO(5)$. $\gk$ can be represented by
$4\times4$ symplectic matrices with a $2$-dimensional Cartan algebra, say 
$\gt={\rm diag}(a,b,-a,-b)$. The charge
lattice consists of those vectors in $\gt$ with integer entries, cf. Fig.\ref{Figure7}.
Let us choose the primitive roots $\alpha_1=\tr(H_1\,\cdot\,)$
and $\alpha_2=\tr(H_2\,\cdot\,)$, where
\beq
H_1=\2{\rm diag}(1,-1,-1,1)
\quad\hbox{and}\quad
H_2=\2{\rm diag}(0,1,0,-1).
\label{9.14}
\eeq
These vectors dual to the primitive roots are
\beq
W_1=\2{\rm diag}(1,0,-1,0)
\quad\hbox{and}\quad
\oW_2=\2{\rm diag}(1,1,-1,-1).
\label{9.15}
\eeq
Then the discussion of  Sec. \ref{Liealgebra} shows that only
$\oW_2$ is minimal: For example, only $\oW_2$ exponentiates into the non-trivial element $(-\II)$ of Sp$(4)$: 
$$\exp2\pi W_1=1,
\qquad
\exp2\pi\oW_2=-\II.
$$
 In other words, while $W_1$ is already a charge, $\oW_2$ is only half-of-a-charge. 
Alternatively, the two remaining positive roots are $\varphi=\alpha_1+\alpha_2$ and the highest root is $\theta=2\alpha_1+\alpha_2$. 

Let the integer $m$ label the topological sectors. For $m$ even,
$m=2k$, the unique minimal charge belongs to the centre,
\beq
\oQ^{(2k)}=k\,\zeta
\label{9.16}
\eeq
where $\zeta$ is a generator of the centre normalized so that
$2\zeta$ is a charge.

For $m$ odd, $m=2k+1$, the unique minimal charge is rather
\beq
\oQ^{(2k+1)}=({k+\2})\,\zeta+\2\oW_2\,.
\label{9.17}
\eeq

\section{Finite energy solutions: the GNO charge}\label{solutions}

Now we return to our monopole investigations.

The only condition imposed on the YMH configurations $(\bA,\Phi)$ up to this point is that the energy be finite. But it is obviously of interest to consider the special case of finite energy configurations that are also \textit{solutions of the YMH field equations} (\ref{YMHeq1}) -- (\ref{YMHeq2}),
 i.e.,
\beq \label{feqbis}
\bD^2\Phi=\frac{\delta U}{\delta\Phi},
\qquad
\bD\times\bB= ie[\bD\Phi,\Phi],
\eeq

\vskip2mm
Finite energy \emph{solutions} may be classified using data
referring to the magnetic field alone.
 For this it is sufficient to consider the field equations (\ref{feqbis}) 
for large $r$. Remembering that $$
\bB(\bx)\to\frac{\bb(\theta,\varphi)}{r^2}
=\frac{b(\theta,\varphi)}{r^2}\,\hat{\bx},
$$
cf. (\ref{asgaugefields}) and putting
$
\Phi(\bx)\approx\Phi(\theta,\varphi)+\eta(r,\theta,\varphi),
$
our field equations reduce to 
\beq
\bigtriangleup\eta_\alpha=\left(\frac{\p^2U}{\p\Phi_\alpha\p\Phi_\beta}\right)\eta_\beta
\quad\hbox{and}\quad
\bD\times\bb=0
\label{4.2}
\eeq
in the generic  case.
 The first equation here shows that, for solutions  the generic finite-energy condition $\eta\to0$ is sharpened to an exponential fall-off of $\eta$ \footnote{$\bigtriangleup\eta=0$ and
$\bD\times\bb=0$ in the Bogomolny case.
Then 
$\bD^2\eta=0$ is consistent with $\eta=b(\Omega)/r$.)}. 

Since $\Phi(\theta,\varphi)$ and $b(\theta,\varphi)$ are the only components of the
field configuration that survive in the asymptotic region,  
 the only possible asymptotic classification of the configurations; within each topological sector, is by $b(\theta,\varphi)$. The conditions
satisfied by $b(\theta,\varphi)$ are then contained in the second equation in
(\ref{4.2}), which may be written as
\beq
D_i b\equiv \p_i b+i[A_i,b]=0.
\label{covconstcond}
\eeq
This equation shows that $b(\theta,\varphi)$ is {\it covariantly constant} on the sphere at infinity and thus takes it values lie on an $H$-orbit. Therefore 
$$
b(\theta,\varphi)=\left\{\barray{ll}
h_N(\theta,\varphi)\,\IQ\,h_N^{-1}(\theta,\varphi) \qquad
&\hbox{in}\,\,\, N
\\[10pt]
h_S(\theta,\varphi)\,\IQ\,h_S^{-1}(\theta,\varphi)&\hbox{in}\,\,\,  S
\earray\right.,
$$ 
where 
$\IQ=b(E)$, the value of the magnetic field at the ``east pole'', belongs to $\gh$. Plainly, $\IQ$ is unique up to 
global gauge rotations, and there is thus no loss of generality in choosing it in a given Cartan algebra.

 In the singular gauge where 
$b(\theta,\varphi)=\IQ$, the loop in (\ref{paralleltransport}) 
i.e. $h^A={\cal P} \big(\exp\displaystyle\oint \bA\big)$
can readily be evaluated: it is simply
\beq
h(\varphi)=\exp[2i\IQ\,\varphi],
\qquad
0\leq\varphi\leq2\pi,
\label{4.4}
\eeq
and the periodicity of $\varphi$ provides us with the
\emph{quantization condition} 
$$
\exp[4\pi i\IQ]=1
$$
cf. (\ref{GNOquant}).
Using the terminology introduced in Sect. \ref{Liealgebra}, 
\beq
2\IQ=Q
\eeq is a \emph{charge}.
Conversely, any quantized $\IQ$ defines an asymptotic solution, namely 
\beq
\bA=\bA^D\IQ
\qquad\hbox{i.e.}\qquad
A_\theta=0,\quad
A_\varphi^{\pm}=(\pm1-\cos\theta)\,\IQ,
\label{imbDmon}
\eeq
in the Dirac gauge, so that 
 (\ref{4.4}) is in fact the transition function. (The $\pm$ superscript refers to the northern and southern hemispheres, respectively.) Hence
 \beq
b=\IQ
\label{GNOgauge}
\eeq
in this gauge, 
and Eqns. (\ref{imbDmon}) -- (\ref{GNOgauge}) say that at infinity,  the magnetic field can be gauge-transformed into a fixed direction in $\gh$:
 asymptotically,
any  't~Hooft--Polyakov monopole is an imbedded Dirac monopole. In mathematical terms, 
\emph{the holonomy of pure YM on $\IS^2$ is
one-dimensional} \cite{HPAMonop} \footnote{The statement
generalizes to Riemann surfaces \cite{ABOTT}.}.

 Solutions can thus be classified by \emph{charges of $H$.} The (half)charge $\IQ$ was introduced by Goddard, Nuyts and Olive \cite{GNO}.

According to (\ref{3.9}), for solutions of the field equations
the expression for the ``Higgs'' quantum numbers $m_a$
reduces to
\beq
m_a=2\frac{\tr(\IQ\,\zeta_a)}{\tr(\zeta_a^2)},
\qquad
a=1,\dots,p.
\label{4.5}
\eeq
Here we would like to mention that the 
``topological quantum numbers'' $m_k$'s are 
related to but still distinct from
 the magnetic charge: the latter can, in fact, only be defined by identifying the ``electromagnetic'' direction, which requires using
a covariantly constant direction field which generates and ``internal symmetry''  \cite{HRcolor}. Then, suitably defining the electric
charge operator provides us with
generalized [fractional] Dirac quantization conditions \cite{GORPP,HRCMP,GOCQ,HRcolor}.

\goodbreak

We are, at last, ready to identify the  minimal charge in each topological
sector.
Let us indeed consider a GNO charge $\IQ$ and denote its topological sector by
$m$. let us decompose $\IQ$ into central and semisimple parts
$\IQ_{||}$ and $\IQ_{\perp}$, respectively,
$
\IQ=\IQ_{||}+\IQ_{\perp}.
$
By (\ref{4.5}),
$$
2\IQ_{||}=2z(\IQ)=\sum_{k=1}^pm_k\zeta_k.
$$
Observe that
\beq
z=\exp[4\pi i\IQ_{||}]=\exp[-4\pi i\IQ_{\perp}]
\label{4.6}
\eeq
lies simultaneously in $Z(H)_0$ (the connected component 
of the centre of $H$) and in the semisimple subgroup $K$, and thus also in $Z(K)$, the centre of
$K$. Let us decompose $\gk=[\gh,\gh]$ into simple
factors, 
$$
\gk=\gk_1\oplus\dots\oplus\gk_s,
$$
and denote by $\widetilde{K}_j$ the simple and simply connected group, whose algebra is $\gk_j$. As explained in
Sec. 2, $K$ is of the form $\widetilde{K}/C$, where
$C=C_1\times\dots\times C_s$ is a subgroup of the
centre $Z=Z(\widetilde{K})$ of $\wK=\big[\wK_1\times\dots\times\wK_s\big]$, $C_j$ being a subgroup of $Z(\wK_j)$.

The situation is particularly simple when $K$ is simply connected,
$K=\wK$, when the central part $\IQ_{||}$ contains all
topological information. Indeed, $z$ is uniquely
written in this case as
\beq
z=z_1\cdots z_s
\qquad\hbox{where}\qquad
z_j\in Z(\wK_j).
\label{4.7}
\eeq
However, as emphasized in Sec. \ref{Liealgebra}, the central elements of a simple and simply connected group are in one-to-one correspondence with the minimal
$\oW$'s and thus, for each $z$ in the centre, there exists a unique set of $\oW_j$'s (where $\oW_j$ is
either zero or a minimal vector of $\gk_j$) such that
\beq
z=\big(\exp[-2\pi i\oW_1]\big)\dots
\big(\exp[-2\pi i\oW_s]\big)=
\exp\big[-2\pi i\sum_{k=1}^s\oW_k\big]
=
\exp\big[-2\pi i\oW{}^{(m)}\big].
\label{4.8}
\eeq
$\oW{}^{(m)}$ depends only on the sector (and not on $\IQ$ itself), because all charges of a sector have
the same $\IQ_{||}$. Hence the {entire sector}
can be characterized by giving
\beq
2\oIQ{}^{(m)}=\sum_km_k\zeta_k+\oW{}^{(m)}.
\label{4.9}
\eeq

By (\ref{4.8}) $2\oIQ{}^{(m)}$ is again a charge, 
$\exp[4\pi\oIQ{}^{(m)}]=1$, and it obviously belongs
to the sector $m$. Furthermore, 
$$
\exp[4\pi i(\IQ-\oIQ)]=\exp[4\pi i\IQ]\exp[-4\pi i\oIQ]=1
$$
shows that $2\IQ'=2(\IQ-\oIQ)$ is in the charge lattice of $K$.

The situation is slightly more complicated if $K$ is
not simply-connected, so that the semisimple part also contributes to the topology. Since $C$ is now non-trivial,
the expansion (\ref{4.7}) is not unique, and $z_j$ can be replaced rather by
$z_j^*=z_jc_j$, where $c_j$
belongs to the subgroup $C_j$ of $Z(\wK_j)$. 
But $z^*_j$ is just another element of
$Z(\wK_j)$, so it is uniquely $z^*_j=\oW_j{}^*$
for some minimal $\oW_j{}^*$ of the simple factor $\wK_j$.
Equation (\ref{4.8}), with all $\oW_j$'s replaced by the $\oW_j{}^*$'s, is still valid, so that (\ref{4.9}) is a charge also now. However, since $\pi_1(K)=C=C_1\times\dots\times C_s$, those loops generated by $Q$ and
$Q^*$  now belong to different topological sectors.

We conclude that a topological sector contains a unique charge $\IQ$ of the form (\ref{4.9}) also in this case, and that, in full generality, any other monopole charge
is uniquely of the form
\beq
\IQ=\oIQ+\IQ'=\oIQ+\2\sum_i^rn_iP_i,
\label{4.10}
\eeq
where the $n_i$ are integers, and the $P_i,\,i=1,\dots r$
are the primitive charges of $K$ i.e. those which generate contractible loops in all cases. (Obviously, the $P_i$ are sums of primitive charges taken for the simple factors $K_j$).
The integers $n_i$ could be regarded as secondary quantum
numbers which supplement the Higgs charge $m$ but
do not contribute to the topology.

In Sec. \ref{negmodes} we shall show that $\oIQ{}^{(m)}$ is the
\emph{unique stable monopole charge} in the sector $m$.

The situation is conveniently illustrated on the Bott
diagram, see Section \ref{Liealgebra}. 

The classification of finite energy solutions according to the secondary quantum numbers or, equivalently the
matrix-valued charge $Q$ is convenient and illuminating,
but in contrast to the classification of finite energy configurations according to the Higgs charge $m$, it is
\emph{not} mandatory, in the sense that (for fixed $m$) the different charges $Q$ are separated only by \emph{finite} energy barriers~\cite{HORR88}.

\section{Stability analysis}\label{stabanal}

\subsection{Reduction from $\IR^3$ 
to $\IS^2$}\label{reduction}

Now we show that those monopoles for which
$\IQ'\neq$ are unstable. More precisely, we show that for a restricted class of variations the stability problem reduces to a corresponding Yang--Mills problem on
$\IS^2$. This will allow us to prove that with respect to our variations there are
\beq
\nu=2\sum_{q<0}\left(2|q|-1\right)
\label{MorseIndexFormula}
\eeq
independent negative modes \cite{HoRa,FrHab,NAUH}
where $q$ is a negative eigenvalue of a certain operator
involving the GNO charge, (\ref{starcomm}) below.
To this end, let us first introduce the
notation
\beqa
\big(\ba\times\bb\big)_i&=&\varepsilon_{ijk}a_jb_k,
\label{5.2}
\\[6pt]
\big[\ba\times\bb\big]&=&\ba\times\bb+\bb\times\ba
\quad\hbox{i.e.}\quad
\left(\big[\ba\times\bb\big]\right)_i=
\varepsilon_{ijk}\big[a_j,b_k\big].
\nn
\eeqa
Note that $\ba\times\ba$ may be different from zero if 
$\gh$ is non-Abelian.

For $\gh$-valued variations of the \emph{gauge potentials alone}
of the ``Brandt--Neri--Coleman type'', i.e.,  for
\beq
\delta\Phi=0,\quad
 \delta\bA=\ba\in\gh\,,
\eeq
the variations of the gauge field and covariant derivative are easily seen to be 
\beq
\delta\bB=\bD\times\ba,
\quad
\delta^2\bB=-i[\ba\times\ba],
\quad
\delta(\bD\Phi)=-i[\ba,\Phi].
\eeq
where, once again, we assumed for simplicity that the Higgs
field belongs to the adjoint representation. 
 All higher-order variations $\delta^3\bB$ etc. are zero.

The first variation of the energy functional (\ref{YMHEner})  is zero since $(\bA,\Phi)$ is a solution of the field equations. The higher order variations are
\beq\begin{array}{lll}
\delta^2E&=&\displaystyle\int d^3\bx\big\{
\tr(\bD\times\ba)^2-i\tr(\bB\cdot[\ba\times\ba])
-\tr([\ba,\Phi])^2\big\},
\\[8pt]
\delta^3E&=&-3i\displaystyle\int d^3\bx\tr\big\{
(\bD\times\ba)\cdot(\ba\times\ba)\big\},
\\[8pt]
\delta^4E&=&-3\displaystyle\int d^3\bx\tr(\ba\times\ba)^2,
\end{array}
\label{5.3}
\eeq
all higher-order variations being zero. We shall assume that all variations are square-integrable and have non-zero norm, 
\beq
0\neq(\ba,\ba)=\displaystyle\int d^3\bx\tr(\ba)^2<\infty.
\label{normcond}
\eeq

There are some general points worth noting.

 First, since $\delta\Phi=0$,
the only terms in (\ref{5.2}) involves the Higgs field is
$\tr([\ba,\Phi])^2$ and, in the 't~Hooft--Polyakov case
$U\neq0$, $\ba$ must be in the little
group of $\Phi(\theta,\varphi)=\lim_{r\to\infty}\Phi(r,\theta,\varphi)$,
this term vanishes asymptotically. Thus, if we only consider asymptotic variations \cite{BN,COL} i.e. such that $\ba(r,\theta,\varphi)=0$
for $r\leq R$ where $R$ is `sufficiently large' so  that  
the fields assume their asymptotic forms (\ref{asgaugefields}), 
 the Higgs terms  can then dropped in  (\ref{5.3}) and we
 shall only consider the pure Yang--Mills variations,
\beq
\delta^2E=\int d^3\bx\big\{
\tr(\bD\times\ba)^2-i\tr(\bB\cdot[\ba\times\ba])
\big\}\ .
\label{5.4}
\eeq

Second, the only term in (\ref{5.4}) that involves radial
derivatives is the $(\p_r\ba)^2$ term in
$(\bD\times\ba)^2$ and this contribution may be shown 
 to be 
\beq
\delta^2E_r=\int d^3\bx\tr(\p_r\ba)^2=
m^2(\ba,\ba),
\qquad 
m^2=\smallover1/4+\delta^2
\label{5.5}
\eeq
where  $\delta^2>0$  whose  infimum of is
$0$ \cite{COL,HORR88}. Thus, although $\delta^2E_r$ is not negligible, it can be
 regarded as a mass term. Therefore, for each value of $r$, \textit{the variations (\ref{5.4})
are essentially variations on the $2$-sphere at infinity, $\IS^2$}.

Finally, it should be noted that the variations $\ba=\bD\chi$ where $\chi$ is
any scalar, are simply gauge transformations of the background field $\bA$ and leave the energy unchanged. In particular, it is easy to verify that, because $\bA$ satisfies the field equations, the second
variation $\delta^2E$ is zero for the infinitesimal variations $\delta\bA=\bD\chi$. It is therefore convenient to define the `physical' variations
$\ba$ as those which are orthogonal to the $\bD\chi$. 
Using  partial integration,
\beq
0=\displaystyle\int d^3\bx\tr(\ba\cdot\bD\chi)=
-\displaystyle\int d^3\bx\tr(\bD\cdot\ba\,\chi)\quad\Rightarrow\,\quad
\bD\cdot\ba=0,
\label{gaugecondi}
\eeq
since $\chi$ is arbitrary.
The physical variations may also be characterized as those which are divergence-free. As a consequence of the gauge condition $A_r=0$ our variations satisfy also $a_r=0$.

It will be convenient to split (\ref{5.4}) into two terms,
\beqa
\delta^2E
=\underbrace{\displaystyle\int d^3\bx\tr\left\{(\bD\times\ba)^2
+(\bD\cdot\ba)^2\right\}}_{\delta^2E_1\geq0}
+\underbrace{
\displaystyle\int d^3\bx\tr\left\{-i\bB[\ba\times\ba]-(\bD\cdot\ba)^2\right\}}_{\delta^2E_2},
\label{Hessiansplit}
\eeqa
bearing in mind that $\bD\cdot\ba$ is unphysical and may be gauged to zero.

Let us first consider $\delta^2E_1$. From the identity
\beq
\big(\bD\times(\bD\times\ba)\big)_j=-\bD^2a_j+
D_j(\bD\cdot\ba)-i\big([\bB\times\ba]\big)_j
\label{5.8}
\eeq
we have
\beq
\delta^2E_1=\int d^3\bx\tr\left\{(-\bD^2\ba-i[\bB\times\ba])\cdot\ba\right\},
\eeq
and putting $\bb=\lim_{r\to\infty}r^2{\bB}$  yields
\beq
\delta^2E_1=\delta^2E_r+\underbrace{\int
 d^3\bx\, r^{-2}
\tr\left\{\big(\bL^2\ba-
i[\bb\times\ba]\big)\cdot\ba\right\}}_{\delta^2E_{S}}\,,
\label{E1split}
\eeq
where $\bL=-i\bx\times\bD$ is the orbital angular momentum.
$\bL$ is neither conserved nor does it satisfy the $\so(3)$ algebra.
For spherically
symmetric (and hence for asymptotic) fields, the components of the angular momentum for a spinless particle
\footnote{For a particle $\psi$ in the adjoint representation for example, $b\cdot\psi$ means  $[b,\psi]$.},
\beq
M_i=L_i-\frac{x_i}{r}\,b ,
\label{5.10}
\eeq
satisfy the $\so(3)$ algebra
$
\big[M_i,M_j\big]=\varepsilon_{ijk}M_k.
$ 
For arbitrary variations $\ba$ the spectrum of $\delta^2E_1$  is conveniently obtained by using instead the spin-$1$ angular momentum operator
\beq
\bJ=\bM+\bS
=-i\bx\times\bD-\bb+\bS
\label{5.11}
\eeq
where $\bS$ is the $3\times3$ spin matrix
 $\big(\bS_i\big)_{jk}=i\varepsilon_{ijk}$.
$\bS$ satisfies the relations
\beqa
[S_i,S_j]=i\varepsilon_{ijk}S_k,
\quad 
(\bb\cdot\bS)\ba=(b_iS_i)\ba=i[\bb\times\ba],
\label{5.12}
\quad
\bS^2=S_iS_i=-2.
\eeqa
Using the gauge conditions $\bD\cdot\ba$ and $\bx\cdot\ba=0$, we see that
\beq
(\bx\times\bD)\times\ba=\bx(\bD\cdot\ba)-r_i\bD a_i=
-r_i\bD a_i=\ba-\bD(\bx\cdot\ba)=\ba,
\label{5.13}
\eeq
i.e., $\bL\cdot\bS=1$. Since $\bx$ and $\bL$ and thus $\bb$ and $\bL$ are orthogonal, this implies,
\beq
\bJ^2\ba=\bL^2\ba+[\bb\times[\bb\times\ba]]-
2i[\bb\times\ba].
\label{5.14}
\eeq
This leads finally to rewriting $\delta^2E_S$ as
\beq
\delta^2E_S=\int
d^3\bx r^{-2}
\tr\Big\{\big(\bJ^2\ba-[\bb\times[\bb\times\ba]]+i[\bb\times\ba]\big)\cdot\ba\Big\}
\,.
\label{5.15}
\eeq
It is convenient to decompose the variation $\ba$ into eigenmodes of the operator
$
[\bb\times\,\cdot\,]
$ 
which combines vector product and Lie algebra commutator, i.e., to write
\beq
-i[\bb\times\ba]=q\,\ba\,,
\label{starcomm}
\eeq
where the $q$'s are the eigenvalues. The $q$'s come in fact
in pairs
of opposite sign and multiplicity $2$ i.e. in
quadruplets $(q,q,-q,-q)$, see the next Section.

On each $q$-sector $\delta^2E_1$ will be
\beq
\delta^2E_1=m^2(\ba,\ba)+\int d^3\bx\tr
\Big\{\big(\{\bJ^2-q(q+1)\}\ba\big)\cdot\ba\Big\}.
\label{5.17}
\eeq
But $\bJ$ is the Casimir of the angular momentum algebra generated by $\bJ$,
so $\bJ^2=j(j+1)$, where $j$ is integer or half-integer,
according as $q$ is integer or half-integer. Now since $\delta^2E_1$ is manifestly positive by (\ref{Hessiansplit}),
we must have
\beq
m^2+\big\{\bJ^2-q(q+1)\big\}=
\big\{\smallover1/4+\delta^2\big\}
+\big\{j(j+1)-q(q+1)\big)\geq0
\label{5.18}
\eeq
and since $\delta^2$ is arbitrarily small, we see that $j\geq|q|-1$. 

Equation (\ref{5.18}) implies that the possible values of $j$ are $|q|-1, |q|, |q|+1,\dots $.
In particular, the value of $j=|q|-1$ can occur only for $q\leq-1$, and as it corresponds
to the case when $\delta^2E_1$ is purely radial
since 
$j(j+1)=(-q-1)(-q)=q(q+1)$, it implies 
 that $\bD\cdot\ba=0$, so that the states corresponding to it are physical. Thus we can write
\beq
\delta^2E_1=m^2(\ba,\ba)
\quad\hbox{for}\quad
j=|q|-1,\quad q\leq-1
\label{jq-1}
\eeq
and
\beq
\delta^2E_1=\big\{m^2 +(j-q)(j+q+1)\big\}(\ba,\ba)
\quad\hbox{for}\quad
j\geq|q|\ .
\label{deltaE1}
\eeq

\vskip2mm
Let us now consider the second term, $\delta^2E_2$, in 
the decomposition (\ref{Hessiansplit})  of the
Hessian. Since $\bD\cdot\ba$ is zero on the physical states,
\beqa
\delta^2E_2&=&(-i)\int d^3\bx\tr\big(\bB\cdot\big[\ba\times\ba\big]\big)=
\int d^3\bx\tr\left\{-i\big[\bB\times\ba\big]\cdot\ba\right\}
\nn
\\[6pt]
&=&\int d^3\bx\, r^{-2}\tr\left\{-i\big[\bb\times\ba\big]\cdot\ba\right\}
=q\int d^3\bx\,r^{-2}\tr\big(\ba^2\big)=q\big(\ba,\ba\big).
\label{5.20}
\eeqa
From the positivity of $\delta^2E_1$ we then see that 
\emph{the Hessian 
$\delta^2E$ will be positive unless $q$ is negative}.
Furthermore, when $q$ is negative, (\ref{deltaE1}) becomes
\beq
\delta^2E_1=\big\{m^2+(j+|q|)(j-|q|+1\big\}
\big(\ba,\ba\big), 
\label{5.21}
\eeq
and hence
\beq
\delta^2E_1\geq2|q|(\ba,\ba)
\quad\hbox{for}\quad j\geq|q|.
\label{E1pos}
\eeq
For $j\geq|q|$, the restriction of $\delta^2E_1$ to
the physical states will therefore dominate $\delta^2E_2$ 
and the Hessian will again be positive. It follows that the only possibility for getting negative
modes is to have
\beq
q\leq-1
\qquad\text{and}\qquad
 j=|q|-1,
\eeq
in which case
\beq
\delta^2E=\big(m^2-|q|\big)\big(\ba,\ba\big)<0.
\label{5.22}
\eeq

In conclusion, observing that the eigenvalues $q$ come in pairs of opposite sign, we have finally the result that the monopole is \emph{unstable} if and only if there is an eigenvalue such that $|q|\geq1$.
 The opposite condition,
\beq
|q|\leq\2\,,
\label{BNcond}
\eeq
is, of course, just the \textit{Brandt--Neri stability condition} \cite{BN,COL,GOCS}. From the discussion of
Sec. \ref{finiteenergy}. we know however that $|q|\leq\2$ if and only if
\beq
\IQ=\oIQ
\eeq
 i.e., $\IQ$ is the [up to conjugation]
\emph{unique stable charge} of the 
given topological sector, cf. (\ref{4.8}).

Note that since, in the case $j=|q|-1$, the first term on the right-hand side of (\ref{Hessiansplit}) vanishes the variation actually satisfies the \emph{first-order} equations
\beq
\bD\times\ba=0,
\qquad
\bD\cdot\ba=0,
\label{1stordereq}
\eeq
where $r\bD=\widetilde{\bD}\to\bD$ is
[with some abuse of notation] the covariant derivative acting on the asymptotic fields defined over the sphere at infinity. 
 (\ref{1stordereq}) says in particular that our $\ba$'s are  true physical modes, which form furthermore a 
\beq
2j+1=2|q|-1
\label{Jcount}
\eeq
dimensional multiplet of the $\bJ$ algebra. We shall see in the next Section that for each $|q|$ there is one and only one such multiplet. Taking into
account the fact that the eigenvalues come in pairs, this proves the index
formula (\ref{MorseIndexFormula}) \footnote{
For \emph{BPS monopoles} the above arguments break down: 
due to the
$b/r$ term in the expansion of the Higgs field, the second variation picks up an extra term $-\tr\big([\ba,b]\big)^2=q^2$ which precisely \emph{cancels} the $-q^2$ in Eq. (\ref{5.17}). The total Hessian is thus manifestly positive,
\beq
\delta^2E=\delta^2E_1+\delta^2E_2=
\big((m^2+\bJ^2)\ba,\ba\big)>0.
\label{5.25}
\eeq
BPS monopoles are therefore \emph{stable} under variations
of the gauge field alone, even if their charge is \emph{not}
of the form (\ref{4.9}). This is no surprise, since they represent the absolute minima of the energy.}.

The simplest way of counting the number of instabilities for $j\geq |q|$ is to use the Bott
 diagram (see the examples of Sec. \ref{examples}): the Morse index $\nu$ in
(\ref{MorseIndexFormula}) is \emph{twice the number of times
the straight line drawn from $2\IQ$ to the origin 
intersects the root planes} \cite{Bott}.

In the sequel, we will work on the sphere at infinity and,
with some abuse, we use the word ``monopole'' for a solution of the pure Yang--Mills equations  with gauge group $H$  on $\IS^2$. 

Let us note for further reference that the first-order equations in
(\ref{1stordereq}) are plainly consistent with the vanishing of the first term in the decomposition (\ref{Hessiansplit}), 
 and then the negative value of the Hessian
$\delta^2E$ comes from the second term, $\delta^2E_2$.
We also stress that all our investigations assume 
that $\ba$ is a variation with non-zero norm, cf. 
(\ref{normcond}). The meaning of this subtle
condition will be clarified in the next Subsection.

\subsection{Negative modes}\label{negmodes}

Our strategy for finding our negative modes is 
therefore:

\begin{enumerate}
\item
First, we find the  eigenmodes of 
the combined operator $-i[\bb\times\,\cdot\,]$ in
(\ref{starcomm}) with eigenvalues
$q\leq-1$;

\item
Next, we solve the two coupled first-order equations
(\ref{1stordereq}) which set the first term in
(\ref{Hessiansplit}) to zero.

\end{enumerate}
This amounts to finding the negative eigenmodes of the
\emph{linear} second variation operator,
\beq
{\cal K}\ba=\bJ^2\ba-[\bb\times[\bb\times\ba]]+i[\bb\times\ba].
\label{2ndvarop}
\eeq

From the technical point of view, our goal can conveniently be achieved
by complexifying the Lie algebra $\gh$. But then we have
to make sure that our eigenmodes are indeed
real and have non-vanishing norm, cf. (\ref{normcond}) \footnote{
For zero-norm states, $(\ba,\ba)=0$, the \emph{value} of the second variation
on the sphere,
\beq
\delta^2E_S=\int_{\IS^2} d^2\bx \tr({\cal K}\ba,\ba)
\eeq
[with some abuse of notation]
would then \emph{vanish}, despite (\ref{2ndvarop}) 
having a negative eigenvalue. When added to the radial part, 
$\delta^2E_1=\delta^2E_r+\delta^2E_S$ would then be
\emph{positive}, even for the lowest angular momentum state.}.

To find our negative modes, it is convenient to use the stereographic coordinate $z$ on $\IS^2$, 
\beq
z=x+iy=e^{i\varphi}\tan\theta/2.
\label{stereocoord}
\eeq
 In stereographic coordinates the background gauge-potential and field strength become
\beq
A_z=
-i\IQ\displaystyle\frac{\bz}{\varrho},\quad
A_{\bz}=
i\IQ\displaystyle\frac{z}{\varrho},\quad
b=2i \displaystyle\frac{\IQ}{\varrho^2},
\quad
\varrho=1+z\bz,
\label{6.1}
\eeq
where we treated $z$ and its conjugate $\bz$ as independent variables.
Set $\p=\p_{z},\ \bp=\p_{\bz}$, and  let us define
\beq\begin{array}{lllllll}
D_z&=&\p-iA_z=\p-\IQ\displaystyle\frac{\bz}{\varrho},
&\qquad
&a_z&=&\2(a_x-ia_y)
\\[12pt]
D_{\bz}&=&\bp-iA_{\bz}=
\bp+\IQ\displaystyle\frac{z}{\varrho},
&\qquad
&a_{\bz}&=&\2(a_x+ia_y)\end{array}\,.
\label{6.2}
\eeq
In complex coordinates the eigenspace-equations (\ref{starcomm}) 
decouple and indeed become \footnote{Remember that $\IQ$ acts on $a_\alpha$ by commutation.}
\beq\barr{cc}
\IQ&0
\\
0&-\IQ
\earr
\barr{c}a_{\bz}\\
a_z\earr
=q\barr{c}a_{\bz}\\
a_z\earr .
\label{cstarcomm}
\eeq
 The general solution of (\ref{cstarcomm}) is
\beq
\barr{c}a_{\bz}\\
a_z\earr=
f\barr{c} E_\alpha\\ 0\earr
+
g\barr{c} 0\\ E_{-\alpha}\earr
\qquad
\text{with eigenvalue}\;\;
\alpha(\IQ)
\label{bzEalpha}
\eeq
where
$f$ and $g$ are arbitrary functions 
of $z$ and $\bz$. Similarly,
\beq
\barr{c}a_{\bz}\\
a_z\earr=
h\barr{c} E_{-\alpha}\\ 0\earr
+
k\barr{c} 0\\ E_\alpha\earr
\qquad
\text{with eigenvalue}\;\;
 -\alpha(\IQ)
\label{zEalpha}
\eeq 
where $h$ and $k$ are again arbitrary. 
These equations 
 show that the eigenvalues come indeed in pairs as stated earlier.
 Then we should select those pairs  for which the eigenvalue, $q$, is negative.

 As discussed in Sec. \ref{reduction}, for each fixed 
eigenvalue $q\leq-1$,  the negative modes are solutions to the two coupled first-order equations in (\ref{1stordereq}) which are,
furthermore, equivalent to
\beq
D_zh=(\varrho\,\p+|q|\bz)h=0
\quad\hbox{and}\quad
D_{\bz}k=(\varrho\,\bp+|q|z)k=0
\label{6.5}
\eeq
($q=-|q|$ because $q$ is negative). One sees that $h$ and $k$ must be
of the form
\beq
h(z,\bz)=\varrho^{-|q|}\Phi(\bz),
\qquad
k(z,\bz)=\varrho^{-|q|}\Psi(z),
\label{6.6}
\eeq
where $\Phi(\bz)$ and $\Psi(z)$ are arbitrary antiholomorphic (respectively holomorphic)
functions. They can be therefore expanded into power series,
$$
\Phi(\bz)=\sum c_n{\bz}^n, \quad\hbox{and}\quad 
\Psi(z)=\sum d_mz^m.
$$
 But they
are also square integrable functions. Now, since
in stereographic coordinates, the inner product for two vector fields is
\beq
(\ba,\bb)=\int dzd\bz\sqrt{g}g^{\alpha\beta}a_\alpha\bar{b}_\beta=\int dzd\bz\, a_\alpha\bar{b}_\alpha=
\int dzd\bz\,(a_zb_{\bz}+a_{\bz}b_z),
\label{6.7}
\eeq
because $\sqrt{g}g^{\alpha\beta}$ is unity, one sees that $\Phi(\bz)$ will be square integrable if, and only if,
$c_n,\ d_m=0$ except for $n,m=0,1,\dots,2|q|-2$. Thus,
assuming 
\beq
\alpha(\IQ)>0,
\eeq
for definiteness, it is the combination (\ref{zEalpha}) that has to be chosen;
 the negative modes of the operator ${\cal K}$ are linear combinations of the 
$
2(2|q|-1)
$
 variations
\beqa
\barr{c}a_{\bz}\\ 0\earr=
\frac{{\bz}^n}{(1+z\bz)^{|q|}}
\barr{c} E_{-\alpha}\\ 0\earr,
\qquad
\barr{c}0\\ a_{z}\earr=\frac{{z}^m}{(1+z\bz)^{|q|}}
\barr{c} 0\\ E_{\alpha}\earr,
\label{Negmode}
\eeqa
$n,m=0,1,\dots,2|q|-2$ \footnote{If $\alpha(\IQ)<0$ then it is (\ref{bzEalpha}) that should be 
chosen; then $E_\alpha$ is paired with $\bz$, and 
$E_{-\alpha}$ is paired with $z$.}. 

But are these the modes we were looking for? To answer this question, we must
remember
our conditions listed at the beginning of Sec. \ref{negmodes}:
firstly, they should be real, and secondly, they should have nonzero norm.

The modes in (\ref{Negmode}) satisfy \emph{neither} of these conditions: they
belong to the complexified Lie algebra $\gh^c$ and not to its real part.
And they also have zero norm, since they lie in the $E_\alpha$ direction, and
$
\tr(E_\alpha^2)=0
$
for any root $\alpha$. 
Happily enough, both defects can be cured
by \emph{mixing} our modes: it is easy to check that
\beq
\left(\barray{c}
a^+
\\[6pt]
a^-
\earray\right)
=
\left(\barray{c}
a_z+a_{\bz}
\\[6pt]
-i(a_z-a_{\bz})
\earray\right)
=\frac{1}{(1+z\bz)^{|q|}}\left(\barray{c}
z^nE_{\alpha}+\bz^nE_{-\alpha}
\\[6pt]
-i(z^nE_{\alpha}-\bz^nE_{-\alpha})
\earray\right)
\label{asumdiff}
\eeq
are both real and have nonzero norm, as it follows from the Lie algebra
relations (\ref{cLiealgrel}).

In conclusion, these are the negative modes we were looking for.

We only mention that the remaining eigenspace of the Hessian
can also be determined \cite{HORR88}. 

\subsection{Supersymmetric interpretation of the negative modes}\label{SUSY}

We conclude this section by mentioning that the Morse [instability] index $2|q|-1$ is also the Witten index for supersymmetry and the Atiyah-Singer index for the Dirac operator. Indeed, let us consider the  part of $\delta^2E_S$ of the Hessian, which played a central role in Sec. \ref{reduction}. From Eq. (\ref{E1split}) one may write,
after some transformations,
\beqa
\delta^2E_S=\displaystyle\int r^2dr\,K,
\quad 
 K=\int dzd\bz 
 \,\varrho^2\tr(\Psi,({\cal H}\Psi)^\dagger),
\quad 
{\cal H}=-\2\Big\{\cQ^+,\cQ^-\Big\},
\eeqa
where
\beq
\cQ^+=\barr{cc}0&D_\bz\\ 0&0\earr,
\quad
\cQ^-=\barr{cc}0&0\\ D_z&0\earr,
\quad
\Psi=\barr{c}a_\bz\\ a_z\earr.
\label{6.17}
\eeq
The multiplicity $\nu$ of the ground state, the latter being a square integrable solution of
\beq
\cQ^+\Psi=\barr{c}D_\bz a_z\\ 0\earr=0,
\qquad
\cQ^-\Psi=\barr{c}0\\ D_z a_\bz\earr=0,
\label{6.18}
\eeq
is called the \emph{Witten index}. But these are just the negative-mode equations (\ref{6.5}). The result 
$\nu=2|q|-1$ is consistent with that found in Ref. \cite{Spiegel} for supersymmetric QM on the sphere. 
Observe that the supersymmetric Hamiltonian ${\cal H}$ can also be
written as
\beq
{\cal H}=-\2{\Dir}^2
\quad\hbox{where}\quad
\Dir=D_\bz\sigma_++D_z\sigma_-=
\barr{cc}0&D_\bz\\ D_z&0\earr
\label{6.19}
\eeq
is a Dirac-type operator, and the negative modes are exactly those satisfying
\beq
\Dir\Psi=0.
\label{6.20}
\eeq 

The number of solutions of (\ref{6.20}) is the \textit{Atiyah-Singer (AS) index}\footnote{
Since $\ba$ is a $2$-vector, the instability index is the $AS$ index for vectors.}.
The result  $2|q|-1$ is obtained by the same calculation as the one in Atiyah and Bott \cite{ABOTT}, which is valid for an arbitrary Riemann surface. See also \cite{Speight}.

It is worth mentioning that supersymmetry is
a useful tool to describe both the fluctuations 
around a self-dual (BPS) monopole 
\cite{BPSSUSY}, as  well as the scattering of BPS
monopoles \cite{MonScattSUSY}.

\section{The geometric picture: YM on $\IS^2$}\label{geom}

The aim of this section is to present an alternative approach \cite{HoRa} devised for more  geometrically-minded readers and close to the spirit of Atiyah and Bott's  generalization to Riemann surfaces \cite{ABOTT}. 
To make this Section self-contained,
we present full (and somewhat redundant) proofs.

\vskip2mm
Consider indeed pure YM theory of the two-sphere at infinity
with gauge group $H$. Geometrically, the YM potential
is a connection $1$-form still denoted by 
$A$ on a principal $H$-bundle
$P$ over $\IS^2\equiv\IS^2_\infty$. The YM action is
\beq
{\cal A}=\int_{\IS^2}\tr(F\wedge\star F),
\label{infYMaction}
\eeq
where $F=DA$ is the curvature $2$-form, and $\star$ is the Hodge duality operator on $\IS^2$. The associated
field equation reads
\beq
D^{\star}F=0.
\label{infYMeq}
\eeq
Now we can state again the geometric form of the GNO theorem,

\Thm (GNO \cite{GNO}): \textit {In a suitable gauge the
general solution of the Yang-Mills equations
(\ref{infYMeq}) is
\beq
F=i\IQ\,\omega\,,
\label{GNOthm}
\eeq
where $\omega$ is the canonical surface form
of the two-sphere and $\IQ$ is a constant
vector in the Lie algebra $\gh$, quantized 
as in (\ref{GNOquant}),}
\beq
\exp[4\pi i\IQ]=1.
\label{GNOquantbis}
\eeq
\kikezd{Proof}:  Our clue is 
that the YM equation (\ref{infYMeq}) can indeed be written as
\beq
D{\star}F=0
\label{infYMeqbis}
\eeq
since the adjoint operator $D^{\star}$ is
$-\star D\star$ on the two-sphere. 
The field strength 
$F=\2F_{\mu\nu}dx^\mu\wedge dx^\nu$ is a two-form
on $\IS^2$ and therefore
\beq
F=b(\theta,\varphi)\,\omega
\label{Fform}
\eeq
for some $\gh$-valued \emph{function} [zero-form] $b=b(\theta,\varphi)$ on the sphere.
Then the Hodge dual of $F$ is precisely $b$,
\beq
\star F=b(\theta,\varphi),
\label{dualform}
\eeq
because the
 dual of the canonical surface form of the sphere is 
$
\star\omega=1
$,
as it can be verified writing $\omega$ either in
 polar or complex coordinates, 
\beq
\omega=\2\sin\theta\, d\theta\wedge d\varphi=
\frac{1}{i}\frac{dz\wedge d{\bz}}{(1+z\bz)^2},
\eeq
respectively.  Then (\ref{infYMeqbis}) requires that
$b$ is covariantly constant (\ref{covconstcond}) --- and this is
precisely the equation we solved in Sec. \ref{solutions}. 
Firstly, there exists, as explained before, a
(singular) gauge transformation $g$ such that 
 \beq
g^{-1}bg=i\IQ.
\label{GNOgaugebis}
\eeq
To find the gauge potential, let us decompose the latter into
components,
$$
A=\alpha\,i\IQ+\beta,
$$
where $\alpha$ and $\beta$ are $1$-forms ($\alpha$ real).

Now we prove that $\alpha$ is a Dirac monopole
potential and $\beta$ can be gauged away.

 The curvature of $A$ is in fact
$$
dA-\frac{i}{2}[A\wedge A]=
(d\alpha)i\IQ+d\beta
-\frac{i}{2}[(\alpha i\IQ+\beta)\wedge(\alpha i\IQ+\beta)].
$$
But it is also
$
\omega i\IQ.
$
Comparing the two expressions allows us to infer that
\beq
\omega=d\alpha,
\qquad
d\beta-\frac{i}{2}[\beta\wedge\beta]=0.
\label{GNOp1}
\eeq
$\beta$ is hence flat and can therefore be gauged away 
by a suitable $\gk$-valued transformation $k$,
$\beta=-dk k^{-1}$,
where $\gk=[\gh,\gh]$. Then putting $h=gk$,
$$
h^{-1}Ah+h^{-1}dh=k^{-1}Ak+k^{-1}dk=
\alpha i\IQ+k^{-1}\beta k+k^{-1}dk=
\alpha i\IQ\,.
$$
Hence $h^{-1}Fh=i\IQ$, and applying the gauge transformation backwards yields (\ref{GNOthm}). 
The first relation in (\ref{GNOp1}) also proves that
 $\alpha$ is in fact the gauge potential of
 a Dirac monopole in the Dirac gauge,
\beq
\alpha=(\pm1 -\cos\theta)d\varphi,
\eeq
where
the $\pm$ signs refer to the N/S hemispheres.
The gauge potentials are consistently defined therefore
when the transition function
\beq
h(\varphi)=\exp[2\pi i\IQ\varphi]
\label{trfunc}
\eeq
is well-defined, i.e., when $\IQ$ is quantized as
in (\ref{GNOquant}).

Now we construct the non-Abelian YM bundle
$\cP$ from the Abelian bundle
$\cY\equiv\cY_1$ on $\IS^2$ whose Chern class
is $c(\cY)=1$. The latter describes a
Dirac monopole of unit strength and is
given by the Hopf fibration $\IS^3\to\IS^2$.
Now 
$$
\chi(e^{it})=\exp[4\pi i\IQ t]
$$
 is a homomorphism of $\UN(1)$ into
$H$ allowing us to form the associated bundle,
$\cY\times_{\chi}H$. 
 In fact,
$\cY\times_{\chi}H$ is isomorphic to our YM bundle $\cP$.

Conversely, the $H$-bundle with its YM connection
can be reduced to the $\UN(1)$ bundle $\cY$.

\vskip2mm
Having  proved the GNO theorem,
we turn to the stability problem. Our investigations follow the same lines
as  in Sec. \ref{stabanal}, the main difference being that instead of vectors
we use rather differential forms and a more geometric language.
The stability properties of the YM solution are determined by the Hessian, which is now
\beq
\2\delta^2{\cal A}(a,a)=\int_{\IS^2}
\tr\big(D^{\star}Da+\star\big[\star F,a\big],a\big),
\label{infHessian}
\eeq
where the variation $a$ is an $ad P$-valued $1$-form on $\IS^2$  i.e. a section of $\Omega^1(\IS^2; \adP)$.
 Those $a$'s of the form $a=D\chi$ are 
 infinitesimal gauge transformations and are therefore
zero modes for (\ref{infYMaction}). Variations which are orthogonal to gauge transformations satisfy hence
\beq
D\star a=0.
\label{Ggaugecond}
\eeq
These are the physical modes cf. (\ref{gaugecondi}).
The Hessian (\ref{infHessian}) can be
completed by adding 
\beq
DD^{\star}a=-D\star D\star a=0
\eeq
to the integrand, which combines with the first term
to yield the \emph{gauge covariant Laplacian}
\beq
\bigtriangleup_A=D^{\star}D+DD^{\star},
\eeq
The Hessian is therefore
\beq
\2\delta^2{\cal A}(a,a)=\int_{\IS^2}
\tr\big(\bigtriangleup_Aa+\hat{F}(a),a\big),
\qquad
\hat{F}(a)=\star\big[\star F,a\big].
\label{Hessian2}
\eeq
Our strategy, analogous to the one we followed in Section \ref{negmodes}, will
then be to show that the only way of producing a negative mode is by
diagonalizing the second term, $\hat{F}$, and
annihilating the first term in (\ref{Hessian2}).

To this end we note that the square of Hodge star operator $*$ on the two-sphere is
minus the identity,  ${*}^2=-\II$; its eigenvalues are therefore $\pm\, i$. 
The space of [complex] $1$-forms on $\IS^2$, $\Omega$, decomposes according to the eigenvalues of the Hodge star,
\beq
\Omega=\Omega^{(1,0)}\oplus \Omega^{(0,1)},
\qquad
*\,\Omega^{(1,0)}=-i\,\Omega^{(1,0)},
\quad
*\,\Omega^{(0,1)}=i\,\Omega^{(0,1)}.
\eeq
The decomposition 
$\gh^{\IC}=\gt^{\IC}\oplus\sum_\alpha\gh_\alpha$
of the complexified Lie algebra where  $\gh_\alpha$ denotes the root spaces
generated by the $E_\alpha$'s
 implies the analogous decomposition 
\beq
\adP^{\IC}=\cP_0\oplus\sum_\alpha\cP_\alpha
\label{bundledecomp}
\eeq
where $\cP_0=\cY\times_{\chi}\gt$ and 
$\cP_\alpha=\cY\times_{\chi}\gh_\alpha$.
Both $\cP_0$ and $\cP_\alpha$ are holomorphic line bundles. $\cP_0$ is trivial, $c(\cP_0)=0$ and
$\cP_\alpha$ has Chern class $c(\cP_\alpha)=2\alpha(\IQ)$.

By (\ref{bundledecomp}) the sections of $\adP^{\IC}\otimes\Omega$ are obtained from those of
\beq
\left(\cP_0\otimes\Omega\right)\oplus\sum_{\alpha}
\big(\cP_\alpha\otimes\Omega^{(1,0)}\big)\oplus
\big(\cP_\alpha\otimes\Omega^{(0,1)}\big).
\label{cPdecomp}
\eeq
The sections of each of these bundles are eigenspaces of
\beq
\hat{F}(a)=\star\big[\star F,a\big]
\quad\text{with eigenvalues}\quad 0,\, -\alpha(\IQ), 
\alpha(\IQ),
\eeq
respectively. On the other hand,
the covariant Laplacian preserves the decomposition 
(\ref{cPdecomp}); on $\cP_\alpha\otimes\Omega$ it reduces in fact to the covariant Laplacian of a Dirac monopole of
strength $\alpha(\IQ)$. This is a positive operator,
$\Delta_A\geq0$, whose
first non-zero eigenvalue is $2|\alpha(Q)|$ \cite{HORR88}.

Let us now consider a root $\alpha$ such that
$
\alpha(\IQ)>0.
$
Then the only way of getting a negative mode is by having a section $a$ of 
$\cP_\alpha\otimes\Omega^{(1,0)}$ for which 
\beq
\bigtriangleup_Aa=0.
\eeq
By (\ref{Ggaugecond}) this is the same as $D a=0$.
Splitting $a$ and $D$ as
\beq
a=a'+a'',
\qquad
D=D'+D''
\eeq
according to the eigenvalues of $\star$, the conditions for getting a negative mode reduce to
\beq
D''a'=0.
\label{negmodcond}
\eeq
Now, according to a theorem of Koszul and Malgrange, a complex vector bundle with connection over $\IS^2$
has a unique holomorphic structure whose holomorphic
sections are exactly the solutions of (\ref{negmodcond}). We conclude that the negative modes are the \emph{holomorphic sections} of
$\adP\otimes\Omega^{(1,0)}$.

 It is sufficient to consider the terms in (\ref{bundledecomp}) separately.
 
 The Chern class of a tensor product is the sum of the Chern classes, and $\Omega^{(1,0)}$ has Chern class $(-2)$. Thus
\beq
c(\cP_0\otimes\Omega^{(1,0)})=-2
\label{trivCc}
\eeq
and $\cP_0\otimes\Omega^{(1,0)}$ has \emph{no}
holomorphic sections.
On the other hand,
\beq
c(\cP_\alpha\otimes\Omega^{(1,0)})=2\alpha(\IQ)-2
=n_\alpha.
\label{alphaCc}
\eeq
Then the Riemann-Roch theorem tells us that
a line bundle with \emph{positive} Chern class $n_\alpha\geq0$  admits
\beq
n_\alpha+1=2\alpha(\IQ)-1
\eeq
holomorphic sections. Summing over all roots
provides us therefore with the 
Morse index  formula (\ref{MorseIndexFormula}).

The negative modes are conveniently found  in terms of complex coordinates (\ref{stereocoord}) i.e., 
$
z=e^{i\varphi}\tan(\theta/2)
$
on the northern hemisphere. Then
\beq
\star(dz)=-idz,
\qquad
\star(d\bar{z})=id\bar{z}
\eeq
shows that a holomorphic section is of the form
$a(z,\bz)dz$, where
\beq
\p_{\bz}a+\frac{n}{2}\frac{z}{1+z\bz}\,a=0.
\eeq
The solution is
\beq
a=\frac{f(z)}{(1+z\bz)^{n/2}},
\eeq
where $f(z)$ is an arbitrary holomorphic function.

Similarly, on the southern hemisphere
\beq
w=e^{-i\varphi}\cot(\theta/2)=\frac{1}{z}
\eeq
is a complex coordinate, and the solutions of eqns.
(\ref{Ggaugecond})  and (\ref{negmodcond}) is
\beq
\frac{g(w)}{(1+w\bw)^{n/2}}
\eeq
where $g(w)$ is holomorphic in $w$.

For Chern class $n$ the transition function is
$e^{in\varphi}=(z/\bz)^{n/2}$. Expanding $f(z)$ and $g(w)$ as
$$
f(z)=\sum_i a_iz^i
\quad\hbox{and}\quad
g(w)=-\sum_j b_jw^j,
$$
consistency requires
$$
\left(\frac{1}{(1+z\bz)^{n/2}}\sum_i a_iz^i\right)
dz=\big(\frac{z}{\bz}\big)^{n/2}
\left(\frac{-1}{(1+w\bw)^{n/2}}\sum_j a_iw^i\right)
dw.
$$
It follows that $f(z)$ and $g(w)$ can only be polynomials of degree
at most $n-2$, and $a_i=b_{n-2-i}$.
Returning to the original problem, we see that the root
$\alpha$ contributes
$ 
2\alpha(\IQ)-1
$ 
negative modes, namely
\beq
a_\alpha\equiv 
a_\alpha^{(k)}=
\frac{z^k}{(1+z\bz)^{n_\alpha/2}}\,dz\,E_\alpha,
\quad
0\leq k\leq 2\alpha(\IQ)-2=n_\alpha.
\eeq
Similarly, we also have for the root $-\alpha$
\beq
a_{-\alpha}\equiv 
a_{-\alpha}^{(k)}=
\frac{\bz^k}{(1+z\bz)^{n_\alpha/2}}\,d\bz\,E_{-\alpha},
\quad
0\leq k\leq 2\alpha(\IQ)-2=n_\alpha.
\eeq

Thus we have re-derived, once again, the negative modes
  (\ref{Negmode}) as holomorphic and anti-holomorphic sections of line bundles over the two-sphere with Chern class $2|q|-2$ \footnote{Note that 
the $(-2)$ comes from the fact that our variations are differential $1$-forms rather than merely functions.}.  

In Section \ref{negmodes} these same modes were
obtained using angular momentum.
This is not a coincidence, since the
holomorphic sections  of line bundles are precisely the carrier
spaces of representations of the rotation group $\SU(2)$.

\section{Loops}\label{loops}
Now we deepen and elaborate the intuitive remark of Coleman \cite{COL}
about the \emph{analogy of monopoles and elastic strings}.

Let us consider $\Omega=\Omega(H)$, the space of loops in a compact Lie group $H$ which start and end at the identity element of $H$. The \textit{energy} of a loop $\gamma(t),\, 0\leq t\leq 1$ is given by
\beq
L(\gamma)=\frac{1}{4\pi}\int_0^1\tr\big(\gamma^{-1}
\frac{d\gamma}{dt}\big)^2dt.
\label{7.1}
\eeq
A variation of $\gamma(t)$ is a 2-parameter map
$\alpha(s,t)$ into $H$ such that $\alpha(0,t)=\gamma(t)$. We fix the end points, $\alpha(s,0)=\gamma(0)$ and 
$\alpha(s,1)=\gamma(1)$ for all $s$. For each fixed $t$,
$\p\alpha/\p s$ at $s=0$ is then a vector field $X(t)$ along $\gamma(t)$, $X(0)=X(1)=0$.
$\Omega$ can be viewed then as an infinite dimensional manifold whose tangent space at a 
``point'' $\gamma$ (i.e., a loop $\gamma(t),\ 0\leq t\leq1$)
is a vector field $X(t)$ along $\gamma(t)$, which vanishes
at the end points. Since the Lie algebra $\gh$ of $H$ can
be identified with the left-invariant vector fields on $H$,
it is convenient to consider $\eta(t)=\gamma^{-1}(t)X(t)$ which is a loop in the Lie algebra
$\gh$ s.t. $\eta(0)=\eta(1)=0$.
This is true in particular for $\zeta(t)=\gamma^{-1}(t)\displaystyle\frac{d\gamma}{dt}$~\footnote{It is worth pointing out that the
Morse index reappears in the Maslov correction of the propagator, see  \cite{HPAMaslov}.}.

The first variation of the loop-energy functional (\ref{7.1}) is
\beq
\delta L(\eta)=-\frac{1}{2\pi}\int\tr\left\{\frac{d\zeta}{dt}\,\eta(t)\right\}dt.
\label{7.2}
\eeq
The \textit{critical points} of the energy satisfy therefore 
${d\zeta}/{dt}=0$ and are, hence,
\beq
h(t)=\exp\big[4\pi i\IQ t\big],
\qquad
0\leq t\leq 1,\quad
\IQ\in\gh
\label{7.3}
\eeq
i.e. \textit{closed geodesics} in $H$ which start and end at the
identity element. To be so, $\IQ$ must be quantized as in (\ref{GNOquant}), 
$\exp4\pi i \IQ=1$. The energy of such a geodesic is obviously 
$L(h)=4\pi\tr(\IQ^2)$.

The stability properties are determined by the Hessian. After
partial integration, this is found to be
\beq
\2\delta^2L(\eta,\eta)=-\frac{1}{4\pi}\int\tr\left\{
\big(\frac{d^2\eta}{dt^2}+4\pi i\big[\IQ,\frac{d\eta}{dt}\big]\big)\eta\right\}dt.
\label{7.4}
\eeq
The spectrum of the Hessian is obtained hence by solving
\beq
\frac{d^2\eta}{dt^2}+4\pi i\big[\IQ,\frac{d\eta}{dt}\big]=\lambda\eta,
\qquad
\eta(0)=\eta(1)=0.
\label{7.5}
\eeq
Taking $\eta$ parallel to the step operators $E_{\pm\alpha}$,
(\ref{7.5}) reduces to the scalar equations
\beq
\frac{d^2\eta}{dt^2}\pm4\pi iq\frac{d\eta}{dt}=-\lambda\eta,
\qquad
\eta(0)=\eta(1)=0,
\label{7.6}
\eeq
where $q=q_\alpha=\alpha(\IQ)$, and whose solutions yield
\beq
\eta_\alpha^{\ k}(t)=e^{\mp2\pi iqt}\big(
e^{i\pi(k+1)t}-e^{-i\pi(k+1)t}\big)E_{\pm\alpha},
\qquad
\lambda=-\pi^2\big(4q^2-(k+1)^2\big),
\label{7.7}
\eeq
where $k\geq0$ is an integer. (For $k=-1$, we
would get $\eta=0$, and for $(-k-2)$ we would
get $(-\eta_\alpha^{\ k})$). $\lambda$ is negative
if $0\leq k\leq2|q|-2$, providing us with $2(2|q_\alpha|-1)$ negative modes. The total number of negative modes is therefore the same as for a monopole with
non-Abelian charge $\IQ$, cf. (\ref{MorseIndexFormula}) \cite{ABOTT,FrHab,NAUH}.

Unlike for ``monopoles'' [i.e. YM on $\IS^2$], loops admit \emph{zero modes}.
For $k+1=2|q|$ we get in fact
\beq
\eta(t)=\pm\big(1-e^{-4\pi i|q|t}\big)E_{\pm\alpha},
\label{7.8}
\eeq
while for $|k|\geq2|q|$ (\ref{7.7}) yields positive modes.

If $|q|\leq1$ i.e. $q=0$ or $\pm1$, there are no negative
modes: the geodesic is \textit{stable}. The results of Sec. 4 imply therefore that in each homotopy sector there is a unique stable geodesic.

\vskip2mm
Loops in $H$ can be related to YM on $\IS^2$.
Indeed, the map (\ref{paralleltransport}) i.e.
\beq
h^{\bA}(\varphi)={\cal P}\left(\exp\oint_{\gamma_{\varphi}}\bA\right)
\label{7.9}
\eeq
associates a loop $h^{\bA}(\varphi)$ to each $YM$ field
$\bA$ on $\IS^2$ \cite{COL,GNO,GORPP}.

For a generic connection the notation (\ref{7.9}) is merely symbolical. It can be calculated, however, explicitly if $\bA$ is Abelian. In particular, if it is a solution to the Yang--Mills equations on $\IS^2$, when it is just the geodesic (\ref{7.3}).

 We conclude that 
\begin{itemize}
\item The map
(\ref{7.9}) carries the critical points of the YM functional
into critical points of the loop-energy functional;

\item
the number of negative YM modes is
the same as the number of negative loop-modes; 

\item
The energies of critical points are also the same, namely
\beq
{\cal A}=L=
4\pi\tr(\IQ^2)\,.
\eeq

\end{itemize}

The differential of the map (\ref{7.9}) carries a YM variation $\ba$ into a loop-variation $\eta^{\bf A}(\varphi)$ i.e. a loop in the Lie algebra $\gh$.
Explicitly, let us consider 
\beq
g(\theta,\varphi)={\cal P}\left(\exp\displaystyle{\left\{
\int_0^\theta\right\}
}_{\gamma_{\varphi}}\bA\right).
\label{7.10}
\eeq
A YM variation $\ba$ goes then into
\beq
\eta^{\bA}(\varphi)=-\oint g^{-1}(\theta,\varphi)
a_{\theta}\big(\gamma_{\varphi}(\theta)\big)
g(\theta,\varphi)d\theta\, .
\label{7.11}
\eeq
Remarkably, $\eta^{\bA}(\varphi)$ depends on the choice of the loops $\gamma_\varphi(\theta)$ and even of the stating point. For example, with the choice of Subsec. \ref{finiteenergy}, the image of the YM negative mode $\ba^{(k)}$ is
\beq
\eta^{\bA}(t)=C^k(1-e^{-2\pi i kt}),
\label{7.12}
\eeq
where the numerical factor $C^k$ is,
\beq
C^k=\int_0^\pi\!(\sin\theta/2)^k(\cos\theta/2)^{2|q|-2-k}d\theta
=\frac{\Gamma\big(\frac{k+1}{2}\big)\Gamma\big(\frac{2|q|-k-1}{2}\big)}{
\Gamma(\frac{2|q|+1}{2})}\ .
\label{7.13}
\eeq
(\ref{7.12}) is similar to, but still different from the loop-eigenmodes (\ref{7.7}). If we choose however $\gamma_\varphi(\theta)$ to be the loop which starts from the south pole, goes to the north pole along the meridian at $\varphi/2$, and returns to the south pole along the meridian at
$-\varphi/2$, we do obtain (\ref{7.7}).

The map (\ref{7.9}) YM $\to$ $\big\{$loops$\big\}$
is \emph{not} one-to-one. One possible inverse
of it is given as 
\beq
A_\theta=0,\qquad
A_\varphi=\left\{
\begin{array}{c}\smallover1/4
\displaystyle{(1-\cos\theta)}\,h^{-1}\frac{dh}{d\varphi}
\\[14pt]
-\smallover1/4
\displaystyle{(1+\cos\theta)}\,\frac{dh}{d\varphi}h^{-1}
\end{array}\right.
\qquad\hbox{in}\qquad
\left\{
\begin{array}{c}
N
\\[24pt]
S
\end{array}\right.
\label{7.14}
\eeq

\section{Global aspects}\label{global}

Where can an unstable monopole go? It can not leave its homotopy sector, since this would require infinite energy. But it can go into another configuration in
the same sector, because any two such configurations are separated only by finite energy \cite{HORR88}. 

Yang--Mills--Higgs theory on $\IR^3$ has the same topology as YM on $\IS^2$. The true configuration space ${\cal C}$
of this latter is furthermore the space ${\cal A}$ of all
YM potentials modulo gauge transformations,
\beq
{\cal C}\simeq{\cal A}/{\cal H}
\quad\hbox{where}\quad
{\cal H}=\Big\{\hbox{Maps\;} \IS^2\to H\Big\},
\label{8.7}
\eeq
and the \textit{path components} of ${\cal C}$ are just
the \textit{topological sectors}: 
\beq
\pi_0({\cal C})\simeq
\pi_2(G/H)\simeq \pi_1(H).
\eeq

When studying the topology of ${\cal C}$, we can also use loops. The map (\ref{7.9}) (widely used for describing the topological sectors \cite{COL,GNO,GORPP,HORR88}), is in fact a \emph{homotopy equivalence} between YM on $\IS^2$ and
$\Omega=\Omega(H)/H$, the loop-space of $H$ modulo global
gauge rotation \cite{Singer} \footnote{One has to divide out by $H$ because a gauge-transformation changes the non-integrable
phase factor by a global gauge rotation.}. This correspondence explains also why we could 
count the negative YM modes using the diagram
 introduced by Bott \cite{Bott} for loops.
 
 For this reason, we shall consider, in what follows,  loops in the residual gauge group $H$ and YM on the sphere at infinity as the same theory and use the word  ``monopole'' for a critical point
 of each of the theories.
  
\subsection{Configuration space topology and energy--reducing two-spheres}\label{spheres}

As we mentioned already, saddle-point solutions  in field theories are often associated to \textit{non-contractible loops} \cite{MaWSTop, Taubes83, FHloops,Sibner}. 
There are \emph{no non-contractible loops} in our case,
\beq
\pi_1({\cal C})\simeq \pi_1(\Omega)\simeq 
\pi_2(H)=0.
\eeq
 There are, however, \emph{non-contractible two-spheres}: 
 \beq
 \pi_2({\cal A}/{\cal H})\simeq \pi_1({\cal H})\simeq\pi_3(H).
\eeq
But for any compact $H$, $\pi_3(H)$ is the direct sum of the
$\pi_3$'s of the simple factors $K_j,\, j=1,\dots,s$. On the other hand, for any compact, simple Lie group, $\pi_3\simeq\IZ$. Note that  $\SO(4)$ is not simple and its $\pi_3$ is 
$\IZ\oplus\IZ$.

The role of our spheres is explained by the \emph{Morse theory} \cite{Morse}: a critical point of index $\nu$ of a ``perfect Morse function'' is in fact associated to a class in $H_\nu$, the
$\nu$-dimensional homology group. The Hurewicz isomorphism \cite{BottTu} tells however that, for simply connected manifolds, $\pi_2$ is isomorphic to $H_2$, the second homology group. The K\"unneth formula \cite{BottTu} shows furthermore that the direct product of the $(\nu/2)$ $2$-spheres has a non-trivial class in $H_\nu$.

\subsection{Energy--reducing two-spheres}\label{spheresbis}

Below we associate an energy-reducing two-sphere which interpolates between a given (unstable) monopole and some other, lower-energy monopole to each
intersection of the line $0\longleftrightarrow Q$ with the
root plane. The tangent vectors to these spheres are furthermore negative modes for the Hessian.

Let us first consider a geodesic $\exp[4\pi i\IQ t],\,
0\leq t\leq1$ in $H$ rather than a monopole. Remember that the step operators 
\beq
E_{\pm\alpha} 
\quad\hbox{and}\quad
H_\alpha=[E_{\alpha},E_{-\alpha}]
\eeq
 close into an $\ort(3)$ subalgebra of $\gk\subset\gh$. Denote by $G_\alpha$ the generated subgroup of $K\subset H$. Our two-spheres are associated to these $G_\alpha$'s.

Observe first that, for each root $\alpha$, 
\beq
S_\alpha=\Big\{g^{-1}P_\alpha g,\,g\in G_\alpha\Big\},
\eeq
is a two-sphere in the Lie algebra $\gk\subset\gh$, where
\beq
P_\alpha=\frac{2H_{\alpha}}{\tr(H_{\alpha}^2)}
\eeq
is the primitive charge associated to the root $\alpha$,
cf. (\ref{primitcharge}). 

If 
$\xi$ is an arbitrary vector from $S_\alpha$,
\beq
\exp[\pi i\xi]=\exp[\pi i\,g^{-1}P_\alpha g]=
g^{-1}\big(\exp[\pi iP_\alpha]\big) g=
\pm\II,
\label{8.8}
\eeq
with the sign depending on $G_\alpha$ being $\SU(2)$
or $\SO(3)$, because 
\beq
\exp\big[2\pi iP_\alpha\big]=\II.
\eeq

Our now clue is  to define, for $\xi\in\S_\alpha$,
\beq
h_\xi^{\ k}(t)=e^{\pi it(k+1)\xi}e^{2\pi it
(2\IQ-(k+1)P_\alpha/2)},
\label{loopsphere} 
\eeq
$0\leq t\leq 1$.
Remarkably, 
\beq 
e^{\pi i(k+1)\xi}
e^{2\pi i(2\IQ-(k+1)P_\alpha/2)}
=(\pm\II)^{k+1}\,\big(e^{4\pi i\IQ}\big)\,
(e^{-\pi iP_\alpha})^{k+1}=(\pm\II)^{2(k+1)}
=\II,
\label{hloopprop}
\eeq
using that $\IQ$ and $P_\alpha$ commute since both of them belongs to the Cartan algebra, and that $\IQ$ is quantized. 
It follows from (\ref{hloopprop}) that, for each $\xi$ from $S_\alpha$ and integer $k$,
is a loop in $H$. Equation (\ref{loopsphere}) provides us therefore 
with a \emph{two-sphere of loops} in $H$, parametrized
by $\xi\in\IS^2$. 
Note that 
\beq
R=2\IQ-\2(k+1)P_\alpha,
\label{Rcenter}
\eeq
is a charge or a half-charge depending on the the global structure
[$\SU(2)$ or $\SO(3)$]
of $G_\alpha$. 
Using the shorthand 
$
h=h_\xi^{\ k},
$
 the velocity of the loop (\ref{loopsphere}) is
\beq
h^{-1}\frac{dh}{dt}=e^{-2\pi iRt}\big((k+1)\pi\xi\big)
e^{2\pi iRt}+2\pi R.
\label{loopspeed}
\eeq
To calculate its energy, observe that, for any vector $\zeta$ from the Cartan algebra, $\zeta-\alpha(\zeta)H_{\alpha}/
(\alpha,\alpha)$ commutes with $E_{\pm\alpha}$, because
$$
[\zeta-\alpha(\zeta)\frac{H_{\alpha}}{
(\alpha,\alpha)},E_\alpha]=
\alpha\big(\zeta-\alpha(\zeta)\frac{H_{\alpha}}{
(\alpha,\alpha)}\big)E_\alpha=0,
$$
and so
$$
g\,\zeta\,g^{-1}=g\big(\zeta-\alpha(\zeta)\frac{H_{\alpha}}{
(\alpha,\alpha)}+\alpha(\zeta)\frac{H_{\alpha}}{
(\alpha,\alpha)}\big)g^{-1}
=\zeta-\alpha(\zeta)\frac{H_{\alpha}}{
(\alpha,\alpha)}+
\big(\frac{\alpha(\zeta)}{(\alpha,\alpha)}\big)\,gH_\alpha\,g^{-1}.
$$
Hence
\begin{eqnarray*}
\tr(\xi,\zeta)&=&\tr\big(g^{-1}P_\alpha\,g,\zeta\big)=
\tr\big(P_\alpha,g\zeta\,g^{-1})
\\[6pt]
&=&\tr\big(P_\alpha,\zeta-\alpha(\zeta)\frac{H_{\alpha}}{
(\alpha,\alpha)}\big)+
\frac{\alpha(\zeta)}{(\alpha,\alpha)}\tr(P_\alpha,
g\,H_\alpha\,g^{-1}).
\end{eqnarray*}
Substituting here $\zeta=R$ we get finally, using $\tr(P_\alpha\IQ)=2/\tr(H_\alpha)^2\alpha(\IQ)=q$,
\beq
L(h)=\pi\Big\{2(k+1)(2|q|-k-1)\tr(P_\alpha/2)^2
\cos\tau+\tr\big(2\IQ-(k+1)P_\alpha/2\big)^2\Big\},
\label{8.11}
\eeq
where $\tau$ is the angle between the primitive charge $P_\alpha$ and 
the parameter-vector $\xi=g^{-1}P_\alpha\,g$,
\beq
\cos\tau=\frac{\tr(P_\alpha\xi)}{\tr\,P_\alpha^2}=\smallover1/4\tr(P_\alpha\xi).
\eeq
Hence,
\begin{itemize}
\item
For $\tau=0$ i.e. for $\xi=\pm P_\alpha$ the two factors in (\ref{loopspeed}) commute
 and we get the \emph{``long'' geodesic} $\exp[4\pi i\IQ t]$ 
 with charge 
 \beq
\IQ_{top}= \IQ
\eeq
we started with.
\item
 For
$\tau=\pi$ i.e. $\xi=-P_\alpha$ instead we get another, 
\emph{lower-energy geodesic}, namely
\beq
h_{\alpha}^{\ k}(t)=e^{4\pi it(\IQ-(k+1)\2P_\alpha)},
\qquad
0\leq t\leq 1
\label{8.12}
\eeq
whose charge is
\beq
\IQ_{bottom}=\IQ-(k+1)\2P_\alpha=\2R-\smallover1/4 (k+1)P_\alpha
\eeq

\end{itemize}
We conclude that, for each $0\leq k\leq2|q|-2$, (\ref{loopsphere}) provides us with a
smooth \textit{energy reducing two sphere of loops}, whose
top is the ``long'' geodesic we started with, and whose bottom is the lower-energy loop (\ref{8.12}).

Carrying out this construction for all roots $\alpha$
and all integers $k$ in the range $0\leq k\leq2|q|-2$, we get exactly the required number of two-spheres.

They can also be shown to be non-contractible and to generate $\pi_2$.

The intuitive content of our loop construction (\ref{loopsphere})
will become clear from the
examples in Subsection \ref{examples}.

\vskip2mm
Consider now the tangent vectors to our two-spheres of loops along the curves 
$$
g_s(t)=e^{2\pi i E_{\pm\alpha}s}P_\alpha
e^{-2\pi i E_{\pm\alpha}s}
$$
at $s=0$, the top of the spheres. These tangent vectors are
\beq
e^{4\pi i|q|t}\big(
e^{-2\pi i(2|q|-k-1)t}-e^{4\pi i|q|t}\big)E_{\pm\alpha}.
\label{8.13}
\eeq

The loop-variations (\ref{8.13}) are again negative modes. They are not, however, eigenmodes, but rather mixtures of negative modes
$
(1-e^{-2\pi i(2|q|-k-1)t})E_{\pm\alpha}$
 and the
zero mode 
$(1-e^{-4\pi i|q|t})E_{\pm\alpha}.$

\vskip2mm
The inverse formula (\ref{7.14}) translates finally the whole construction to YM:  
\beq
A_\theta^{\ \xi}=0,\:
A_\varphi^{\ \xi}=\left\{\begin{array}{l}
\,\,\,\smallover1/4
\displaystyle{(1-\cos\theta)}\Big[e^{-2\pi iRt}(k+1)\xi\,e^{2\pi iRt}+2R\Big]
\\[16pt]
-\smallover1/4
\displaystyle{(1+\cos\theta)}\Big[(k+1)\xi+
e^{\pi i(k+1)t\xi}(2R)\,e^{-\pi i(k+1)t\xi}\Big]
\end{array}\right.
\;\hbox{in}\;
\left\{\begin{array}{c}
N
\\[22pt]
S
\end{array}\right.
\label{YMsphere}
\eeq
is a energy-reducing two-sphere of YM
potentials on $\IS^2$. 

The top of the sphere is $\IQ\bA_D$, the monopole
we started with, and the bottom is another, lower-energy
monopole, whose charge is $\IQ-(k+1)P_\alpha/2$.

Once again, the situation is conveniently illustrated on the Bott diagram, (see the examples
in the next Section).

Note finally that our definition (\ref{loopsphere}) can easily be
modified so that the spheres fit to the negative eigenmodes (\ref{7.7}). However, the loops are then no longer of constant speed and do not interpolate in a monotonically energy-reducing manner between the critical points.

\subsection{Examples}\label{examples}

\kikezd{Example 1:~$H=\SU(2)$}

\begin{figure}
\begin{center}
\includegraphics[scale=.84]{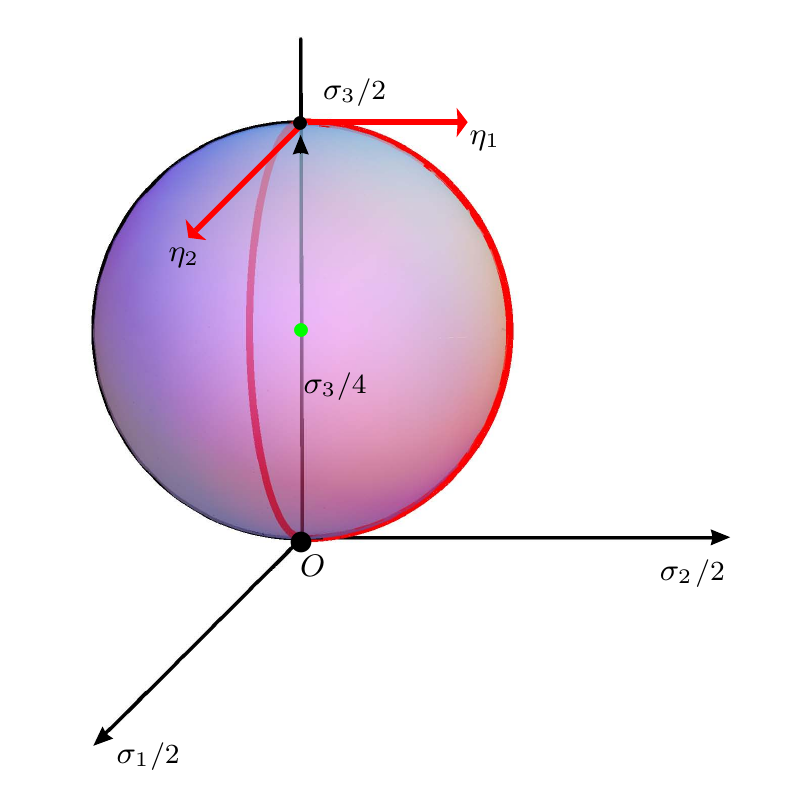}
\end{center}\vspace{-10mm}
\caption{{\it The GNO charge $\IQ=\sigma_3/2$ supports
an unstable solution of the $\SU(2)$ YM theory
over the two-sphere. Its $\nu=2$ negative modes are
recovered as the tangents to the energy-reducing two-sphere
whose top is $\IQ=\sigma_3/2$ is and whose
bottom is the vacuum. The vertical axis is the Cartan algebra of $\su(2)$.}}
\label{SU21S}
\end{figure}

The simplest case of interest is that of residual
group $H=\SU(2)$. The only topological sector is the
trivial one since $\SU(2)$ is simply connected. The ground state is the vacuum
with vanishing energy, $A_i\equiv0$ in a suitable gauge. Higher-energy solutions do arise, though,
and have the form (\ref{imbDmon}) with
$\IQ$ a half-charge, 
\beq
\IQ\equiv \IQ_n=n\,\frac{\sigma_3}{2},
\label{SU2charge}
\eeq
 with $n$ some integer. All monopoles with $n\neq0$
 are, however, unstable with Morse index
$ 
\nu=2(2|n|-1).
$ 
Then our construction 
provides us with $n$ spheres all of whose tops 
are at $\IQ=\IQ_n$ but  whose bottoms are  at some lower-energy charge,
\beq
\hbox{top:}\;\; \IQ=\IQ_n,
\qquad
\hbox{bottom:}\;\;\IQ_k,\,\,k=n-1,\dots,0,\dots, -n+1.
\eeq
The tangent vectors $\eta_{\pm}^k$
to these spheres yield $\nu=2n$
negative modes.
\vskip2mm
Let us consider, for example, the $H=\SU(2)$ monopole with
 GNO charge 
\beq
\IQ=\frac{\sigma_3}{2}\,.
\label{su2ex}
\eeq
According to the general theory, our monopole is a solution is unstable with Morse index $\nu=2$, and
our construction in Sect. \ref{loops} provides us with one energy-reducing
two-sphere which interpolates between the monopole with charge (\ref{su2ex}) and the vacuum,~see Fig.\ref{SU21S}.

Let us first consider loops. According to (\ref{loopsphere}) our two-sphere of loops is
\beq
h_\xi(t)=e^{i\pi t\xi}e^{i\pi t\sigma_3},
\label{su2loopsphere}
\eeq
where $\xi\in S=g^{-1}\sigma_3g$, $g\in\SU(2)$.
 Parametrizing 
 $\SU(2)$  with Euler angles $(\tau,\varrho,\psi)$,
 $0\leq\tau\leq\pi,\, 0\leq\varrho\leq2\pi, \, 0\leq\psi\leq4\pi$,
\begin{eqnarray*}
g=e^{i\varrho\sigma_3/2}e^{i\tau\sigma_2/2}e^{i\psi\sigma_3/2} 
=\left(
\begin{array}{cc}
e^{i\frac{\varrho+\psi}2}\cos\frac{\tau}{2} &e^{i\frac{\varrho-\psi}{2}}\sin
\frac{\tau}{2}
\\[6pt]
-e^{-i\frac{\varrho-\psi}{2}}\sin\frac{\tau}{2} &e^{-i\frac{\varrho+\psi}{2}}%
\cos\frac{\tau}{2}
\end{array}
\right),
\end{eqnarray*}
yields the parametrization with polar angles $(\tau,\psi)$
of the $\xi$-sphere,
\begin{eqnarray*}
\xi=\left(
\begin{array}{ll}
\cos\tau & e^{-i\psi}\sin\tau  
\\[6pt]
e^{i\psi}\sin\theta &-\cos\tau
\end{array}
\right) \in\su(2)\,.
\end{eqnarray*}
 Then
\beq
e^{i\pi t\xi}=\left(
\begin{array}{ll}
\cos\pi t+i\sin\pi t\cos \tau  & ie^{-i\psi }\sin \pi t\sin \tau  
\\[8pt]
ie^{i\psi }\sin\pi t\sin\tau  & \cos\pi t-i\sin\pi t\cos\tau
\end{array}\right),
\quad
e^{i\pi t\sigma_3}=\left(
\begin{array}{ll}
e^{i\pi t}& 0 
\\[8pt]
0 & e^{-i\pi t}
\end{array}
\right)
\eeq
so that our sphere of loops (\ref{su2loopsphere})
reads
\begin{eqnarray}
h_\xi(t) =
\left(
\begin{array}{cc}
\left(\cos\pi t+i\sin \pi t\cos \tau \right) e^{i\pi t} & ie^{-i\left(
\psi +\pi t\right)}\sin \pi t\sin \tau  
\\[12pt]
ie^{i\left(\psi +\pi t\right)}\sin\pi t\sin\tau
&\left(\cos\pi t-i\sin\pi t\cos\tau\right)e^{-i\pi t}
\end{array}
\right).
\label{exploopsphere}
\end{eqnarray}
Calculating the speed of our loops,
\beq
h^{-1}\p_th=i\pi\left(\barray{cc}
1+\cos\tau&e^{-i(\psi+2\pi t)}\sin\tau
\\[8pt]
e^{i(\psi+2\pi t)}\sin\tau
&-1-\cos\tau
\earray\right),
\eeq
[$h\equiv h_\xi$] allows us to infer that
\emph{the energy of the loop} with parameter $\xi$ is
simply the \emph{height function on the unit $\xi$-sphere},
\beq
L^{(\tau,\psi)}=\pi(1+\cos\tau).
\label{su2loopen}
\eeq
This result is also consistent with (\ref{8.11}).
The intuitive picture is that of Fig. \ref{shrink} with the
``sphere'' meaning now $\SU(2)\approx\IS^3$.

\vskip2mm
In YM terms,  (\ref{YMsphere}) yields in turn
 the $2$-sphere of YM configurations  reads
\beqa
A_\theta^{\ \xi}=0,
\;
A_\varphi^{\ \xi}&=&\smallover1/4(1-\cos\theta)\Big\{
(1+\cos\tau)\sigma_3-\sin\tau(e^{-i(\varrho+\varphi)}\sigma_++
e^{i(\varrho+\varphi)}\sigma_-)\Big\}
\label{9.7}
\\[8pt]
&=&\smallover1/4(1-\cos\theta)\Big\{
(1+\cos\tau)\sigma_3-\sin\tau\big(\cos(\varrho+\varphi)\sigma_1
+\sin(\varrho+\varphi)\sigma_2\big)\Big\}\quad
\nn
\eeqa
in $N$, and similarly in $S$. 

$\bullet$ For $\tau=0,\ \xi=\sigma_3$ and so we get $\bA=(\sigma_3/2)\bA^D$ i.e. the monopole we started with. 

$\bullet$ For $\tau=\pi,\ \xi=-\sigma_3$, and we get 
instead $\bA=0$,
the vacuum. The energy  field strength tensor reads
\beq
F_{\theta\varphi}^\xi=\smallover1/4\sin\theta\Big\{
(1+\cos\tau)\sigma_3-\sin\tau\big(\cos(\varrho+\varphi)\sigma_1
+\sin(\varrho+\varphi)\sigma_2\big)\Big\},
\eeq
whose energy is 
\beq
E^{(\tau,\psi)}=\pi(1+\cos\tau),
\label{9.8}
\eeq
the same as  the loop energy (\ref{su2loopen}).

\begin{figure}
\begin{center}
\includegraphics[scale=1.2]{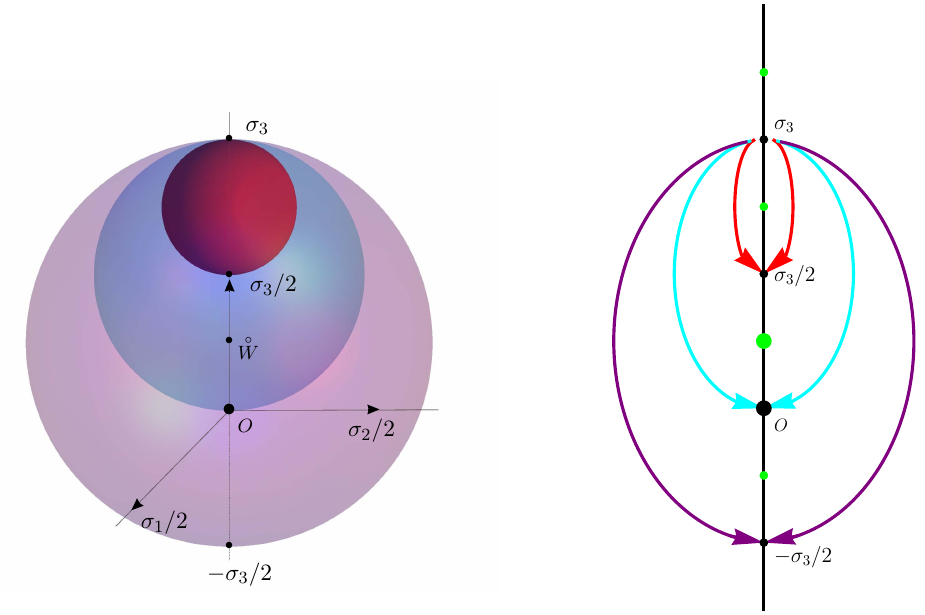}
\vspace{-3mm}
\end{center}
\caption{{\it The $\SU(2)$ monopole with
GNO charge $Q=\sigma_3$ has $\nu=6$ negative modes.
It sits at the top of $3$ energy-reducing  two-spheres whose
bottoms are the lower-lying charges $\sigma_3/2,\, 0$
and $-\sigma_3/2$.  
}}
\label{SU23S}
\end{figure}

\vskip2mm
Higher charges support more instabilities.
Fig.\ref{SU23S} shows, for example, what happens
for the GNO charge $\IQ=\sigma_3$, which  has $\nu=6$ negative modes and 
 sits at the top of $3$ energy-reducing  two-spheres. 

\kikezd{Example 2:~$H=\SO(3)$}

For $H=\SO(3)$ the topologically trivial sector is
the same as for $H=\SU(2)$. Non-trivial topology 
 arises for 
 \beq
 \oIQ{}^{(1)}=\frac{\sigma_3}{4},
\eeq
which is the unique stable charge with $m=1\in\IZ_2$.

\begin{figure}\vspace{-15mm}
\begin{center}
\includegraphics[scale=1.2]{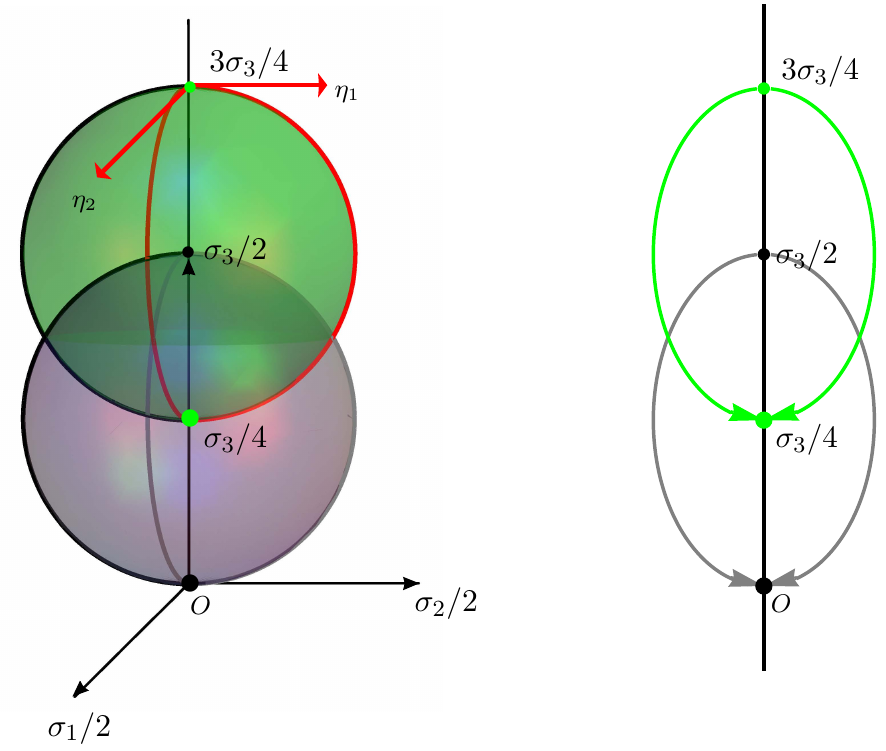}
\end{center}\vspace{-4mm}
\caption{{\it For $H=\SO(3)$ the vacuum sector is that of $\SU(2)$. In the non-trivial $m=1\in\IZ_2$ sector  the only stable monopole  has GNO
charge $\oIQ{}^{(1)}=\sigma_3/4$. The monopole with charge $\IQ=\frac{3}{4}\sigma_3$ is unstable with  $\nu=2$ negative modes, which are
 tangent to the energy-reducing two-sphere
whose top is $\IQ=\frac{3}{4}\sigma_3$ is and whose
bottom is $\oIQ{}^{(1)}$.}}
\label{SO31S}
\end{figure}

The $\SO(3)$ monopole with GNO charge 
\beq
\IQ=\frac{3}{4}\sigma_3,
\eeq
for example, is
unstable. Its $\nu=2$ negative modes are
 tangent to the energy-reducing two-sphere 
 \beq
 A_\theta^{\ \xi}=0
\qquad
{\left(A_\varphi^{\ \xi}\right)}_{\pm}=\smallover1/4
\displaystyle{(\pm1-\cos\theta)}\big(e^{-i\varphi\2\sigma_3}
\,\xi\,e^{i\varphi\2\sigma_3}+2\sigma_3\big),
\label{SO3SPHERE}
 \eeq
which interpolates  between
$\IQ=\frac{3}{4}\sigma_3$ and the stable monopole
with $\oIQ{}^{(1)}=\frac{1}{4}\sigma_3$,
see Fig.\ref{SO31S}. This sphere is in fact obtained from 
(\ref{9.7}) by shifting its bottom from the origin to
$\oIQ{}^{(1)}=\sigma_3/4$.

\goodbreak
\kikezd{Example 3:~$H=\UN(2)$}

The case 
$H=\UN(2)$ has already been studied in Section \ref{Liealgebra}.
The Bott diagram of $\UN(2)$ is as on Fig.\ref{U2Bott}. The horizontal lines are the topological sectors labelled by $m$. In each sector, 
the [up to conjugation unique] stable charge is the one which is the closest to the central $\un(1)$, namely $2\oIQ{}^{(m)}=\oQ{}^{(m)}$ in (\ref{9.2}).
\begin{figure}
\begin{center}\vskip-10mm
\includegraphics[scale=.88]{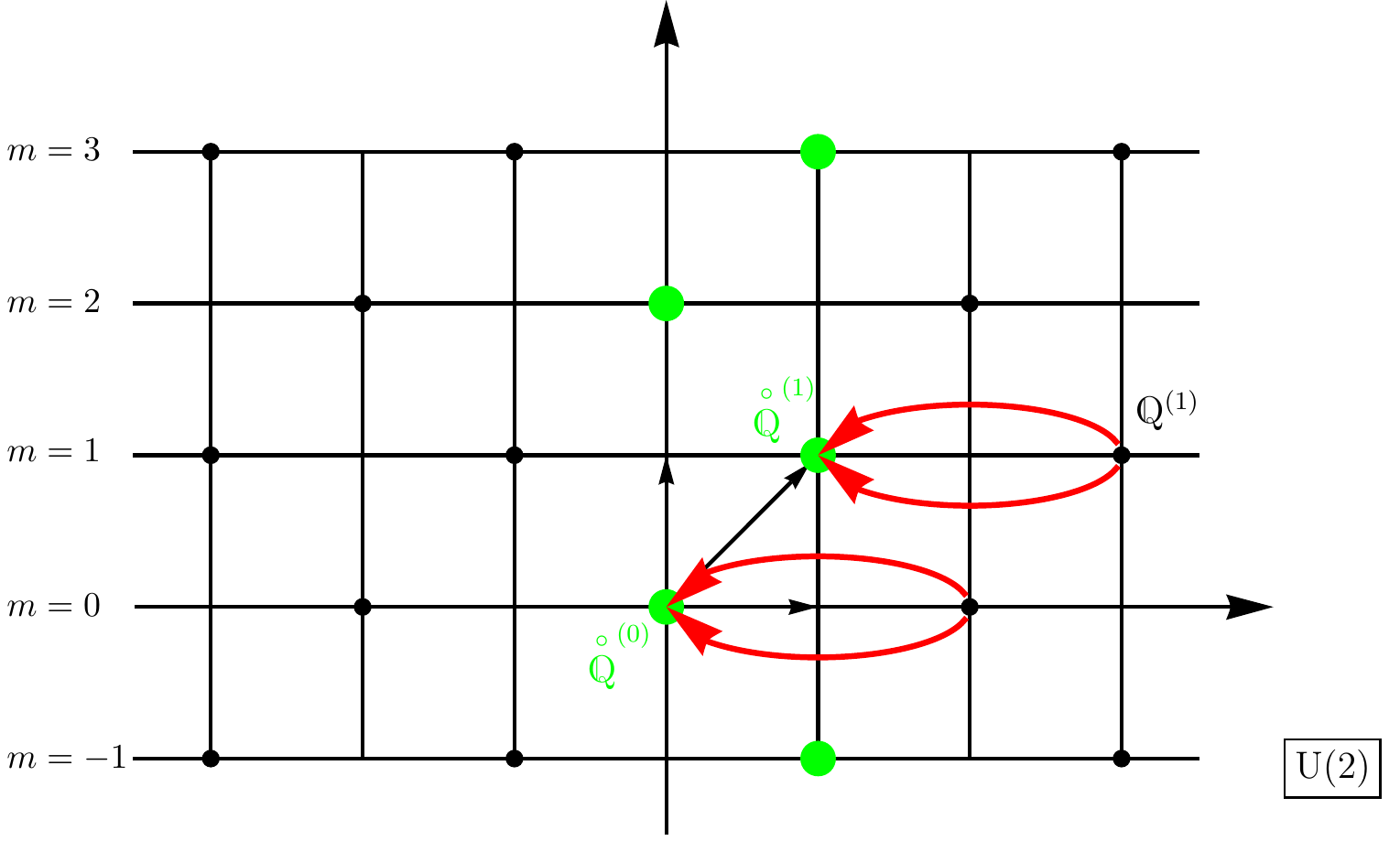}
\end{center}\vspace{-7mm}
\caption{{\it  For $\UN(2)$ 
the vacuum sector $m=0$ is that of $\SU(2)$ imbedded into of $\UN(2)$.  
The $m=2k+1$ odd sector is 
obtained from that of $m=0$
by shifting the origin to $\oIQ{}^{(m)}=\frac{m}{4}\sigma_3$.
Alternatively, it is identical to the
$m=1\in\IZ_2$ sector of $\SO(3)$.
For example, the monopole
with charge $\IQ=\frac{1}{4}{\rm diag}(3,-1)$ lies in the
$m=1$ sector and is unstable with Morse index $\nu=2$. 
An energy-reducing $2$-sphere  links it to the stable
monopole of the sector with GNO charge $\oIQ{}^{(1)}$. 
}}
\label{U2Bottbis}
\end{figure}
Any other monopole charge of  sector $m$ is 
$$
\IQ^{(m)}=\2\oQ{}^{(m)}+\frac{n}{2}P_1=\2\oQ{}^{(m)}+\;
\frac{1}{2}{\rm diag }(n,-n).
$$
Those monopoles for which $n\neq0$ are unstable with index $\nu=2(2n-1)$ for 
$m$ even, and $\nu=4n$ for $m$ odd (Fig. \ref{U2Bottbis}). 

\vskip2mm
For example,
when $G=\SU(3)$ is broken to $H=\UN(2)$ by an adjoint Higgs $\Phi$ \cite{COFN}, the vacuum sector contains a configuration whose non-Abelian charge is [conjugate to] 
\beq
\IQ={\rm diag\ }(\2,-\2,0),
\label{imbedSU2charge}
\eeq
where we considered $H=\UN(2)$ as imbedded into $G=\SU(3)$.
Note that (\ref{imbedSU2charge})  is precisely in the 
$\SU(2)$ monopole imbedded into the topologically trivial
sector of $\UN(2)$ considered in Example 1 and depicted on Fig.\ref{SU21S}. 
This configuration is therefore unstable
as conjectured in \cite{COFN}, and has indeed $2$ negative modes, namely
\beq
a_\theta^{\pm}=e^{\mp i\varphi}\,\frac{\sigma_{\pm}}{2},
\qquad
a_\varphi^{\pm}=\mp e^{\mp i\varphi}\sin\theta\,\frac{\sigma_{\pm}}{2}\,.
\label{9.4}
\eeq
tangent to  the energy-reducing two-sphere
(\ref{9.7}).

\goodbreak
\kikezd{Example 4:~$H=\UN(3)$.}

The physically most relevant example is when the Higgs little
group is $H=\UN(3)$ i.e. locally $\su(3)_c+\un(1)_{em}$, the symmetry group of  strong and electromagnetic interactions. The Bott diagram is shown on Fig. 
\ref{U3Bott}.

The topological sectors are labelled by an integer $m$; the unique stable charge in the $m$-sector is
$2{\oIQ}{}^{(m)}={\oQ}{}^{(m)}$ in (\ref{9.9}).
Any other monopole in the sector has charge 
\beq
2{\IQ}^{(m)}={\oQ}{}^{(m)}+Q'=
{\oQ}^{(m)}+n_1P_1+n_2P_2=
{\oQ}{}^{(m)}+{\rm diag}(n_1,n_2-n_1,-n_2).
\label{9.10bis}
\eeq
Those configurations with $Q'\neq0$ are unstable. 

\begin{figure}
\begin{center}
\includegraphics[scale=0.8]{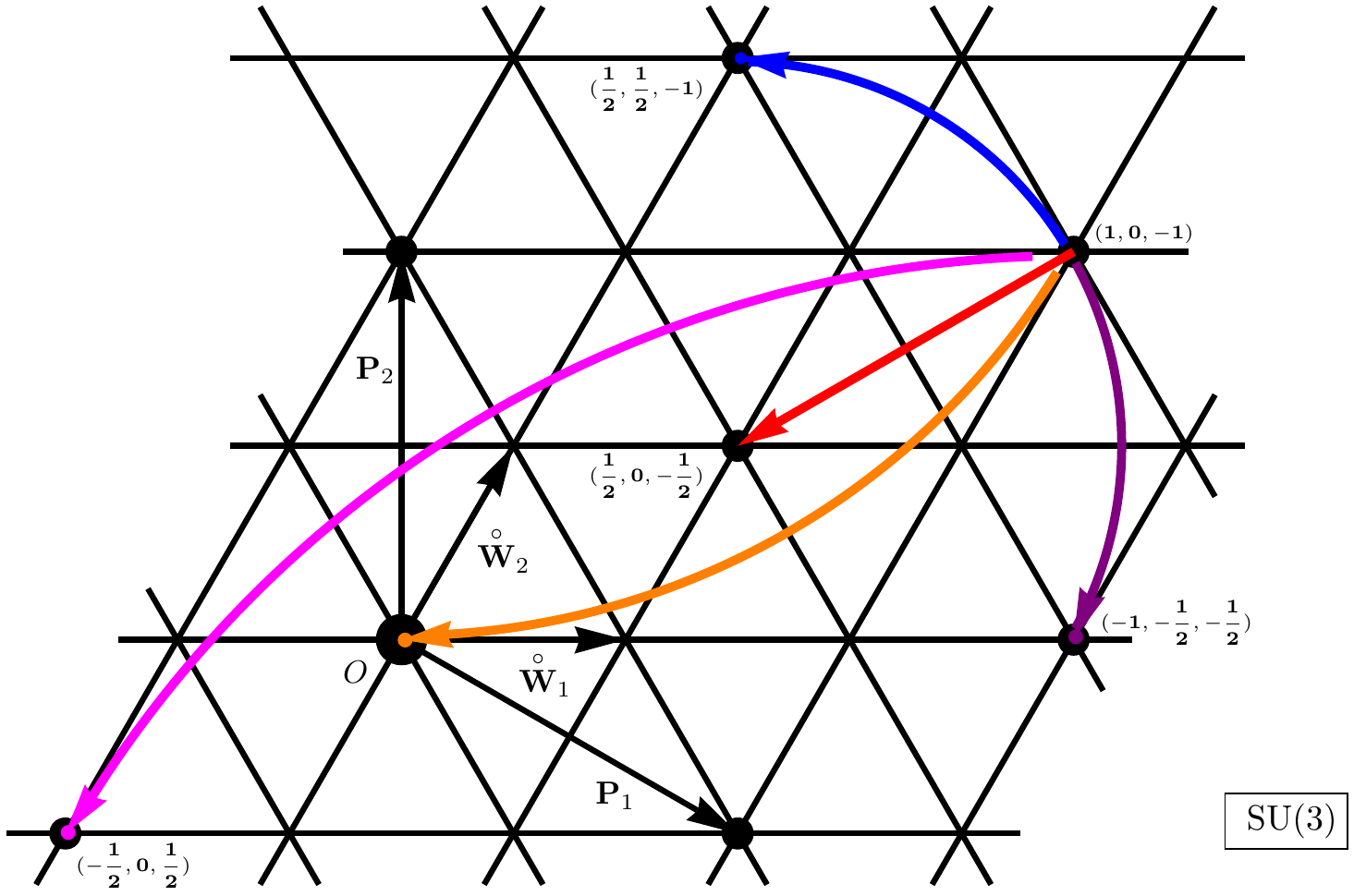}
\end{center}
\caption{\it The $H=\SU(3)$ monopole with GNO charge $\IQ={\rm diag}(1,0,-1)$ is unstable with 10 negative modes, tangent to $5$ energy-reducing spheres, which 
connect $\IQ$ to $5$ lower-energy monopoles. 
}
\label{SU3Bottbis}
\end{figure}

\vskip2mm
For example, the $\IQ={\rm diag}(1,0,-1)$ (Fig.\ref{SU3Bottbis})
belongs to the vacuum sector, because its charge is in $\gk=\su(3)$. 
$$
\alpha_1(2\IQ)=2,\quad
\alpha_2(2\IQ)=2,\quad
\theta(2\IQ)=4,
$$
and so there are $10$ negative modes, given by (\ref{6.18}).
Equation (\ref{loopsphere}) yields in turn $5$ energy-reducing
$2$-spheres, which end at
\beqa
\IQ_{\alpha_1}={\rm diag\ }(1,-1/2,-1/2),
\qquad
\IQ_{\alpha_2}={\rm diag\ }(1/2,1/2,-1),
\label{9.11}
\\[8pt]
\IQ_{\theta}^{(1)}={\rm diag\ }(1/2,0,-1/2),\quad
\IQ_{\theta}^{(2)}=0,\quad
\IQ_{\theta}^{(3)}={\rm diag\ }(-1/2,0,1/2),
\nn
\eeqa
respectively.

\vskip5mm
\goodbreak
\kikezd{Example 5:~$H=\SU(3)/\IZ_3$}

In Ref. \cite{R-DSSt} the authors consider a $6$-dimensional pure
$\SU(3)/\IZ_3$ Yang--Mills model, defined over $M^4\times\IS^2$,
where $M^4$ is Minkowski space. They claim that any
(Poincar\'e)$\times\SO(3)$ symmetric configuration is unstable against the formation of tachyons.
This is not so, however; a counterexample is given by Forg\'acs \textit{et al}. \cite{FHP84},
who show that the ``symmetry-breaking vacuum''
\beq
A_i=0,\quad i=1,\dots4,
\qquad
\bA=\smallover1/6\,{\rm diag}(2,-1,-1)\,\bA^D
\label{9.12}
\eeq
(where $\bA$ is a $2$-vector on the extra-dimensional $\IS^2$), is indeed stable.

These observations have a simple explanation: the assumption of spherical symmetry in the extra dimensions leads to asymptotic monopole configurations on $\IS^2$ with gauge group 
$H=\SU(3)/\IZ_3$. Since $\pi_1(\SU(3)/\IZ_3)\simeq\IZ_3$,
there are three topological classes corresponding to the three central elements
$z^*_0=1,\, z^*_1=e^{2\pi i/3},\,z^*_2=e^{4\pi i/3} $ of
$H=\SU(3)$ (Fig. \ref{SU3ZBott}).
\begin{figure}
\begin{center}\vspace{-2mm}
\includegraphics[scale=0.8]{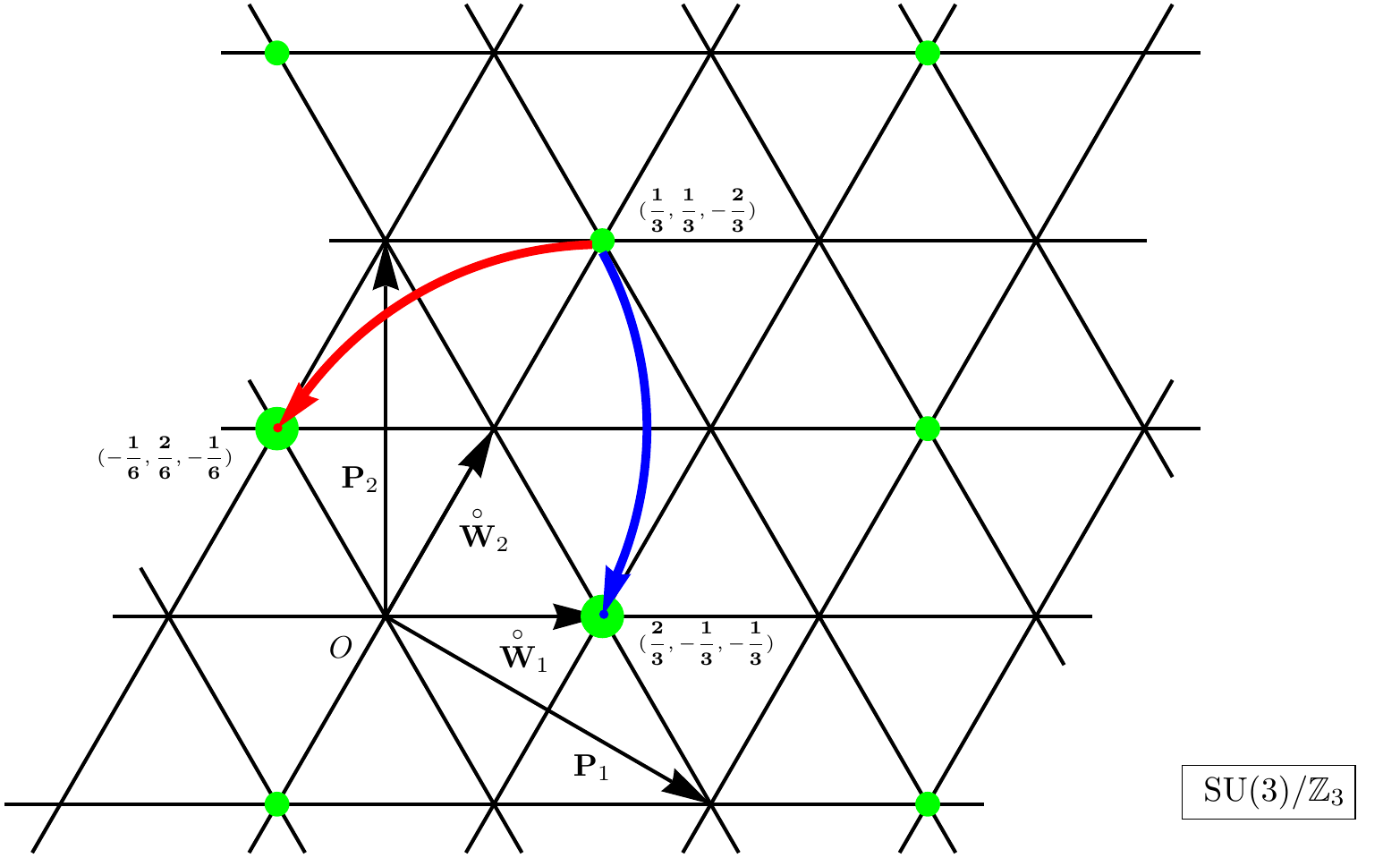}
\end{center}
\caption{{\it  The YM theory on $\IS^2$ with
gauge group $H=\SU(3)/\IZ_3)\simeq \IZ_3$ has $3$ topological sectors labelled by the central element
$z_a\in Z$ of $SU(3)$. Each sector contains exactly
one minimal charge
$\oW_j,\ j=0,1,2,$ representing the [up to conjugation] only stable 
monopole of the sector.  
The one with charge $2\IQ=2\oW_2={\rm diag}(1,-\2,-\2),$
for example, belongs to
the sector the sector labelled by $z_1=e^{4\pi i/3}$, 
and  is unstable with $4$ independent negative modes. It lies
at the top of two energy-reducing $2$-spheres whose bottoms are $\oW_1$ and its conjugate.}}
\label{SU3ZBottbis}
\end{figure}

The configuration (\ref{9.12}) is indeed stable, because it is an [asymptotic] monopole with charge 
\beq
\oIQ=\2\oW_1,
\eeq
which is the unique stable charge of the Sector characterized by $z^*_2$.

 On the other hand, any other configuration, e.g. \cite{FHP84}
\beq
\bA=\smallover1/3{\rm diag}(1,1,-2)\,\bA^D
\label{9.13}
\eeq
is unstable. Counting the intersections with the root planes shows that there are $\nu=4$ negative modes.

Both configuration (\ref{9.12}) and (\ref{9.13}) belong to the
same sector, and the construction of Sec. \ref{loops} provides us with two energy-reducing two-spheres from the
monopole (\ref{9.13}) to those
with charges  (\ref{9.12})) and its conjugate,
\beq
\smallover1/6{\rm diag}(2,-1,-1)
\qquad\hbox{and}\qquad
\smallover1/6{\rm diag}(-1,2,-1)
\eeq

Choosing instead $\oIQ=\2\oW_2$ would obviously lead
to another stable configuration.

\kikezd{Example 5:~$H={\rm Spin}(5)\simeq\SO(5)$.}

As a final example, let us consider
$H=\SO(5)$ in Example 5 of Section \ref{Liealgebra}.

It may be worth noting that, in contrast to the $K=\SU(N)$ case,
$\IQ=\2W_1$ is an \emph{unstable} monopole in the vacuum sector which has index $2(\theta(W_1)-1)=2$ \footnote{Remark that if $W_1$ was the
charge of a Prasad-Sommerfield monopole, it would be stable \cite{HORR88}.}.

The negative modes are expressed once more by (\ref{9.4}), but this time
$\sigma_{\pm}$ mean
\beq
\sigma_+=\2\barr{cccc}
0&1&&\\
0&0&&\\
&&0&-1\\
&&0&0
\earr,
\qquad
\sigma_-=\2\barr{cccc}
0&0&&\\
1&0&&\\
&&0&0\\
&&-1&0
\earr\ .
\label{9.18}
\eeq

\begin{figure}
\begin{center}\hspace{-4mm}
\includegraphics[scale=1.1]{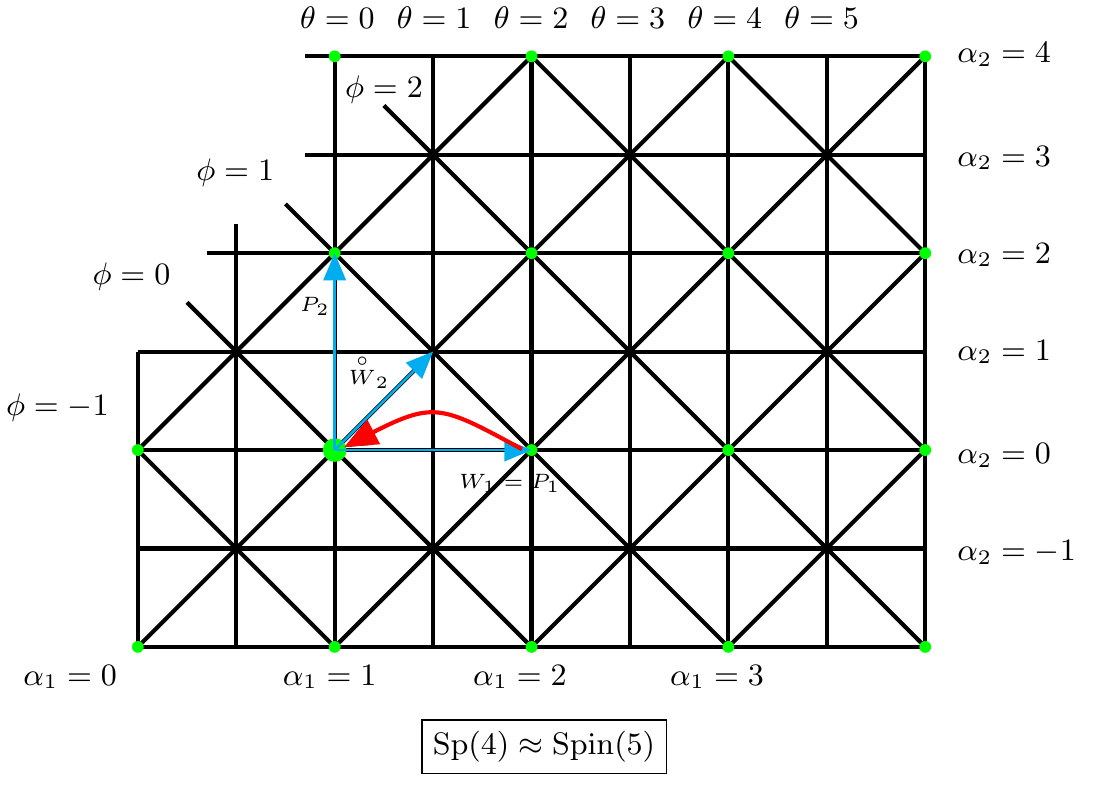}\vspace{-4mm}
\end{center}
\caption{\it For ${\rm Spin}(5)$, the two co-weights  are 
$W_1={\rm diag}(1,0,-1,0)$ and  
$\oW_2={\rm diag}(\2,\2,-\2,-\2)$ but only $\oW_2$ is minimal.  The monopole with charge
$\IQ=\2W_1$ is unstable with two negative modes. It lies on the top of an energy-reducing $2$-sphere which ends at the vacuum.
}
\label{Figure7bis}
\end{figure}

\section{Conclusion}

This review is devoted to the study of various aspects of ``Brandt--Neri--Coleman''  instability of monopoles \footnote{After  posting this review to ArXiv we became aware of a paper of A. Bais \cite{Bais} which uses similar ideas and techniques.}. Our clue is to reduce the problem to
pure YM theory on the ``sphere at infinity" with the residual group $H$
as gauge group.  Studying the Hessian we have proved the Theorem announced by Goddard an Olive \cite{GOCS}, and by Coleman \cite{COL}, which says that
\emph{each topological sector admits a unique stable monopole} whereas all
other solutions of the YM equations are unstable with an \textit{even} Morse index [=number of negative modes].

Turning to the global aspects, we have shown that to each such unstable monopole with Morse index $\nu=2n$ sits on the top of
$n$ energy-reducing two-sphere whose bottom is some lower-energy monopole \footnote{For mathematicians: these two-spheres generate the homology group $H_{2n}$ of
the space of finite-energy configurations.}.

An unstable monopole should decay by radiating away its energy. 
Describing such a process would require 
solving the time-dependent YM(H) field equations with initial
conditions close to a static solution cf. \cite{GuoWe}. 
But this is beyond the reach of present technical knowledge.

An approximate approach would be to argue first
that, under suitable conditions,  a monopole can be considered
a classical object which preserves its identity during the process; then radiation could act as a sort of effective potential.

Intuitively, our monopole
could ``roll down'' to some lower-lying critical point, from which it 
would continue to ``roll'' further down
 as a sort of ``cascade'' until it ends up at the 
 stable lowest-lying state. 
 It is tempting to figure that our energy-reducing spheres could
 provide us with possible \textit{decay routes for an unstable monopole}. 

Realizing this picture
in a physical framework is quite challenging an we have not
been able to carry it out yet. 
All what we did so far has been to construct a sort ``monopole landscape'': no dynamics has been considered.

Would like to
hint at an analogy, namely that with \emph{monopole scattering} following 
Manton's ideas \cite{MonScatt}. 
Remember first that the space of static self-dual monopoles of the
Bogomolny--Prasad--Sommerfield type \cite{GORPP,COL,BPSmon,HPAMonop,Konishi} form a finite dimensional submanifold called the \textit{moduli space}, whose dimension is  the number of independent zero modes. Every point in the moduli space labels such a solution which saturates the Bogomolny bound of the energy. All of them have therefore the same [namely the lowest possible] energy, determined by the topological charge.
 
Then the  kinetic term in the Yang--Mills--Higgs action defines a metric on the moduli space, and slowly moving monopoles follow approximately geodesics  \cite{MonScatt,MonScattSUSY}.

Intuitively, the moduli space
is  the horizontal ``bottom'' on Fig. \ref{flat}, 
and monopole scattering corresponds to a point
  ``rolling'' along such a flat direction with no resistance.  

\kikezd{Acknowledgments}.
The investigations presented in this
review started quarter of a century ago \cite{HoRa,HORR88}, and were revived
at the occasion of a Lecture presented by one of us (PAH) at the meeting {\it ``Nonlinear phenomena: a view from mathematics and physics''}, organized by the {\it National Taiwan University} and the {\it Taida Institute for Mathematical Sciences}. Taipei, Jan.  2011. We dedicate our paper to the memory of late L. O'Raifeartaigh in collaboration with whom some of the results presented here were obtained.
 P.A.H and P.M.Z are indebted to the \textit{
Institute of Modern Physics} of the Lanzhou branch of
the Chinese Academy of Sciences and to the \textit{Laboratoire de 
Math\'ematiques et de Physique Th\'eorique} of Tours University for hospitality, respectively. This work has been partially supported by the National Natural Science Foundation of 
China (Grant No. 11035006) and by the Chinese Academy of Sciences visiting 
professorship for senior international scientists (Grant No. 2010TIJ06).




\begin{thebibliography}{99}

\bibitem{tHooftPolyakov}
G.  't~Hooft, 
``Magnetic Monopoles in Unified Gauge Theories,''
Nucl. Phys. {\bf B79}, 276 (1974);
A. M. Polyakov, 
``Particle Spectrum in the Quantum Field Theory,''
JETP Lett. {\bf 20}, 194 (1974)

\bibitem{GORPP}
P. Goddard and D. Olive, 
``New Developments in the Theory of Magnetic Monopoles,''
Rep. Prog. Phys. {\bf 41} (1978) 1357

\bibitem{COL}
S. Coleman, 
``The magnetic monopole fifty years later,''
in \textit{The Unity of Fundamental Interactions},
ed. A. Zichichi (Plenum New York, 1983);
\textit{Aspects of Symmetries}. Selected Erice Lectures. Cambridge UP (1885).

\bibitem{HPAMonop}
  P.~A.~Horv\'athy,
  {\it Introduction to monopoles},
 Naples, Italy: Bibliopolis (1988).  
 Monographs and Textbooks in Physical Science. Lecture Notes, 6.

\bibitem{MaSu}
N. Manton and P. Sutcliffe,
{\it Topological solitons},
Cambridge Monographs on Mathematical Physics 
(2004) 
Book DOI: 10.1017/CBO9780511617034.

\bibitem{Konishi}
  K.~Konishi,
  ``The magnetic monopoles seventy-five years later,''
  Lect.\ Notes Phys.\  {\bf 737} (2008) 471
  [arXiv:hep-th/0702102].
   
\bibitem{BN}
R. A. Brandt and F. Neri,
``Stability analysis for singular non-Abelian magnetic monopoles,''
Nucl. Phys. {\bf B161} (1979) 253


\bibitem{HoRa}
P. A. Horv\'athy and J. H. Rawnsley, 
``On the stability of monopoles,'' in
\textit{Proceedings of the International Conference on
Differential Geometric Methods in Theoretical Physics}. 
Clausthal'1986. p.108
(World Scientific, Singapore);
 P.~A.~Horv\'athy,
 ``Monopole geography,''
Proc. {\it Non-perturbative Methods in QFT},
ed. Z. Horv\'ath, L. Palla and A. Patk\'os. 
p.39. Singapore: World Scientific (1987).
  
\bibitem{HORR88}
 P.~A.~Horv\'athy, L. O'Raifeartaigh and J.~H.~Rawnsley,
``Monopole-charge instability'',   
  Int. Journ.  Mod. Physics {\bf A3} (1988) 665-702.
 Available as arXiv:0909.2523 [hep-th].

\bibitem{GOCS}
P.~Goddard and D.~I.~Olive,
 ``The magnetic charges of stable selfdual monopoles,'' 
 Nucl. Phys. {\bf B191} (1981) 528.
   
\bibitem{FrHab}
Th. Friedrich and L. Habermann,
``Yang--Mills equations on the two-dimensional sphere,'
Commun. Math. Phys. {\bf 100} (1985) 231;

\bibitem{NAUH}
W. Nahm and K. Uhlenbeck,
``The equivalence of quantized gauge fields on $\IS^2$ and
the quantum mechanics of a particle moving on the group manifold.''.
Unpublished notes, Chicago (1988).

\bibitem{Morse}
J. Milnor, \textit{Morse Theory}, Ann. Math. Studies {\bf 51}. Princeton University Press, New Jersey (1963); R. Bott, Bull. Amer. Math. Soc. 
{\bf 7} (3) (1982) 331

\bibitem{Arafune}
  J.~Arafune, P.~G.~O.~Freund and C.~J.~Goebel,
 ``Topology of Higgs fields,''
  J.\ Math.\ Phys.\  {\bf 16} (1975) 433.
  
\bibitem{Schwarz}
A. S. Schwarz,
``Magnetic monopoles in gauge theories,''
Nucl. Phys. {\bf B112}, 358 (1976)
  
\bibitem{Taubes81}
C. H. Taubes,
``Surface integrals and monopole charges in
non-Abelian gauge theories,''
Commun. Math. Phys. {\bf 81}, 299 (1981)

\bibitem{HRCMP}
P. A. Horv\'athy and J. H. Rawnsley, 
``Topological charges in monopole theories''
Commun. Math. Phys. 
{\bf 96} (1984) 497; 
``Monopole charges for arbitrary compact gauge groups and Higgs fields in any representation''
Commun. Math. Phys. {\bf 99} (1985) 517;
  P.~A.~Horvathy and J.~H.~Rawnsley,
``Monopole invariants,''
  J.\ Phys.\ A  {\bf 20} (1987) 747.
In a rather peculiar way, the same topological charge
arises  for \emph{non-topological} 
 ``Jackiw-Pi'' vortex solutions, see 
P.~A.~Horvathy,
 ``Topology of non-topological Chern-Simons vortices,''  
Lett.\ Math.\ Phys.\  {\bf 49},  67 (1999).

  
\bibitem{Lubkin}
  E.~Lubkin,
  ``Geometric definition of gauge invariance,''
  Annals Phys.\  {\bf 23} (1963) 233.

\bibitem{GNO}
P. Goddard, J. Nuyts, and D. Olive,
``Gauge Theories and Magnetic Charge,''
Nucl. Phys. {\bf B125} (1977) 1;

\bibitem{ABOTT}
M. F. Atiyah and R. Bott, 
``The Yang--Mills equations over Riemann surfaces,'' Philos. Trans. Roy. Soc. London Ser. A 308 (1983) 523.

\bibitem{Speight}
  J.~M.~Speight,
  ``The ground state energy of a charged particle on a Riemann surface,''
  arXiv:1012.3337 [hep-th].

\bibitem{Bott}
R. Bott, in \textit{Representation Theory of Lie Groups}. Proc. Oxford'77,
ed. Atiyah (Cambridge University Press, 1979) 
p. 63.

\bibitem{Singer}
I. Singer,
``The Geometry of the Orbit Space for Non-Abelian Gauge Theories.'' Phys. Scripta {\bf 24} (1984) 817

\bibitem{MaWSTop}
N. S. Manton, 
``Topology in the Weinberg-Salam theory,''
Phys. Rev. {\bf D28} (1983) 2019;

\bibitem{Taubes83}
C. Taubes, 
``Stability in Yang--Mills theories,''
Commun. Math. Phys. 
 {\bf 91}, 235 (1983).

\bibitem{FHloops}
P. Forg\'acs and Z. Horv\'ath, 
``Topology and saddle points in field theories,''
Phys. Lett. {\bf 138B} 397 (1984);

\bibitem{Sibner}
L.~M.~Sibner, J.~Talvacchia,
 ``The Existence of nonminimal solutions of the Yang-Mills Higgs equations over R**3 with arbitrary positive coupling constant,''
  Commun.\ Math.\ Phys.\  {\bf 162}, 333-351 (1994).

\bibitem{BPSmon}
E. B. Bogomolny, 
``Stability of classical solutions,''
Sov. J. Nucl. Phys. {\bf 24} (1976) 449,
M. K. Prasad and C. M. Sommerfield, 
``An exact classical solution for the 't~Hooft monopole and the Julia--Zee dyon,'' Phys. Rev. Lett. {\bf 35} (1975) 760;

\bibitem{Dirac1931}
  P.~A.~M.~Dirac,
  ``Quantized singularities in the electromagnetic field,''
  Proc.\ Roy.\ Soc.\ Lond.\  A {\bf 133} (1931) 60.
    
\bibitem{Steenrod}
N. Steenrod,
{\it Topology of fibre bundles},
Princeton Univ. Press (1951)

\bibitem{Kobayashi}
S. Kobayashi and K. Nomizu,
{\it Foundations of differential geometry},
Vol. I N.Y. Interscience (1963);
Vol. II N.Y. Interscience (1969)

\bibitem{BottTu}
 R. Bott and L. Tu, \textit{Differential forms in algebraic topology}.
(Springer Verlag, New York, Heidelberg, Berlin 1982);

\bibitem{WuYa1} 
T.~T.~Wu and C.~N.~Yang,
``Concept of nonintegrable phase factors and global formulation of gauge fields,''
  Phys.\ Rev.\  D {\bf 12}, 3845 (1975).
  
\bibitem{WuYa2}
  T.~T.~Wu and C.~N.~Yang,
``Dirac's monopole without strings: Classical Lagrangian theory,''
  Phys.\ Rev.\  D {\bf 14} (1976) 437.
  
\bibitem{WYHPA}
P. A. Horv\'athy,
``Classical action, the Wu-Yang phase factor and prequantization,''
\hfill\break	
Proc. Int. Coll. on {\it Diff. Geom. Meths. in Math. Phys}., Aix-en Provence '79.
Ed. Souriau.\\ Springer Lecture Notes in Math. {\bf 836}, 67 (1980).

\bibitem{Sniatycki}
J. Sniatycki,
``Prequantization of charge,''
Journ. Math. Phys. {\bf 15}, 619 (1974)

\bibitem{Petry}
W. Greub and H. R. Petry,
``Minimal coupling and complex line bundles,''
Journ. Math. Phys. {\bf 16}, 1347 (1975)

\bibitem{TrautmanHopf}
A. Trautman,
``Solutions of the Maxwell and Yang-Mills Equations Associated with Hopf Fiberings,''
Int.J.Theor.Phys. {\bf 16} 561 (1977).

\bibitem{HPADirac}
  P.~A.~Horv\'athy,
``Rotational Symmetry And Dirac's Monopole,''
  Int.\ J.\ Theor.\ Phys.\  {\bf 20} (1981) 697.
  
\bibitem{TrautmanDG}
A. Trautman,
{\it Differential geometry for physicists}.
Stony Brook lectures. Monographs and Textbooks in Physical Sciences. Bibliopolis: Napoli (1984).

\bibitem{WS}
  A.~Salam,
  ``Weak And Electromagnetic Interactions,''
{\it In the Proceedings of 8th Nobel Symposium, Lerum, Sweden, 19-25 May 1968, pp. 367-377}.
 ed. Alqvist and Wicksell, Stockholm (1968);

\bibitem{GUT}
  H.~Georgi and S.~L.~Glashow,
  ``Unified weak and electromagnetic interactions without neutral currents,''
  Phys.\ Rev.\ Lett.\  {\bf 28} (1972) 1494.
  ``Partial Symmetries Of Weak Interactions,''
  Nucl.\ Phys.\  {\bf 22} (1961) 579.

\bibitem{GUTmonopoles}
  Z.~Horv\'ath and L.~Palla,
  ``Monopoles and Grand Unification Theories,''
  Phys.\ Lett.\  B {\bf 69} (1977) 197;
  ``On the structure of generalized monopole solutions in gauge theories,''
  Nucl.\ Phys.\  B {\bf 116}  (1976) 500.
\\
  C.~P.~Dokos and T.~N.~Tomaras,
 ``Monopoles and dyons in the SU(5) model,''
  Phys.\ Rev.\  D {\bf 21} (1980) 2940.

\bibitem{Humphreys}
J.E. Humphreys, \textit{Introduction to Lie Algebras and Representation Theory}.
(Berlin, Springer Verlag 1972). 
    
\bibitem{HRcolor}
P. A. Horv\'athy and J. H. Rawnsley:
``Internal symmetries of non-Abelian gauge field configurations.'' 
 Phys. Rev. {\bf D32}, 968 (1985). 

\bibitem{GOCQ}
P.~Goddard and D.~I.~Olive,
  ``Charge quantization in theories with an adjoint representation Higgs mechanism,''
 Nucl.\ Phys.\  B {\bf 191} (1981) 511;

\bibitem{WuYa3} 
T. T. Wu and C. N. Yang, 
``Dirac monopole without strings: monopole harmonics,''
Nucl. Phys. {\bf B107} (1976) 365;

\bibitem{MonScatt}
  N.~S.~Manton,
  ``A remark on the scattering of BPS monopoles,''
  Phys.\ Lett.\  B {\bf 110} (1982) 54;
  ``Monopole interactions at long range,''
  Phys.\ Lett.\  B {\bf 154} (1985) 397
  [Erratum-ibid.\  {\bf 157B} (1985) 475].
\\
  M.~F.~Atiyah and N.~J.~Hitchin,
  ``Low-energy scattering of non-Abelian magnetic monopoles,''
  Phil.\ Trans.\ Roy.\ Soc.\ Lond.\  A {\bf 315} (1985) 459.
\\
  M.~F.~Atiyah and N.~J.~Hitchin,
  ``Low-Energy Scattering Of Non-abelian Monopoles,''
  Phys.\ Lett.\  A {\bf 107} (1985) 21.
  \\
  M.~F.~Atiyah and N.~J.~Hitchin,
  ``The geometry and dynamics of magnetic monopoles.'' 
  M. B. Porter Lectures.
{\it  Princeton, USA, Univ. Press (1988)}.
  
\bibitem{Spiegel}
M. Spiegelglass, 
``Supersymmetric quantum mechanics on the sphere,''
Phys. Lett. {\bf 166B} (1986) 160.

\bibitem{BPSSUSY}
L.~Feh\'er,  P.~A.~Horv\'athy, and L.~O'Raifeartaigh,
``Applications of chiral supersymmetry 
for spin fields in self-dual backgrounds.''
 Int.\ J.\ Mod.\ Phys. {\bf A4} 5277 (1989).
  
\bibitem{MonScattSUSY}
E.~J.~Weinberg and P.~Yi,
``Magnetic monopole dynamics, supersymmetry, and duality,''
  Phys.\ Rept.\  {\bf 438} (2007) 65
  [arXiv:hep-th/0609055];
  \\
E.~J.~de Vries and B.~J.~Schroers,
  ``Supersymmetric Quantum Mechanics of Magnetic Monopoles: A Case Study,''
  Nucl.\ Phys.\  B {\bf 815} (2009) 368
  [arXiv:0811.2155 [hep-th]].

\bibitem{HPAMaslov}
P. A. Horv\'athy,
``Extended Feynman formula for harmonic oscillator.''
Int. Journ. Theor. Phys. {\bf 18}, 245 (1979);
``The Maslov correction in the semiclassical
Feynman integral.''
C. Eur. J. Phys. {\bf 9}  (2011) 1 [quant-ph/0702236].

\bibitem{FoMa}
  P.~Forg\'acs and N.~S.~Manton,
 ``Space-time symmetries in gauge theories,''
  Commun.\ Math.\ Phys.\  {\bf 72}, 15 (1980).

\bibitem{COFN}
E. Corrigan, D. Olive, D. B. Fairlie J. Nuyts, 
``Magnetic monopoles in SU(3) gauge theories''
Nucl. Phys. {\bf B106} (1976) 475.

\bibitem{R-DSSt}
S. Randjbar-Daemi, A. Salam and J. Strathdee, 
``Instability of higher dimensional Yang--Mills systems,''
Phys. Lett. {\bf 124B} (1983) 345;

\bibitem{FHP84}
P. Forg\'acs, Z. Horv\'ath and L. Palla, 
``Stable compactifying Einstein Yang--Mills systems,'' 
Phys. Lett. {\bf 147B} (1984) 311

\bibitem{Bais}
  F.~A.~Bais,
 ``To be or not to be? Magnetic monopoles in non-Abelian gauge theories,''
  In *'t Hooft, G. (ed.): 50 years of Yang-Mills theory* pp. 271-307.
  [hep-th/0407197].

 
\bibitem{GuoWe}
  H.~Guo and E.~J.~Weinberg,
``Instabilities of chromodyons in SO(5) gauge theory,''
  Phys.\ Rev.\  D {\bf 77} (2008) 105026
  [arXiv:0803.0736 [hep-th]].
  

\end{thebibliography}
\end{document}